\documentclass[aps,pra,superscriptaddress,twocolumn,showpacs,titlepage=false]{revtex4-1}
\usepackage{graphicx}
\usepackage{amsmath}
\usepackage{dsfont}
\usepackage{amssymb}
\usepackage{subfigure}
\usepackage{natbib}
\usepackage{color}
\usepackage{blkarray}
\usepackage{enumerate}
\usepackage[breaklinks=true,colorlinks=true,linkcolor=blue,urlcolor=blue,citecolor=blue]{hyperref}
\def\bra#1{{\left\langle #1 \right|}}
\def\ket#1{{\left| #1 \right\rangle}}

\definecolor{amber}{rgb}{1.0, 0.4, 0.05}
\usepackage[dvipsnames]{xcolor}

\newcommand{\bs}[1]{\boldsymbol{#1}}

\begin{document}

\title{Networked quantum sensing}
\author{T. J. Proctor}
\thanks{The first two authors contributed equally to this work. Corresponding author: tjproct@sandia.gov}
\affiliation{Sandia National Laboratories, Livermore, CA 94550, USA}
\affiliation{Department of Chemistry, University of California, Berkeley, CA 94720, USA}
\author{P. A. Knott}
\thanks{The first two authors contributed equally to this work. Corresponding author: tjproct@sandia.gov}
\affiliation{Department of Physics and Astronomy, University of Sussex, Brighton BN1 9QH, UK}
\author{J. A. Dunningham}
\affiliation{Department of Physics and Astronomy, University of Sussex, Brighton BN1 9QH, UK}
\date{\today}

\begin{abstract}
We introduce a general model for a network of quantum sensors, and we use this model to consider the question: when do correlations (quantum or classical) between quantum sensors enhance the precision with which the network can measure an unknown set of parameters? We rigorously answer this question for a range of practically important problems. When each sensor in the network measures a single parameter, we show that correlations between sensors cannot increase the estimation precision beyond what can be achieved with an uncorrelated scheme, regardless of the particular details of the estimation problem in question. We also consider the more general setting whereby each sensor may be used to measure multiple parameters, e.g., the three spatial components of a magnetic field. In this case, we show that correlations between sensors can only provide, at best, a small constant precision enhancement, over uncorrelated estimation techniques. Finally, we consider optimizing the network for measuring a single linear function of the unknown parameters, e.g., the average of all of the parameters. Here quantum correlations between the sensors \emph{can} provide a significant precision enhancement over uncorrelated techniques, and this enhancement factor scales with the number of sensors. To illustrate the broad implications of this work, we apply our results to a wide range of estimation problems of practical interest, including multi-mode optical interferometry, networks of atomic sensors, and networks of clocks. Our findings shed light on a number of results in the literature, provide a rigorous general framework for future research on networked quantum sensors, and have implications for both quantum multi-parameter estimation theory, and quantum sensing technologies.
\end{abstract}
\maketitle

\section{Introduction}
Networked quantum information is an integral part of the quantum technology revolution. Quantum computers with a network architecture are highly promising for realizing a scalable quantum computer \cite{kimble2008quantum,nickerson2013topological}, and quantum cryptography networks already span cities \cite{Sasaki2011field} and are now being tested on a global scale using satellites \cite{wang2013direct}. It has also been suggested that quantum sensing and metrology may benefit from a spatially distributed network architecture employing entangled states and measurements \cite{komar2014quantum,eldredge2016optimal}, with potential applications to mapping magnetic fields \cite{steinert2010high,hall2012high,pham2011magnetic,seo2007fourier,baumgratz2016quantum}, phase imaging \cite{humphreys2013quantum,liu2014quantum,yue2014quantum,knott2016local,gagatsos2016gaussian,ciampini2015quantum} and precision clocks \cite{komar2014quantum}. However, although there have been a variety of results relating to the usefulness of entangled states or measurements for enhancing precision in multi-parameter estimation (MPE) \cite{humphreys2013quantum,liu2014quantum,yue2014quantum,ciampini2015quantum,knott2016local,gagatsos2016gaussian,kok2017role,baumgratz2016quantum,szczykulska2016multi}, there is currently no general framework that demonstrates when such resources are advantageous. Given the immense challenges faced in the creation and manipulation of entangled states, a complete understanding of when entanglement is (and is not) critical to optimizing estimation precision is of paramount importance.

  \begin{figure}
\includegraphics[width=5.5cm]{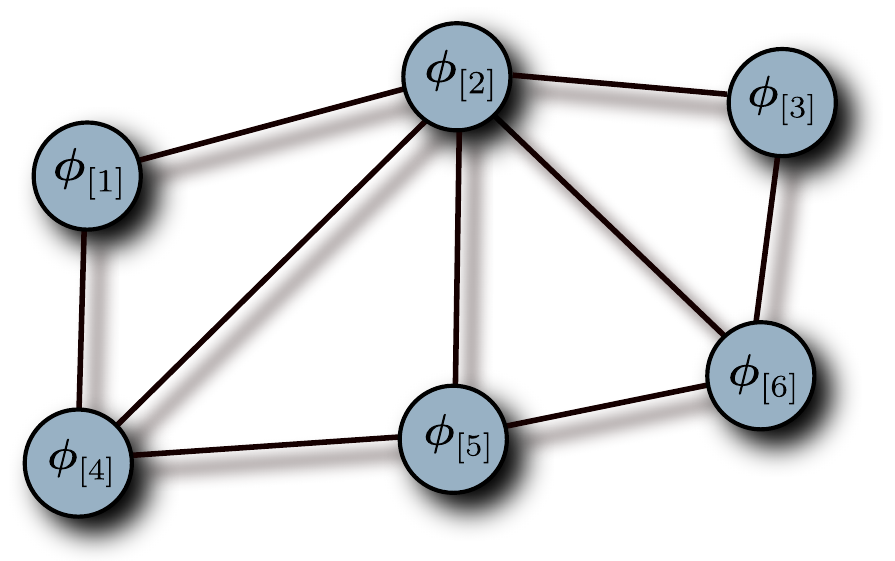}
\caption{A network of quantum sensors. Each node in the diagram represents a generic ``sensor'' (e.g., an ensemble of two-level atoms or an optical mode) into which the parameter denoted there is encoded via a local unitary evolution. Each locally-encoded parameter could be either (1) a vector, e.g., the three spatial components of a magnetic field, or (2) a scalar parameter, e.g., an optical phase shift. The connections between the nodes denote that, in general, the initial state in which the network is prepared can be entangled between the sensors. Moreover, the measurement of the sensors (to determine the parameters) may project onto a state that is entangled between sensors. }
\label{fig:quantum-sensing-networks}
\end{figure}

A number of results have began to probe the role of entanglement, but no overall conclusions can yet be drawn. For example, for the task of measuring $d$ optical phase differences, it has been shown that a $d$-fold enhancement in the estimation precision, over the well-known Heisenberg limit, can be obtained if $d$-mode entangled states are input into the multi-mode interferometer \cite{humphreys2013quantum,liu2014quantum,yue2014quantum}. However, very recently it has been shown that an equal estimation precision is available with equivalent mode-separable states \cite{knott2016local}, suggesting that mode-entanglement is perhaps not the source of the enhanced estimation precision demonstrated in Refs.~\cite{humphreys2013quantum,liu2014quantum,yue2014quantum}. Moreover, in other MPE applications it is known that entanglement can be detrimental, such as when measuring coupled phases \cite{kok2017role}. Similarly, although entanglement is useful for estimating a multi-dimensional field, too much entanglement can be detrimental to the estimation precision \cite{baumgratz2016quantum}.

In this paper we investigate the circumstances under which entangled states and/or entangled measurements can provide enhancements in the estimation precision for generic MPE problems using a ``network of quantum sensors''. In our model, $d$ unknown parameters are unitarily encoded into a set of quantum systems (the sensors). We may wish to estimate the unknown parameters themselves, or alternatively we may wish to estimate some function of the parameters, such as the average. In both scenarios, we provide a rigorous analysis of exactly when quantum or classical correlations (e.g., entanglement) \emph{between} the sensors can provide an enhancement in the precision. This is illustrated in Fig.~\ref{fig:quantum-sensing-networks}, and we introduce the model more formally in Section~\ref{sec:basic_formalism}. Our construction encompasses many of the quantum MPE problems that have already been studied, e.g., multi-mode optical sensing \cite{humphreys2013quantum,liu2014quantum,yue2014quantum,knott2016local,gagatsos2016gaussian,ciampini2015quantum}, networks of clocks \cite{komar2014quantum}, and estimating parameters in certain many-qubit Hamiltonians \cite{eldredge2016optimal}.
 
Denoting the collection of $d$ unknown parameters by the vector $\bs{\phi} = (\phi_1,\phi_2,...,\phi_d) $, a very natural problem is that of estimating the elements of $\bs{\phi}$ to the best precision possible (we consider this in Section~\ref{sec:est-phi}). For example, we may wish to use a network of sensors to map an unknown field. We first look at the case in which each sensor is encoded with a single parameter only, i.e., parameter $\phi_k$ is encoded into sensor $k$. In this setting we demonstrate that initial states or measurements that are entangled \emph{between sensors} are detrimental to the parameter estimation precision. 

We then consider estimation problems in which each sensor is encoded with more than one parameter, i.e., sensor $k$ is now encoded with a vector of parameters, denoted $\phi_{[k]}$. We show that, whenever the generating operators for all of the parameters commute, the conclusion above still holds: entanglement \emph{between sensors} reduces parameter estimation precision. Hence, in \emph{any} set of systems in which parameters are encoded locally by generators that all mutually commute, neither quantum nor classical correlations between the systems can directly enhance the precision with which the parameters can be measured. This has significant implications for multi-mode optical sensing \cite{humphreys2013quantum,liu2014quantum,yue2014quantum,ciampini2015quantum,knott2016local,gagatsos2016gaussian}, and it substantially strengthens the very recent optics-specific results of these authors and others \cite{knott2016local} (see Section~\ref{sec:optics}). 
 
In some important estimation problems the parameter generators do not commute \cite{baumgratz2016quantum,ballester2004estimation}, e.g., if each sensor is estimating the local magnitude and direction of an unknown magnetic field. Here we demonstrate that the ideal estimation strategy in this scenario is slightly more subtle, and depends on the relevant definition of ``resources'' for the problem at hand (in quantum sensing, we are normally interested in optimizing the estimation precision as a function of some resource, such as the total number of particles used). We show that the estimation precision-per-resource is either always improved by using separable probe states and local measurements or is, at worst, only reduced by a factor of two, in comparison to any strategy using entangled states and measurements. This complements the recent work of Baumgratz and Datta \cite{baumgratz2016quantum}, who have shown that entanglement is helpful for estimating a multi-dimensional field: in the language used herein, they have shown that entanglement \emph{within} a sensor can be beneficial for measuring a set of parameters encoded into that sensor via non-commuting generators. The relationship between our results and those of Ref.~\cite{baumgratz2016quantum} are discussed in Section~\ref{sec:atomic_sensing}.

In some sensing problems it may not actually be necessary to estimate the elements of $\bs{\phi}$. Instead, the objects of interest could be some functions of the elements of $\bs{\phi}$, such as the average of all of the $\phi_i$, or the difference between adjacent elements (e.g. in a two-mode interferometer the parameter of interest is the phase difference between the two paths). In these cases the aim would be to optimize the sensing network for estimating these functions to the best precision possible.

In Section~\ref{Sec:functions} we consider the problem of estimating arbitrary linear functions of $\bs{\phi}$. In this setting, entangled states and measurements can, in many cases, provide significant enhancements over the best possible estimation precision that can be obtained without the aid of quantum correlations between sensors. Moreover, the enhancement factor can scale with the number of sensors. We largely focus on the problem of estimating a \emph{single} linear function of $\bs{\phi}$. Here, we show that the maximal enhancement obtainable via entanglement depends on the form of this linear function, and it appears to be greatest when the function is an equally weighted sum of the elements of $\bs{\phi}$ (e.g., an average). We will relate our conclusions to some well-known single-parameter quantum metrology results, and also to the recent MPE work of Eldredge \emph{et al.}~\cite{eldredge2016optimal} on linear function estimation.

Our main results are quite general and are presented with the minimum assumptions about the details of the estimation problem of interest. Hence, we explicitly demonstrate the practical relevance of our findings by applying them to a variety of problems of experimental interest, with a particular emphasis on important problems in multi-mode optical interferometry (Section~\ref{sec:optics}) and atomic sensing (Section~\ref{sec:atomic_sensing}). Sections~\ref{sec:optics} and~\ref{sec:atomic_sensing} are both reasonably self-contained, with readers interested exclusively in either of these topics in mind.

\section{Multi-parameter estimation}\label{MPE-intro}
We begin with a review of the relevant material from classical and quantum estimation theory. 

\subsection{The classical Cram\'er-Rao bound}
Consider the problem of estimating an unknown $d$-element vector $\bs{\phi}= (\phi_{1},\phi_{2},\dots,\phi_{d})^T  \in \mathbb{R}^d$ from a sequence of data $\bs{m}=(m_1,....,m_{\mu})\in \mathbb{R}^\mu$, where each data point has been drawn from the probability density function (PDF) $p(m|\bs{\phi})$. An \emph{estimator} of $\bs{\phi}$ based on the data $\bs{m}$, denoted $\bs{\Phi} \in \mathbb{R}^d$, has a covariance matrix defined by \cite{kay1993fundamentals,helstrom1976quantum}
\begin{equation}
\text{Cov}(\bs{\Phi}) := \mathbb{E}\left[ \left(\bs{\Phi} -  \mathbb{E}[ \bs{\Phi} ]\right)\left(\bs{\Phi} -  \mathbb{E}[\bs{\Phi}]\right)^T \right] ,
\end{equation}
where $\mathbb{E}(V)$ denotes the expectation value of the random variable $V$.

 The covariance matrix is a common figure of merit for the precision of the estimator for $\bs{\phi}$ \cite{kay1993fundamentals,helstrom1976quantum,humphreys2013quantum,liu2014quantum,yue2014quantum,ciampini2015quantum}. Any locally unbiased \footnote{An estimator, $\hat{\bs{\phi}}$, is a locally unbiased estimator for $\bs{\phi}$ if $ \mathbb{E}[\hat{\phi}_l ] = \phi_{l}$ \cite{helstrom1976quantum}.} estimator for $\bs{\phi}$ has a covariance matrix that obeys the (classical) Cram\'er-Rao bound (CRB) \cite{kay1993fundamentals,helstrom1976quantum}
\begin{equation}
\text{Cov}(\bs{\Phi})  \geq \frac{F^{-1}}{\mu} ,
 \end{equation}
 where $\mu$ is the number of independent data points and $F^{-1}$ is the inverse of the (classical) Fisher information matrix (FIM), $F$, for the PDF $p(m|\bs{\phi})$. The FIM for $p(m|\bs{\phi})$ is defined by \cite{kay1993fundamentals,helstrom1976quantum}
\begin{equation}F_{kl} := \mathbb{E}\left[\frac{\partial  \ln p(m| \bs{\phi}) }{\partial \phi_{k}} \frac{\partial  \ln p(m| \bs{\phi}) }{\partial \phi_{l}}\right] ,
\end{equation}
and it is positive semi-definite \cite{ciampini2015quantum}. 

The CRB holds under the assumption that the FIM is invertible, which is when it is positive definite. In the CRB and throughout this paper the relation $A\geq B$ ($A>B$), between two matrices $A$ and $B$, should be interpreted to mean that $A-B$ is a positive semi-definite (positive definite) matrix. The CRB can be saturated in the limit of an asymptotic amount of data (large $\mu$) with an appropriate choice of estimator, e.g., the maximum likelihood estimator \cite{vallisneri2008use}.

\subsection{The quantum Cram\'er-Rao bound}
\label{Sec:QCRB}
In quantum multi-parameter estimation (MPE) theory the parameter to be estimated is encoded in a quantum state $\rho_{\bs{\phi}}$, which may in general be mixed. A measurement procedure on $\rho_{\bs{\phi}}$ is described by some positive-operator valued measure (POVM), $M$, which is a set of positive operators $M=\{\Pi_m\}$ such that $\int d m \, \Pi_m = \mathds{1}$. The probability of the outcome $m$ given $\bs{\phi}$ is the PDF $p(m|\bs{\phi})=\text{Tr} ( \rho_{\bs{\phi}} \Pi_m)$. Hence, the classical FIM and associated CRB can be calculated from this PDF, with the resulting precision bound for an estimator of $\bs{\phi}$ depending on both the state and the measurement procedure used.

The quantum Fisher information matrix (QFIM) for a quantum state, denoted $\mathcal{F}$, is defined by \cite{fujiwara1995quantum,matsumoto2002new,helstrom1976quantum,paris2009quantum,helstrom1967minimum}
\begin{equation}
\mathcal{F}_{l m} := \frac{1}{2} \text{Tr} [ \rho_{\bs{\phi}} \hat{L}_{l}\hat{L}_{m} +  \rho_{\bs{\phi}} \hat{L}_{m}\hat{L}_{l}],
\end{equation}
where the $\hat{L}_{l}$ are the symmetric logarithmic derivatives (SLDs). The $l$\textsuperscript{th} SLD is defined implicitly by \cite{fujiwara1995quantum,helstrom1967minimum}
\begin{equation}
\frac{\partial \rho_{\bs{\phi}} }{\partial \phi_{l}}  = \frac{1}{2} (\rho_{\bs{\phi}} \hat{L}_{l} + \hat{L}_{l} \rho_{\bs{\phi}}). 
\end{equation}
The QFIM is a real, symmetric and positive semi-definite matrix \cite{paris2009quantum} and, for a given state $\rho_{\bs{\phi}}$, it is always at least as large as any FIM obtained from $\rho_{\bs{\phi}}$. That is, $\mathcal{F} \geq F$ where $F$ is found with respect to any POVM. This directly implies that $\mathcal{F}^{-1} \leq  F^{-1}$, where $F$ is found with respect to any POVM.

Hence, the QFIM may be used to bound the covariance matrix of any unbiased estimator for $\bs{\phi}$, irrespective of the particular measurement procedure. This is known as the \emph{quantum} Cram\'er-Rao bound (QCRB) and is given by \cite{fujiwara1995quantum,matsumoto2002new,helstrom1976quantum,paris2009quantum,helstrom1967minimum}
\begin{equation}
\text{Cov}(\bs{\Phi}) \geq \frac{F^{-1}}{\mu} \geq \frac{\mathcal{F}^{-1}}{\mu}  ,
\label{Eq:QCRB}
 \end{equation}
 where $\mathcal{F}$ is the QFIM for $\rho_{\bs{\phi}}$ and $\mu$ is the number of independent repetitions of the experiment. 
 
For single-parameter estimation ($d=1$) a measurement can always be found that saturates the QCRB in the limit of large $\mu$ \cite{braunstein1994statistical,paris2009quantum,demkowicz2014quantum}. However, this is not the case when $d>1$, as there is not necessarily any POVM where the FIM of $\rho_{\bs{\phi}}$ with respect to this POVM is such that $F =\mathcal{F}$ \cite{fujiwara1995quantum,vidrighin2014joint,helstrom1976quantum}. Hence, for the multi-parameter QCRB it is important to assess whether this bound can be saturated in each problem of interest. A necessary and sufficient condition for QCRB to be saturable (with large $\mu$) is that
\begin{equation} 
\text{Tr}(\rho_{\bs{\phi}} [L_{l},L_{m}])=0, 
\label{sat-QCRB-cond}
\end{equation}
for all $l$ and $m$ \cite{fujiwara2001estimation,ragy2015resources,ragy2016compatibility}. When this holds, all $d$ parameters may be simultaneously estimated at the optimal precision allowed by the QCRB.

\subsection{Importance weighting}
\label{Sec:weight-resource}
In MPE with $d>1$, there are quantum states that cannot be unambiguously ranked in terms of their QFIM for estimating $\bs{\phi}$. This is in the sense that there are $\rho_{\bs{\phi}}$ and $\rho'_{\bs{\phi}}$ with associated QFIMs, $\mathcal{F}$ and $\mathcal{F}'$, such that neither $\mathcal{F} \leq \mathcal{F}'$ nor $\mathcal{F}' \leq \mathcal{F}$ \footnote{For example, in a problem with $d>1$ it is possible for two states to have diagonal and positive QFIMs, $\mathcal{F}$ and $\mathcal{F}'$, with $\mathcal{F}_{11}>\mathcal{F}_{11}' $ and $\mathcal{F}_{22}'>\mathcal{F}_{22}$. This implies that neither $\mathcal{F}-\mathcal{F}'$ nor $\mathcal{F}'-\mathcal{F}$ is positive semi-definite.}. One way to fully specify the problem of interest is to introduce a $d\times d$ diagonal, real and positive semi-definite \emph{weighting matrix}, denoted $W$, and consider minimizing the scalar quantity \cite{genoni2013optimal,vaneph2013quantum,fujiwara1995quantum}\footnote{In the literature the weighting matrix is not necessarily restricted to being diagonal (e.g., see \cite{genoni2013optimal,vaneph2013quantum}). In this paper we consider only this case and estimating functions of parameters is considered explicitly using the Jacobian formalism.}
\begin{equation}
E_{\bs{\Phi}}:= \text{Tr}(W\text{Cov}(\bs{\Phi})) = \sum_{l=1}^{d} W_{ll}\text{Cov}(\bs{\Phi})_{ll}.
 \end{equation}
This then specifies the relative importance assigned to minimizing the variance for the estimator of each parameter (the variances are the diagonals of the covariance matrix). 

To avoid trivial $W$, such as $W=0$, and trivial differences between alternative choices for $W$, such as $W$ and $W'$ with $W=cW'$ for scalar $c$, we demand that $\text{Tr}(W)=1$. Minimizing $E_{\bs{\Phi}}$ for a given $W$ will be considered the figure of merit for the precision of the parameter estimation herein. The QCRB of Eq.~(\ref{Eq:QCRB}) implies that \footnote{The diagonal elements of a positive semi-definite matrix are non-negative \cite{bhatia2009positive,bobrovsky1987some}. This implies that the diagonal elements of the covariance matrix are bounded below by the diagonal elements of the inverse QFIM, i.e., $\text{Cov}_{kk} \geq [\mathcal{F}^{-1}]_{kk} /\mu$. Hence $ \text{Tr}(W\text{Cov}) \geq \text{Tr}( W \mathcal{F}^{-1}) / \mu$ for any weighting matrix, $W$.}
\begin{equation}
E_{\bs{\Phi}}= \text{Tr}(W\text{Cov}(\bs{\Phi})) \geq \frac{1}{\mu}\text{Tr}( W \mathcal{F}^{-1}) ,
\label{Eq:error-QCRB}
  \end{equation} 
and hence $E_{\bs{\Phi}}$ is bounded by a weighted sum of the diagonal elements of the inverse QFIM. $W\propto \mathds{1}$ represents the situation in which estimating each of the $d$ parameters is considered to be equally important, and this has often been considered as a figure of merit in the quantum MPE literature (sometimes implicitly), e.g., see \cite{humphreys2013quantum,liu2014quantum,yue2014quantum,ciampini2015quantum,knott2016local}. However, we will largely refrain from restricting ourselves to this particular sub-case.

\subsection{Unitary estimation problems}
\label{sec:uni-est}
In this paper we will consider estimation problems in which the unknown parameters are imprinted via a unitary evolution $U(\bs{\phi})$ that acts on an initial state $\rho$ of an experimentalist's choosing. Let
\begin{equation} 
U(\bs{\phi})=\exp (-i  \bs{\phi}^{T} \hat{\bs{H}} ),
\end{equation}
with $\hat{\bs{H}} = (\hat{H}_1,\dots,\hat{H}_d)^T$ for some Hermitian operators. The parameters to be estimated will be some functions of $\bs{\phi}$, e.g., we may wish to estimate $\bs{\phi}$ itself, or just the average of the elements of $\bs{\phi}$.  

Define the \emph{generator} of $\phi_k$ by $ \hat{G}_k := -i (\partial U^{\dagger} / \partial_{\phi_k}) U$. When the elements of $\hat{\bs{H}}$ all mutually commute, that is when $[\hat{H}_k,\hat{H}_l]=0$ for all $k$ and $l$, then $\hat{G}_k=\hat{H}_k$, but more generally $\hat{G}_k \neq \hat{H}_k$. Moreover, for mutually commuting $\hat{H}_k$, then the QCRB may always be saturated, as it may be shown that Eq.~(\ref{sat-QCRB-cond}) holds \cite{matsumoto2002new}. However, note that the QCRB can also be saturated outside of this setting \cite{fujiwara2001estimation,baumgratz2016quantum}.
  
The generic estimation procedure is that some measurement is performed on the $U(\bs{\phi})$-evolved input state, this entire experiment is repeated many times, and finally an estimate of the parameters of interest is obtained. Hence, an \emph{estimation strategy} can be considered to consist of choosing a probe state, $\rho$, a measurement to be performed on $\rho_{\bs{\phi}} = U(\bs{\phi}) \rho U(\bs{\phi})^{\dagger}$, and an estimator for $\bs{\phi}$ from the data.

\section{A general model for networked quantum sensors}
\label{sec:basic_formalism}
We now introduce our general model of \emph{networked quantum sensors}. This model encompasses many important problems in quantum parameter estimation, which we will analyze throughout this paper.

\subsection{A network of quantum sensors}
\label{sec:network}  
Consider a set of quantum subsystems, $\mathbb{S}=\{1,2,\dots, |\mathbb{S}|\}$, where the $k$\textsuperscript{th} subsystem has the Hilbert space $\mathcal{H}_k$. We do not assume each Hilbert space is identical, nor that each Hilbert space is associated with one ``particle''. E.g., $\mathcal{H}_k$ could be the Hilbert space of one or more field modes, a single two level atom, or an arbitrary number of two-level atoms. The total Hilbert space is
\begin{equation}
 \mathcal{H}_{\mathbb{S}} = \mathcal{H}_1 \otimes  \mathcal{H}_2 \otimes \cdots \otimes \mathcal{H}_{|\mathbb{S}|}.
 \label{Hilb-1}
 \end{equation}
We will refer to the physical system associated with each Hilbert space $\mathcal{H}_k$ as a \emph{quantum sensor}, and the entire set of sensors as a \emph{quantum sensing network}.

Now, consider any unitary $U(\bs{\phi})=\exp(-i \bs{\phi}^T\hat{\bs{H}})$ that acts on $\mathcal{H}_{\mathbb{S}}$, with $ \hat{\bs{H}}=(\hat{H}_1,\dots,\hat{H}_d)$ and where each $\hat{H}_l$ is a Hermitian operator that acts non-trivially on only \emph{one} of these Hilbert spaces. That is, denoting the non-trivial action of $\hat{H}_l$ by $\hat{h}_l$, then  $\hat{H}_l$ has the form
\begin{equation}
\hat{H}_l =  \mathds{1} \otimes \mathds{1} \otimes \cdots \otimes \hat{h}_l \otimes \cdots  \otimes \mathds{1} ,
\label{H_k-com-gen}
 \end{equation}
where $\mathds{1}$ denotes the identity operators of the appropriate dimensions. Hence, we have that
\begin{equation}
\label{unitary_separable}
 U(\bs{\phi}) = U_1\big(\bs{\phi}_{[1]}\big)\otimes U_2\big(\bs{\phi}_{[2]}\big)\otimes \cdots \otimes  U_{|\mathbb{S}|}\big(\bs{\phi}_{[|\mathbb{S}|]}\big) ,
 \end{equation}
where $\bs{\phi}_{[k]}$ is some $d_k$-dimensional sub-vector of $\bs{\phi}$ with $d_1+d_2 +\dots + d_{|\mathbb{S}|} = d$, and therefore the $k$\textsuperscript{th} sensor has $d_k$ parameters unitarily encoded into it.

Due to the tensor product structure, an evolution of this sort on $\mathcal{H}_{\mathbb{S}}$ may be viewed as a model for a collection of spatially distributed sensors, with unknown parameters encoded locally into each sensor. This is demonstrated schematically in Fig.~\ref{fig:quantum-sensing-networks}. This is the motivation for the ``quantum sensing network'' terminology that we use for our model, but this framework encompasses any sensing problem with the tensor product structure introduced above. As such, it also encompasses sensing systems which might not normally be termed ``sensing networks''. Throughout this paper, we will be considering MPE problems that are within this very general framework.

In quantum sensing and metrology, the aim is normally to minimize the uncertainty in the estimate of the parameters of interest for a given amount of ``resources''. In the majority of practical estimation problems this resource can be defined as the expectation value, taken with respect to the the initial state of the sensors $\rho$, of some Hermitian operator $\hat{R}$. That is, the resources used in a single experiment of a sensing protocol may be considered to be
    \begin{equation}
  \langle  \hat{R} \rangle = \text{Tr}(\rho \hat{R} ) .
    \end{equation}
By performing $\mu$ classical repeats of the experiment, the total resources are then $\mu   \langle  \hat{R} \rangle$.
For the majority of problems of practical interest we have that
    \begin{equation} 
    [\hat{R},U(\bs{\phi})]=0 ,
\label{RUcom}
    \end{equation}
 for all possible $\bs{\phi}$. This is a very natural condition for the resource operator to satisfy, as otherwise the resources in the output state are not equal to those in the input, and they will depend on the actual value of $\bs{\phi}$. In this paper, we will only consider MPE problems in which Eq.~(\ref{RUcom}) holds.

This definition for the resources used in an estimation may seem rather abstract. A natural example is when the total amount of resources used in the $|\mathbb{S}|$-sensor state is the sum of local properties of the sensors, that is
 \begin{equation} 
 \hat{R}_{\text{sum}} = \hat{R}_1+ \hat{R}_2 + \dots+ \hat{R}_{|\mathbb{S}|},
\end{equation}
where $\hat{R}_k$ acts non-trivially only on the $k$\textsuperscript{th} sensor, and where each $\hat{R}_k$ commutes with $\hat{H}_l$ for all $l$. Physically relevant specific examples are discussed below. Alternatively, in some situations we may only be concerned with finding the best state to optimize the precision with no consideration for any ``resources'' used. Estimation problems of this sort can be encoded in our formalism by the trivial case of $\hat{R}=\mathds{1}$.

The most obvious estimation problem within our framework is the optimization of the network for estimating $\bs{\phi}$ to the best precision. We will consider this MPE problem, but other problems are also of practical interest. The general estimation problem we consider herein is that of choosing a strategy to minimize the uncertainty in the estimates of some linear functions of $\bs{\phi}=(\phi_1,\dots,\phi_d)^T$, for some arbitrary amount of resources. That is, the aim is to estimate an $m\leq d$ dimensional vector $\bs{\theta} = (f_1(\bs{\phi}), f_2(\bs{\phi}), \dots, f_m(\bs{\phi}))$ where $f_k$ is a linear function of $\phi_1$, $\phi_2$, $\dots$, and $\phi_d$, and we wish to find the strategy that minimizes $E_{\bs{\Theta}}$ for a given weighting matrix, $W$, and a certain amount of resources. This encompass a range of problems of practical interest, such as: measuring the strength of a magnetic or electric field at a range of locations via atomic sensors \cite{steinert2010high,hall2012high,pham2011magnetic,seo2007fourier,baumgratz2016quantum} (see Section~\ref{sec:atomic_sensing}), networked clocks \cite{komar2014quantum} and Hamiltonian estimation \cite{eldredge2016optimal}, or estimating functions of optical linear or non-linear phase shifts on a set of $d$ modes \cite{humphreys2013quantum,liu2014quantum,yue2014quantum,ciampini2015quantum,knott2016local,gagatsos2016gaussian} (see Section~\ref{sec:optics}). 

For clarity, we now briefly describe how two problems of practical interest can be described as sensing networks. For multi-mode optical MPE \cite{humphreys2013quantum,liu2014quantum,yue2014quantum,ciampini2015quantum,knott2016local,gagatsos2016gaussian}, each optical mode can be considered to be a ``sensor'', the generator of each parameter is, in many cases, the number operator, and the resource is the total number of photons in all of the modes. In distributed magnetic field sensing with two (or more) level atoms, each sensor is an ensemble of atoms, for mapping out a one-dimensional field the parameters are generated by the collective spin operator around some axis, and the standard resource function is the total number of atoms (given a fixed time of evolution \cite{huelga1997improvement}). This can be expressed as the expectation value of a matter-excitation number operator using an appropriate Hilbert space description.


\subsection{Global and local estimation strategies}
\label{sec:global-local}

Entangled probe states and entangling measurements (meaning measurements that project the system onto entangled states) have previously been shown to be well-suited to high-precision sensing for a variety of tasks that are encompassed by our networked quantum sensors formalism \cite{komar2014quantum,eldredge2016optimal,humphreys2013quantum,baumgratz2016quantum,liu2014quantum,yue2014quantum,ciampini2015quantum,knott2016local,gagatsos2016gaussian}. However, it has been an open question as to the source of the precision enhancements apparently available in quantum MPE, as noted by Humphreys \emph{et al.}~\cite{humphreys2013quantum}. Given that creating sensor-entangled states and implementing sensor-entangling measurements are generically extremely challenging experimental procedures, it is important to ask: ``when do such operations truly enhance the precision of our sensors?''

We will give rigorous answers to this question for a range of MPE problems, and in doing so we will clarify a range of results in the literature. Moreover, our networked sensing model provides the framework to tackle the parts of this question we leave unanswered. To achieve this, it is convenient to introduce the concepts of ``global'' and ``local'' estimation strategies, to encompass those estimation strategies that require entangled resources, and those that do not, respectively. 

Let us be more precise. We define an estimation procedure (the choice of the probe state, the measurement and the estimator calculation) to be a \emph{local estimation strategy} if both
 \begin{enumerate}[i.]
 \item The input probe state is \emph{separable} with respect to the given decomposition into sensors.
\item The measurement of the state and the construction of the estimator can be implemented with only \emph{local operations and classical communication} (LOCC), along with local (classical) computations.
\end{enumerate}
In contrast to this, we define a \emph{global estimation strategy} to simply be any estimation procedure that is \emph{not} local. Hence, a global estimation strategy either (1) must use a sensor-entangled probe state, or (2) must use a measurement that requires non-local quantum operations. Obviously, global estimation strategies may use both sensor-entangled states and sensor-entangling measurements. Note that the notions of local and global estimation strategies are only meaningful when considered with respect to some given decomposition of an estimation problem into different sensors.

As the concepts introduced above are central to our results, we wish to be absolutely clear. In the above definitions, and throughout this paper unless otherwise stated, ``local'', ``separable'' and ``entangled'' should all be considered to be defined with respect to the given decomposition of the complete sensing network into sensors. E.g., a separable state is only necessarily separable with respect to the decomposition into sensors. It may be entangled with respect to some further decomposition of each sensor. An operation is considered to be local if it acts on only a single sensor.

Finally, before turning to our main results, it is important to mention some alternative terminology that has been used previously. In the literature, comparisons between \emph{simultaneous} and \emph{individual} (or \emph{independent} or \emph{separate}) estimation strategies have often been considered, e.g., see \cite{humphreys2013quantum,liu2014quantum,yue2014quantum}. Simultaneous estimation means that all of the parameters are estimated at the same time, and individual, independent and separate estimations refer to processes in which each parameter is estimated in its own individual experiment. These concepts are less convenient for assessing whether entanglement provides an enhanced estimation precision. Therefore, we will largely consider the notions of local and global estimation defined herein. However, we do note that, in a rough sense, a global estimation strategy has much in common with how the term `simultaneous estimation' has been used in the literature, and a local estimation strategy has much in common with previous usage of the `individual estimation' terminology.

\section{Estimating locally encoded parameters}
\label{sec:est-phi}

In this section we analyze all estimation problems in which we wish to estimate some or all of the elements of $\bs{\phi}$ to the best precision possible. Note that, in our framework, each element of $\bs{\phi}$ is a parameter encoded into one, and only one, of the sensors. Hence, this is not the most general MPE problem within our model of quantum sensing networks (in Section \ref{Sec:functions} we look at the more general case in which we wish to estimate some \emph{functions} of the elements of $\bs{\phi}$).

We begin this section by studying a network of sensors in which each sensor is encoded with a single scalar parameter (i.e., sensor $k$ is encoded with parameter $\phi_k$). In this setting we show that there is always a local estimation strategy that is better than a given global estimation strategy, in the sense that it has a lower estimation uncertainty whilst using the same amount of resources. 

Next, we analyze the case in which a vector of parameters is encoded into each sensor (i.e., vector $\phi_{[k]}$ is encoded into sensor $k$). We show that, whenever the generating operators for all of the parameters commute, as above, entanglement \emph{between sensors} again reduces parameter estimation precision. However, in many of the examples where each sensor is estimating a vector of parameters, the parameter generators do not all commute. For such estimation problems, we show that the precision obtainable by a global estimation strategy can always be equalled or bettered by a local strategy \emph{if} each sensor has access to a local ancillary sensor with which it may be entangled. The resource counting is more subtle in this case, and in some situations a global estimation strategy \emph{might} possibly facilitate a reduction in the estimation uncertainty by a factor of $\frac{1}{2}$ for a fixed amount of resources. 

\subsection{Estimating local scalar parameters}
\label{Sec:proof}
Consider any estimation problem, within our quantum sensing networks framework, in which:
\begin{enumerate}[i.]
\item The aim is to estimate each $\phi_k$.
\item One parameter is encoded into each sensor.
\end{enumerate}
More precisely, the first condition is that we wish to find an estimation strategy that minimizes $E_{\bs{\Phi}}=\text{Tr}(W\text{Cov}(\Phi))$ with $W_{kk} \neq 0$ for all $k$. The second condition is that there are the same number of sensors as parameters to estimate ($|\mathbb{S}|=d$), and that the $k$\textsuperscript{th} generator $\hat{H}_k$, acts non-trivially only in Hilbert space $\mathcal{H}_k$. That is
\begin{equation}
\hat{H}_k =  \underbrace{\mathds{1} \otimes \mathds{1} \otimes \cdots}_{k-1\,\, \mathds{1} \,\,\text{operators}} \otimes \hat{h}_k \otimes \underbrace{\cdots \otimes \mathds{1} \otimes \mathds{1}}_{d-k\,\, \mathds{1} \,\,\text{operators}}.
\label{H_k-com}
 \end{equation}
Hence, we may write the total unitary evolution as
 \begin{equation}
 U(\bs{\phi}) = U_1(\phi_{1})\otimes U_2(\phi_{2})\otimes \cdots \otimes  U_{d}(\phi_{d}) .
 \end{equation}
A practical example that fits into this setting is the estimation of $d$ optical phases (with respect to a reference beam) that each act on one of $d$ modes \cite{humphreys2013quantum,liu2014quantum,yue2014quantum,knott2016local} (see Section~\ref{sec:optics}). 

Because each parameter is encoded into a single sensor we have $[ \hat{H}_l,\hat{H}_m]=0$ for all $l$ and $m$. Therefore the QCRB can always be saturated for problems of this sort (see Section~\ref{sec:uni-est}). We now show that global estimation strategies are fundamentally sub-optimal for MPE problems of this sort, in the sense that a local estimation strategy can always be found that has a smaller estimation uncertainty (smaller $E_{\bs{\Phi}}$) for a fixed amount of resources. One implication of this is that entanglement between sensors is detrimental to the estimation precision.

\subsubsection{A general bound on the QFIM inverse \label{sec:scalar-inv}}

In Appendix~\ref{App:Finequality} we show that for \emph{any} invertible QFIM
\begin{equation}
 [\mathcal{F}^{-1}]_{kk} \geq \frac{1}{ \mathcal{F}_{kk}},
 \label{Eq:F>1/F}
 \end{equation}
 with the equality obtained for a particular $k$ if and only if the $k$\textsuperscript{th} column and row of $\mathcal{F}$ are such that the only non-zero entry is on the diagonal. Hence, the equality is obtained for all $k$ if and only if $\mathcal{F}$ is diagonal. Note that a closely related result for the classical FIM has been shown previously in Refs.~\cite{bobrovsky1987some,ciampini2015quantum}. This inequality is perhaps unsurprising as $\mathcal{F}_{kk}$ is the QFI for the one-parameter problem that the multi-parameter problem reduces to if all of the other parameters are known \footnote{In the one-parameter setting the QCRB for estimating $\phi_k$ is $\text{Var}(\Phi_k) \geq 1/\mu\mathcal{F}_{kk}$ with the bound saturable.}. It is intuitively clear that the optimal setting for estimating a parameter is when the values of all other variables are known exactly. 

Eqs.~(\ref{Eq:error-QCRB}) and~(\ref{Eq:F>1/F}) imply that
 \begin{equation}
 E_{\bs{\Phi}}\geq \frac{1}{\mu}\text{Tr}( W \mathcal{F}^{-1}) \geq \frac{1}{\mu}\sum_{k=1}^d \frac{W_{kk}}{\mathcal{F}_{kk}}.
 \label{Eq:diag-f-bound}
 \end{equation}
The first of these bounds can be saturated when the QCRB can be saturated, which it can be here for any input state. The second bound is saturated \emph{only} when the QFIM is diagonal, which is because we have assumed that $W_{kk}\neq0$ for all $k$.

 \subsubsection{Pure states \label{sec:ent-sep}}
 
We first of all consider only the case of \emph{pure probe states}, denoting the input probe state $\ket{\psi}$. For pure input states, and when $[\hat{H}_l,\hat{H}_m]=0$ for all $l$ and $m$, the QFIM is given by \cite{knott2016local,baumgratz2016quantum,yue2014quantum}
\begin{equation}
  \mathcal{F}_{lm} =  4 \big( \langle \hat{H}_l\hat{H}_m \rangle-  \langle \hat{H}_l\rangle \langle \hat{H}_m \rangle \big),
\label{QFIM-CovH}
   \end{equation}
where the expectations values are taken with respect to $\ket{\psi}$. Therefore, the diagonal elements of the QFIM are proportional to generator variances and the off-diagonal elements are proportional to generator covariances. The weighted sum of one over the variances provides a lower bound on $E_{\bs{\Phi}}$ via Eq.~(\ref{Eq:diag-f-bound}). To be explicit, by denoting 
\begin{equation} \text{Var}(\varphi,\hat{O}) := \langle \varphi| \hat{O}^2 |\varphi \rangle-  \langle \varphi | \hat{O}| \varphi \rangle^2 ,\end{equation}
for any $\varphi$ and $\hat{O}$, we have that
\begin{equation} 
 E_{\bs{\Phi}} \geq \frac{1}{\mu} \sum_{k=1}^d \frac{W_{kk}}{4\text{Var}(\psi,\hat{H}_k)}.
 \label{Eq:pre-bound-var}
    \end{equation}

This inequality is saturated if and only if the off-diagonal elements in the QFIM are all zero, which here are covariances. Hence, a non-zero covariance in the (pure) probe state between sensors $k$ and $l$ (with respect to $\hat{H}_k$ and $\hat{H}_l$) is detrimental to achieving a minimal $E_{\bs{\Phi}}$, for all $k$ and $l$. For \emph{pure} states these covariances are directly associated with entanglement in the probe state between sensors $k$ and $l$. We now show that, given any pure probe state, $\ket{\psi_{e}}$, a separable pure probe state, $\ket{\psi_{s}}$, may be constructed with the same variances for all $d$ generators and, because it is separable, zero covariances. This will then imply that for \emph{any} pure probe state (which may exhibit entanglement between sensors) there is a separable state with an equal or lower precision bound.

For any operator $\hat{O}_k$ on $\mathcal{H}_{\mathbb{S}}$ with the action of $\hat{o}_k$ on the $k$\textsuperscript{th} sensor and a trivial action on all other sensors then
$\langle \psi_e | \hat{O}_k | \psi_e \rangle = \text{Tr}(\rho_{k} \hat{o}_k),$
where $\rho_k = \text{Tr}_{\mathbb{S}\setminus k} (\ket{\psi_e} \bra{\psi_e})$ is the one-sensor reduced density matrix obtained by tracing out all other sensors, $\mathbb{S}\setminus k$. Denote $\hat{h}_k \ket{\lambda_k}=\lambda \ket{\lambda_k}$, where $\lambda \in \sigma_k$, with $ \sigma_k$ representing the spectrum of $\hat{h}_k$, and where the eigenvectors are orthonormal (or quasi-orthonormal for a continuous spectrum). Note that here $\hat{h}_k$ has been implicitly assumed to have no degeneracies for notational simplicity -- the following results can be easily extended to degenerate operators using a further label to denote the degeneracy for each $\lambda \in \sigma_k$. For any observable which is \emph{diagonal} in the eigen-basis of $\hat{h}_k$, the pure state of sub-system $k$ \footnote{Note that we may add arbitrary phase factors onto each term in this summation, so there is a freedom in the precise state.}
\begin{equation}
  \ket{\psi_k} := \sum_{\lambda \in \sigma_k}  \sqrt{\text{Tr}(\rho_k \ket{\lambda_k}\bra{\lambda_k})}  \ket{\lambda_k},
  \label{Eq:local-state}
 \end{equation}
has identical measurement statistics to $\rho_k$. This is easily confirmed, as it is immediate that $| \langle \lambda_k |\psi_k\rangle |^2 = \text{Tr}(\rho_k\ket{\lambda_k}\bra{\lambda_k}) $ for all $\lambda \in \sigma(\hat{h}_k)$. It then follows that
 \begin{equation}
  \text{Var}(\psi_k,\hat{h}_k) = \text{Var}(\rho_k,\hat{h}_k)=\text{Var}(\psi_e,\hat{H}_k).
 \end{equation}
 
To construct a probe state for the full estimation problem, consider the \emph{separable} state of all $d$ sensors
 \begin{equation}
  \ket{\psi_{s}(\psi_e) } :=\ket{\psi_1} \otimes \ket{\psi_2}\otimes \cdots \otimes \ket{\psi_d}.
  \label{eq:psi-s}
 \end{equation}
By construction $\text{Var}(\psi_{s},\hat{H}_k) = \text{Var}(\psi_e,\hat{H}_k)$
 for all $k=1,\dots,d$. That is, the two states have the same variances for the generator of all $d$ parameters. Furthermore, as $\ket{\psi_{s} }$ is a separable pure state, the covariances between any two $\hat{H}_l$ and $\hat{H}_m$ with respect to this state are zero for $l\neq m$ (as the generators act non-trivially on different sensors). Therefore, for \emph{any} weighting matrix, $W$, we have
 \begin{equation}
 \text{Tr}( W \mathcal{F}^{-1}(\psi_e)) \geq \text{Tr}( W \mathcal{F}^{-1}(\psi_s)),
  \end{equation}
with the equality holding only when $\ket{\psi_e}$ is separable between the $d$ sensors. In all cases the separable state $\ket{\psi_s}$, constructed from $\ket{\psi_e}$, saturates the precision bounds in Eq.~(\ref{Eq:pre-bound-var}).

So far we have shown that a separable state can always be found with a smaller estimation uncertainty than any given entangled state. However, we are not interested in minimizing this uncertainty in isolation, but instead we wish to minimize it for a given fixed amount of resources. These resources are calculated as the expectation value with respect to the input probe state of some Hermitian operator $\hat{R}$ (see Section~\ref{sec:network}). Now, as $\hat{R}$ is assumed to commute with $U(\bs{\phi})$ for all $\bs{\phi}$, it must commute with $\hat{H}_k$ for all $k$. This means that $\hat{R}$ is diagonal in the simultaneous eigen-basis of all the generators, and so $\ket{\psi_s}$ and $\ket{\psi_e}$ contain the same amount of resources (on average).

 \subsubsection{Optimal measurements \label{sec:opt-for-p}}
To confirm that a local estimation strategy can always be found with a smaller $E_{\bs{\Phi}}$ than any given pure-state global estimation strategy, it is also necessary to show that an optimal estimator for $\bs{\phi}$ can be extracted from the probe state using only LOCC. The QFIM for the separable state $\ket{\psi_s}$ is diagonal, and therefore an estimation procedure using this state can be treated as a collection of single parameter estimation problems. 

In single parameter estimation, the optimal measurement is a projection onto the eigenstates of the SLD \cite{baumgratz2016quantum,paris2009quantum,humphreys2013quantum}. For pure states, the $k$\textsuperscript{th} SLD is given by
 \begin{equation} \hat{L}_k=2(\ket{\partial_k \psi_{\bs{\phi}}} \bra{\psi_{\bs{\phi}}}+\ket{\psi_{\bs{\phi}}}\bra{\partial_k \psi_{\bs{\phi}}}),
 \end{equation}
 where $\ket{\partial_k \psi_{\bs{\phi}}}= \partial \ket{\psi_{\bs{\phi}}}/ \partial_{\phi_k}$ and $\ket{\psi_{\bs{\phi}}} = U(\bs{\phi})\ket{\psi}$ \cite{baumgratz2016quantum,paris2009quantum}. Using this, it may then be confirmed that, because $\ket{\psi_s}$ is separable between sensors, the SLDs for the input probe state $\ket{\psi_s}$ are given by 
 \begin{equation}
 \hat{L}_k=\left(  e^{-i \phi_j\hat{h}_j} \ket{\psi_j}\bra{\psi_j}  e^{i \phi_j\hat{h}_j} \right)^{\otimes_{j \neq k}} \otimes \hat{l}_k,
 \end{equation}
 where
 \begin{equation}
 \hat{l}_k=2i e^{-i \phi_k\hat{h}_k}\left[\ket{\psi_k}\bra{\psi_k},\hat{h}_k\right]  e^{i \phi_k\hat{h}_k}.
 \end{equation}
 The eigenstates of $\hat{L}_k$ are of the form
 \begin{equation}
\ket{L_k} = \left(  e^{-i \phi_j\hat{h}_j} \ket{\psi_j} \right)^{\otimes_{j \neq k}} \otimes \ket{l_k},
 \end{equation}
 where the $\ket{l_k}$ are the eigenstates of $\hat{l}_k$. 

The optimal measurement on the evolved probe state is to act with the projectors made from these eigenstates. We see that
\begin{equation}
\bra{\psi_{\bs{\phi},s}}( \ket{L_j}\bra{L_j})^{ \otimes_j} \ket{\psi_{\bs{\phi},s}} = \prod_{j=1}^d |\bra{l_j} e^{-i \phi_j\hat{h}_j} | \psi_j \rangle |^2,
\end{equation}
where $\ket{\psi_{\bs{\phi},s}} \equiv U(\bs{\phi})\ket{\psi_s}$. The optimal measurement can therefore be performed with a local measuring device for each sensor, as at sensor $k$ the optimal measurement is a projection onto the eigenstates of $\hat{l}_k$. An estimator for $\phi_k$ may then be constructed locally at each sensor, and an estimator for $\bs{\phi}$ may be constructed via classical communication -- this is a local estimation strategy.

\subsubsection{Mixed states}
\label{sec:pure-optimal}
We now wish to show that pure probe states provide better estimation precision than mixed states. To do this, we first prove something slightly different, from which we can infer that pure probe states are preferable. Here we allow for the possibility of ``ancillary'' sensors on which the parameter-imprinting unitary acts only trivially. As such we now consider the enlarged Hilbert space $\mathcal{H}_{\mathbb{S}} \to  \mathcal{H}_{\mathbb{S}} \otimes \mathcal{H}_{\mathbb{A}}$, with $\mathcal{H}_{\mathbb{A}}$ the Hilbert space of some ancillary sensors, and a unitary evolution imprinting the parameters given by 
\begin{equation} U(\bs{\phi}) =\exp(-i \bs{\phi}^T \hat{\bs{H}}) \otimes \mathds{1},
\end{equation}
 with $\bs{\hat{H}}=(\hat{H}_1,\dots,\hat{H}_d)$ still satisfying Eq.~(\ref{H_k-com}), and where $\mathds{1}$ is the identity operator on $\mathcal{H}_{|\mathbb{A}|}$.

 To fully define the problem it is also necessary to specify how the resource operator is extended to the larger Hilbert space. We take
  \begin{equation}
   \hat{R} \to \hat{R} \otimes \mathds{1} + \mathds{1} \otimes \hat{R}_{\mathbb{A}},
   \end{equation}
   where $\hat{R}_{\mathbb{A}}$ is some positive operator (meaning that it has non-negative eigenvalues, e.g., a number operator). Note that this construction for the extended resource operator includes, to our knowledge, all cases of practical interest. Moreover, it is a very natural assumption, as it would be a strange estimation problem if, by simply adding ancillary systems, the resources used could be reduced.
   
Considering this extension of the quantum sensing networks setting to include ancillary sensors, the derivation of Section~\ref{sec:scalar-inv} to Section~\ref{sec:opt-for-p} can be easily adapted to show that entanglement with ancillary sensors is also detrimental to the estimation precision (a brief outline of how this can be achieved is given in Appendix~\ref{app:ancillas}), and a local estimation strategy is still preferable. Indeed, adding the ancillary systems can at best only keep the precision per resource the same. It should be noted that Ballester \cite{ballester2004entanglement}
has already shown that entanglement with ancillary systems can provide no enhancement in MPE with commuting generators. This was in the different context of optimizing the estimation of $d-1$ parameters in a $d\times d$ dimensional unitary acting on a single $d$-dimensional quantum system in the special case where the generators of all of the parameters commute. However, the results of Ref.~\cite{ballester2004entanglement} imply our conclusions on ancillary sensors, at least in some cases.

Ancillary sensors cannot enhance estimation precision, for the MPE problems we are considering here, and this can be used to imply that a mixed probe state is sub-optimal. Any density operator $\rho$ satisfies $\rho = \text{Tr}_{\mathbb{A}} (\ket{\Psi_\rho}\bra{\Psi_\rho})$ from some $\ket{\Psi_{\rho}} \in  \mathcal{H}_{\mathbb{S}}\otimes  \mathcal{H}_{\mathbb{A}}$, known as a purification of $\rho$, with $\mathbb{A}=\mathbb{S}$ always sufficient \cite{nielsen2010quantum}. It is clear that $\mathcal{F}_{kk}(\Psi_{\rho}) \geq \mathcal{F}_{kk}(\rho)$ for all $k$, as one possible measurement strategy with the pure probe $\ket{\Psi_{\rho}}$ is to discard the ancillary sensor(s), which is entirely equivalent to having the probe state $\rho$. Moreover, any such purification $\ket{\Psi_{\rho}}$ has an equal or worse precision bound $E_{\bs{\Phi}}$ than the state $\ket{\Psi_{\rho,s}} = \ket{\psi_s(\Psi_{\rho}) } \otimes \ket{\psi_{\mathbb{A}}}$, where $\ket{\psi_s(\Psi_{\rho})} $ is given by Eqs.~(\ref{Eq:local-state}) and~(\ref{eq:psi-s}) and $\ket{\psi_{\mathbb{A}}}$ is any state of the ancillary system(s). Furthermore, the estimation uncertainty bound is at best equal to that achieved by $\ket{\psi_s(\Psi_{\rho})} $ without any ancillary sensors, and they are only actually equal when $\ket{\Psi_{\rho}}$ is a product state between the probe and ancillary sensors, which is true only when $\rho$ is pure. Hence, the bound on $E_{\bs{\Phi}}$ for $\rho$ is greater than or equal to that for $\ket{\psi_s(\Psi_{\rho})}$, with the equality only when $\rho$ is pure. Finally, $\rho$ and $\ket{\psi_s(\Psi_{\rho})}$ contain the same amount of resources, confirming that the separable pure state $\ket{\psi_s(\Psi_{\rho})}$ provides a better estimation precision than $\rho$ for a fixed amount of resources.

\subsubsection{Local estimation should be preferred}

In this subsection we have considered the problem of estimating a $d$-dimensional vector $\bs{\phi}$ where (1) each $\phi_k$ is encoded into a single quantum sensor, and (2) we wish to implement non-trivial estimates of all the elements of $\bs{\phi}$, but they are not necessarily all equally important. We have shown that: for any given mixed state, there is a pure state containing the same or less resources with an equal or smaller QCRB for $E_{\bs{\Phi}}$ (Section~\ref{sec:pure-optimal}); that for any given pure state, there is a state which is separable between sensors that contains the same resources and that has an equal or smaller QCRB for $E_{\bs{\Phi}}$ (Section~\ref{sec:ent-sep}); and that, for any given pure separable state, the optimal measurement saturates the QCRB and can be performed by local measurements at each sensor (Section~\ref{sec:opt-for-p}). 

Combining these results we then have that, for any MPE problem of the sort described above and given any global estimation strategy, it is always possible to find a local estimation strategy that uses the same amount of resources and that has a smaller estimation uncertainty. Hence, in this setting, multi-sensor entangled states and global measurements are not only unnecessary for obtaining a high estimation precision, but the entanglement \emph{reduces} the attainable precision. Moreover, as the ideal probe state is pure, we can conclude that neither quantum nor classical correlations between the sensors enhance the precision in estimating $\bs{\phi}$. 

Our argument has nothing to say on the relative \emph{feasibility} of various estimation strategies. What we have pointed out is that, ideally, it is preferable to use a local estimation strategy. Hence, despite our results, it is possible that in some cases sensor-entangled probe states, or multi-sensor measurements, may still be the best option in practice. For example, preparing an entangled state might be easier than preparing any of the separable states that can obtain a better estimation precision. This is something that can only be assessed in a given physical setting.

\subsubsection{Discussion}
\label{lebge}


There are some further interesting subtleties to our results, and also some important caveats, which we wish to now make clear. For any given global estimation strategy, to provide an equivalent local estimation strategy with a lower estimation uncertainty, our argument maps entangled probe states to separable probe states with similar local properties (see Section~\ref{sec:ent-sep}). This implicitly assumes that we are interested in comparing estimation strategies that use arbitrary probe states in the full sensor Hilbert space. However, we might actually be interested in finding the best estimation strategy out of all those that use a probe state from some given sub-space of the full Hilbert. 

If this is the case, we are asking a different question to the one we have answered in this section. Moreover, in this case it is clear that whether we can always find a local estimation strategy with an equal or better precision than a given global estimation strategy, will unavoidably depend on the particular sub-space under consideration. For example, we could consider a sub-space of only maximally entangled states, in which case it is trivial that a global estimation will be required, as we have explicitly discounted all local estimation strategies from the analysis. This rather trivial example shows that care must be taken if the results we have presented above are to be applied to some sub-space $\mathcal{S}$ of the total Hilbert space. The relevant question is: 

\vspace{0.1cm}
``Can we map an arbitrary state $\Psi \in \mathcal{S}$ to a sensor-separable state $\Psi_s \in \mathcal{S}$ where (1) $\Psi_s$ has the same, or larger, generator variances than $\Psi$, and (2) $\Psi_s$ has the same, or a smaller, value for the resource function than $\Psi$?''

\vspace{0.1cm}
 If yes, our argument may be applied. However, note that the entangled state to separable state mapping, required for the argument, may be more subtle than that in Eq.~(\ref{Eq:local-state}). 

Ultimately, it is a matter of judgement as to what sub-space one is, or should be, interested in. However, in many examples where these results are of practical interest, we do not know of any fundamental physical reason to not consider a sub-space containing the necessary separable probe state (such as the full Hilbert space). This will be covered in detail when we apply these results to optical MPE and atomic sensing, in Sections~\ref{sec:optics} and~\ref{sec:atomic_sensing} respectively. However, some theoretical analyses might implicitly or explicitly exclude the relevant separable states and arrive at different conclusions to ours in this section. Indeed, this will be made explicitly clear in Sections~\ref{m>1func} and~\ref{sec:optics}. 

Similarly, our argument also implicitly relies on the potential availability of arbitrary measurements (as we use the QFIM). That is, we have assumed that we should not \emph{a priori} remove certain types of measurement from the analysis. As above, if one wishes to perform an equivalent analysis with only certain measurements available, it is critical to confirm that this does not invalidate our argument before this result is used. Again, when we apply this result to practical sensing problems, we will address this explicitly in each setting (it cannot be addressed in an entirely abstract setting).


Finally, there is one more important subtlety which is critical to understanding the precise claim we are making. In general, there is no \emph{optimal} estimation strategy. This is why we have refrained from stating that local strategies are ``optimal''. When the generating operators are bounded then our argument \emph{does} imply that the optimal strategy is a local estimation strategy (modulo the discussion on probe state and measurement sub-spaces given above). However, if the generators are unbounded then probe states with arbitrarily large generator variances, which are the diagonal elements of the QFIM, exist. In some such settings the saturable QCRB on $E_{\bs{\Phi}}$ can then be made arbitrarily small for a fixed quantity of resources \cite{knott2016local}. This is a technicality that can also appear in single-parameter estimation \cite{rivas2012sub,hall2012universality,giovannetti2012sub}, where it is known that it is not possible to obtain an arbitrarily high precision-per-resource in practice \cite{hall2012universality,giovannetti2012sub}. This is because (to get close to) saturating the QCRB requires many experimental repeats, and the number of repeats needed is not independent of the probe state. 

One way to at least partially resolve this issue, whilst still considering the QCRB and the QFIM, is to consider only probe states in some physically well-motivated bounded sub-space of the total Hilbert space of each sensor. However, considering such sub-spaces can be problematic if not chosen carefully (see above). We will consider this further in Section~\ref{sec:optics}, when we focus on optical MPE (none of these problems are of practical relevance in atomic MPE). Finally, we note that all of these subtleties also apply to the remainder of this section, but we do not explicitly discuss them again.


\subsection{Estimating local vector parameters \label{Sec:proof2}}
In the previous subsection we proved that a local estimation strategy is preferable to a global estimation strategy when each quantum sensor in the network is estimating a single scalar parameter. This does not cover all cases of practical interest, for example, if each sensors is estimating the strength and direction of a three-dimensional magnetic field \cite{baumgratz2016quantum}, or if each sensor is characterizing a completely unknown $D$-dimensional unitary operator \cite{ballester2004estimation}. We now extend our argument into this setting, with the conclusion significantly more subtle in this case (except when the generating operators all commute).

Before we can formally state the MPE problem of interest, we need to introduce a succinct notation for dividing vectors and matrices into sub-vectors and sub-matrices. Consider ``partitioning'' the $d$-dimensional vector parameter $\bs{\phi}$ into $m$ sub-vectors, where the $k$\textsuperscript{th} sub-vector has a dimension of $d_k$ and $d=d_1+\dots+d_m$. More specifically, let the 1\textsuperscript{st} sub-vector, denoted $\bs{\phi}_{[1]}$, be given by $\bs{\phi}_{[1]} :=(\phi_{1},\dots,\phi_{d_1})^T$, let the 2\textsuperscript{nd} sub-vector be $\bs{\phi}_{[2]} :=(\phi_{1+d_1},\dots,\phi_{d_1+d_2})^T$, and so on. Therefore, by denoting $d_{<k} := d_1 + d_2 + \dots + d_{k-1}$,
the $k$\textsuperscript{th} sub-vector is given by
\begin{equation}
 \bs{\phi}_{[k]}:=(\phi_{(1+d_{<k})},\dots,\phi_{(d_k + d_{<k})})^T.
  \end{equation}
It is then clear that, as desired, we have
\begin{equation}
\bs{\phi} =  \begin{pmatrix}
  \bs{\phi}_{[1]}  \\
 \bs{\phi}_{[2]}  \\
  \vdots   \\
 \bs{\phi}_{[m]}  \\
 \end{pmatrix}.
\end{equation}

Using an analogous notation, for a $d\times d$ matrix $M$ and a given partitioning of $d$ into $d=d_1+\dots+d_m$, we let $M_{[jk]}$ denote the sub-matrix of $M$ obtained by removing the elements that are not both in rows $1+d_{<j}$ to $d_j+d_{<j}$ and columns $1+d_{<k}$ to $d_k+d_{<k}$. Hence, 
\begin{equation}
M =  \begin{pmatrix}
  M_{[11]} & M_{[12]} & \cdots & M_{[1m]} \\
  M_{[21]} & M_{[22]} & \cdots & M_{[2m]} \\
  \vdots  & \vdots  & \ddots & \vdots  \\
  M_{[m1]} & M_{[m2]} & \cdots & M_{[mm]}
 \end{pmatrix}.
\end{equation}
Note that the parentheses in the subscripts of this notation are used to denote that these are sub-vectors and sub-matrices of $\bs{\phi}$ and $M$, respectively, and not just the scalar vector and matrix elements of $\bs{\phi}$ and $M$ (which will still be denoted as normal). It will be useful to define 
\begin{equation}
\mathbb{P}_j := \{ 1+d_{<j}, 2+d_{<j}, \dots, d_j+d_{<j} \},
\end{equation}
 i.e, $\mathbb{P}_j$ contains the labels for the parameters in the $j$\textsuperscript{th} partition.

We are now ready to give a formal construction of the general MPE problem we are going to consider in this subsection. As throughout, we are considering estimation problems within our general quantum sensing networks framework, with the total Hilbert space $\mathcal{H}_{\mathbb{S}} = \mathcal{H}_1 \otimes  \mathcal{H}_2 \otimes \cdots \otimes \mathcal{H}_{|\mathbb{S}|}$. The completely general situation is that a $d$-dimensional vector $\bs{\phi}$ (with $d>|\mathbb{S}|$ in general) is encoded it the sensing network, with each parameter encoded into one sensors. This can be described by some partitioning $d=d_1+\dots+d_{|\mathbb{S}|}$,
specifying that $\phi_k$ has an associated $\hat{H}_k$ operator that acts nontrivially only on the $l$\textsuperscript{th} sensor if $k \in \mathbb{P}_l$. That is, the $\bs{\phi}_{[l]}$ vector parameter is encoded into the Hilbert space of the $l$\textsuperscript{th} sensor, $\mathcal{H}_l$. As always, the parameters are imprinted via $U(\bs{\phi})=\exp(-i \bs{\phi}^T\hat{\bs{H}})$, and now we have
\begin{equation} 
U(\bs{\phi}) = U_1(\bs{\phi}_{[1]}) \otimes \dots \otimes U_{|\mathbb{S}|}(\bs{\phi}_{[|\mathbb{S}|]}).
\label{eq:unit-2}
\end{equation}
 It is clear that $[ \hat{H}_k,\hat{H}_l] =0 $ if $k \in \mathbb{P}_p$ and $l \in \mathbb{P}_q$ with $p \neq q$, but the $\hat{H}_k$ operators that act on the same sensor need not commute. 

As in Section~\ref{Sec:proof}, we now consider the problem of optimizing the network to estimate $\bs{\phi}$ (rather than some functions of the $\phi_k$). We allow the importance weighting for estimating each parameter to be arbitrary, but we assume it is non-zero (i.e., $W_{kk} \neq 0$ for all $k$). Finally, we assume that the resource operator is given by
\begin{equation} 
 \hat{R}_{\text{sum}} = \hat{R}_1+ \hat{R}_2 + \dots+ \hat{R}_{|\mathbb{S}|},
\end{equation}
where $\hat{R}_l$ acts non-trivially only on the $l$\textsuperscript{th} sensor. Note that this is a very natural assumption (as we have already discussed in Section~\ref{sec:network}), but that this assumption was not needed in Section~\ref{Sec:proof}. 

For MPE problems of this sort, we will now show that any global estimation exhibits, at best, very limited precision improvements over an equivalent local estimation strategy that uses the same amount of resources. Moreover, in many settings the local estimation strategy is preferable. Our argument is similar in many ways to that given in Section~\ref{Sec:proof}, and follows the same basic structure.

\subsubsection{A general bound on the QFIM block-wise inverse}
Given an arbitrary QFIM $\mathcal{F}$ for a $d$-dimensional vector $\bs{\phi}$, and an arbitrary partitioning $d=d_1+\dots+d_{|\mathbb{S}|}$, consider the block-diagonal matrix obtained from $\mathcal{F}$ by setting all of the off-diagonal matrices in $\mathcal{F}$ to zero, i.e., the matrix
\begin{equation}
\mathcal{D}(\mathcal{F}) :=  \begin{pmatrix}
  \mathcal{F}_{[11]}  & 0 & \cdots & 0 \\
 0 &   \mathcal{F}_{[22]}  & \cdots & 0 \\
  \vdots  & \vdots  & \ddots & \vdots  \\
  0 &0 & \cdots &  \mathcal{F}_{[|\mathbb{S}||\mathbb{S}|]} 
 \end{pmatrix}.
\end{equation}
In Appendix~\ref{App:Finequality-2} we show that for \emph{any} invertible $\mathcal{F}$, and \emph{any} partitioning,
\begin{equation}
 [\mathcal{F}^{-1}]_{[kk]} \geq [\mathcal{D}^{-1}]_{[kk]} = \left[\mathcal{F}_{[kk]}\right]^{-1}, 
 \label{eq:matrix-in}
\end{equation}
for all $k$. Moreover, the equality is obtained for a given $k$ if and only if $\mathcal{F}_{[jk]}=\mathcal{F}_{[kj]} =0$ for all $j\neq k$. To properly understand this statement, is is important to note that, for two matrices $A$ and $B$, $A\geq B$ and $A\neq B$ does \emph{not} imply that $A>B$. Moreover, note that Eq.~(\ref{eq:matrix-in}) is a generalization of the inequality given in Eq.~(\ref{Eq:F>1/F}). 

Eq.~(\ref{eq:matrix-in}) implies that $ [\mathcal{F}^{-1}]_{kk} \geq [\mathcal{D}^{-1}]_{kk}$ for all $k=1,\dots,d$. It follows from this, and Eq.~(\ref{Eq:error-QCRB}), that
\begin{equation}
 E_{\bs{\Phi}}\geq \frac{1}{\mu}\text{Tr}( W \mathcal{F}^{-1}) \geq\frac{1}{\mu}\text{Tr} ( W \mathcal{D}^{-1}(\mathcal{F})) ,
 \label{Eq:diag-f-bound2}
  \end{equation}
 for any weighting matrix $W$, and any given partitioning. The first bound is saturated when the QCRB may be saturated, which is not always true for the MPE problems we are now considering. The second bound is saturated only when $\mathcal{F}$ is block-diagonal, with respect to the given partitioning, as we have assumed that $W_{kk} \neq 0$ for all $k$.

The elements of the QFIM for a pure probe state are given by \cite{liu2015quantum,liu2014fidelity} 
\begin{equation} 
\mathcal{F}_{mn}(\psi) = 2 \langle  \{\hat{G}_m,\hat{G}_n \} \rangle  -4 \langle \hat{G}_m \rangle \langle \hat{G}_n  \rangle,
\label{eq:QFIM-non-com}
 \end{equation} 
with the expectation values taken with respect to the input state. Here $\hat{G}_k$ is the generator of $\phi_k$, defined by $\hat{G}_k:= -i (\partial U^{\dagger} / \partial_{\phi_k}) U$ (see Section~\ref{sec:uni-est}), and $\{\cdot,\cdot\}$ is the anti-commutator, i.e., $\{A,B\}=AB+BA$.  Eq.~(\ref{eq:unit-2}) implies that $[ \hat{G}_k,\hat{G}_l] =0$ for $k \in \mathbb{P}_p$ and $l \in \mathbb{P}_q$ with $p \neq q$. More importantly, this also implies that $\hat{G}_k$ acts non-trivially only in sub-space $\mathcal{H}_l$ if $k\in\mathbb{P}_l$. That is, if $k\in\mathbb{P}_l$ then
\begin{equation}
\hat{G}_k =  \underbrace{\mathds{1} \otimes \mathds{1} \otimes \cdots}_{l-1\,\, \mathds{1} \,\,\text{operators}} \otimes \hat{g}_k \otimes \underbrace{\cdots \otimes \mathds{1} \otimes \mathds{1}}_{|\mathbb{S}|-l\,\, \mathds{1} \,\,\text{operators}},
\label{eq:g-in-l}
 \end{equation}
for some $\hat{g}_k$.
 
 \subsubsection{Ancillary sensors allow for an equivalent local estimation}
 
We now argue that, for any global estimation strategy, we can find an equivalent local estimation strategy with an equal or smaller estimation uncertainty $E_{\bs{\Phi}}$, if we have access to $|\mathbb{S}|$ ancillary sensors -- one for each probe sensor -- and by ``local'' we now mean local with respect to a probe and ancillary sensor pair. We will term each probe-ancillary pairing a ``duplicated sensor'' \footnote{Note that, in many situations a duplicated sensor could simply be implemented by using half of the particles in a sensor as ancillas which do not undergo the unknown unitary evolution.}. We will then argue that, in all practical settings, this equivalent local estimation strategy (using duplicated sensors) uses no more than twice the resources of the global estimation strategy, and in some cases the most appropriate accounting of resources will imply that the resources in both strategies are equal. Note that, if the generators all mutually commute, we will show that our argument can be adapted to remove the ancillary sensors (see Section~\ref{sec:non-com-dis}), and hence the local strategy is unambiguously preferable. However, more generally, it appears that this is not the case.

Consider providing each sensor with a (spatially local) ancillary copy of itself, or more precisely consider the duplicated Hilbert space
 \begin{equation}
  \mathcal{H}_{\mathbb{S}\cup \mathbb{S}} = (\mathcal{H}_1 \otimes \mathcal{H}_1 ) \otimes \dots \otimes (\mathcal{H}_{|\mathbb{S}|} \otimes \mathcal{H}_{|\mathbb{S}|} ).
\end{equation}
Hence, the unitary operator imprinting the parameters acts non-trivially only on the first of each pair. Now, consider any probe state $\rho$ that is a (possibly pure) density operator on $  \mathcal{H}_{\mathbb{S}}$. This probe state may in general be entangled between sensors. Consider any purification of $\rho$ into the Hilbert space  $\mathcal{H}_{\mathbb{S}\cup \mathbb{S}}$, which we denote $\ket{\Psi_\rho}$ ($\rho$ can always be purified into this duplicated Hilbert space \cite{nielsen2010quantum}). This purified state must have an optimal estimation uncertainty (i.e., $E_{\bs{\Phi}}$ minimized over all measurements) that is equal to or smaller than that of $\rho$. This is for the same reasons as earlier (see Section~\ref{sec:pure-optimal}). Specifically, any measurement strategy for $\rho$ is equivalent to one for $\ket{\Psi_\rho}$ where the additional sensors are discarded. 

Denote the QFIM of $\ket{\Psi_\rho}$ by $\mathcal{F}$.
For any purified probe  $\ket{\Psi_\rho}$, we now show that we may find an alternative probe state $\ket{\Psi_s}  \in  \mathcal{H}_{\mathbb{S}\cup \mathbb{S}}$ that is separable between different duplicated sensors (probe-ancilla sensor pairs) and that has the QFIM $\mathcal{D}(\mathcal{F})$. Note that this state may be entangled between a probe sensor and its ancillary copy. The following argument is very similar to that used in Section~\ref{sec:pure-optimal}. 

For a pure state, the $\mathcal{F}_{[ll]}$ sub-matrix of $\mathcal{F}$ depends only on the reduced density operator $\rho_{l}=\text{Tr}_{\mathbb{S} \setminus l }(\ket{\Psi_\rho}\bra{\Psi_\rho})$. This follows from Eq.~(\ref{eq:QFIM-non-com}), and by noting that $\hat{G}_k$ acts non-trivially only in the Hilbert space on which $\phi_k$ is encoded. That is, it acts non-trivially on $\mathcal{H}_l$ only if $k \in \mathbb{P}_l$, as stated in Eq.~(\ref{eq:g-in-l}). But we can also find a pure state in $\ket{\Psi_l} \in \mathcal{H}_{l} \otimes \mathcal{H}_{l}$ with the same reduced density matrix, $\rho_l$, obtained by tracing over the second ancillary sensor. Therefore the state
 \begin{equation}
 \ket{\Psi_s(\rho)} = \ket{\Psi_1} \otimes \ket{\Psi_2} \otimes \dots \otimes | \Psi_{|\mathbb{S}|}\rangle,
 \end{equation}
where each $\ket{\Psi_k}$ is a purification of $\rho_{k}$ for $k=1,\dots,|\mathbb{S}|$, has a QFIM $\mathcal{F}'$ with $\mathcal{F}'_{[ll]}=\mathcal{F}_{[ll]}$ for all $l$. Furthermore, as $\ket{\Psi_s(\rho)}$ is separable between each pair of duplicated sensors, the off-diagonal sub-matrices of  $\mathcal{F}'$ are zero. Hence $\mathcal{F}'=\mathcal{D}(\mathcal{F})$. By Eq.~(\ref{Eq:diag-f-bound2}), $\ket{\Psi_s}$ has an equal or lower QCRB-derived bound on the estimation uncertainty $E_{\bs{\Phi}}$ (for any weighting matrix) than the purified state $\ket{\Psi_\rho}$, for any $\rho$ and any purification. Therefore, it also has a lower QCRB-derived bound on $E_{\bs{\Phi}}$ than that for $\rho$. These lower bounds on $E_{\bs{\Phi}}$ are equal if and only if $\rho$ is pure and separable (as we are assuming the weighting matrix, $W$, has $W_{kk} \neq 0$ for all $k$).

  If the QCRB is saturable for all probe states then we can conclude that, when the optimal measurement is implemented, $\ket{\Psi_s(\rho)}$ has an equal or lower estimation uncertainty than $\rho$, for any measurement strategy on $\rho$. In this case the optimal measurement for $\ket{\Psi_s(\rho)}$ is local, for exactly the same reasons as in Section~\ref{sec:opt-for-p}. Hence, in any such case, for any global estimation strategy there is a local estimation strategy that obtains a lower uncertainty in the estimator, as quantified by $E_{\bs{\Phi}}$. Note however that we are yet to consider the resources used for each strategy -- before we do this, we turn to those situations when the QCRB cannot always be saturated.

\subsubsection{Local estimation with non-commuting generators}
\label{sec:non-comm-proof}
Outside a setting where the QCRB may be saturated for all input states, simply showing that the QCRB-derived bound on $E_{\bs{\Phi}}$ for the separable state $\ket{\Psi_s(\rho)}$ is smaller or equal to the bound on $E_{\bs{\Phi}}$ for $\rho$, for any $\rho$, is insufficient to prove that a separable state is always preferable. However, we can confirm this is the case, with the following argument. 

The precision with which $\phi_{[l]}$ can be measured is always improved or unaffected if we know $\phi_{[k]}$ for all $k \neq l$. Both $\ket{\Psi_{\rho}}$ and $\ket{\Psi_s}$ have the same QFIM for $\phi_{[l]}$, which is $\mathcal{F}_{[ll]}$, and if all the other parameters are known we may set them to zero (by local known unitaries before the measurement). Therefore, the $\phi_{[l]}$-encoded state in each of these cases is 
\begin{align} 
\ket{\Psi_{\rho}^l} &\equiv (\mathds{1}   \otimes \dots \otimes U_l(\phi_{[l]})    \otimes \dots \otimes \mathds{1} ) \ket{\Psi_{\rho}}, \\
\ket{\Psi_{s}^l} &\equiv    \ket{\Psi_{1}} \otimes \dots \otimes  \ket{\Psi_{l}^l} \otimes \dots \otimes |\Psi_{|\mathbb{S}|}\rangle,
\end{align}
where $\ket{\Psi_{l}^l} \equiv (U_l(\phi_{[l]}) \otimes \mathds{1}) \ket{\Psi_{l}}$.

In Appendix~\ref{app:POVMs} it is shown that, using only $\phi_{[l]}$-\emph{independent} unitary operations and partial traces (on an extended Hilbert space), we may map $\ket{\Psi_{s}^l} \to \ket{\Psi_{\rho}^l}$. Hence, any POVM on $\ket{\Psi_{\rho}^l}$ is exactly equivalent to some POVM on $\ket{\Psi_{s}^l}$. This is in the sense that the POVMs have the same number of POVM effects and each measurement outcome, $m$, is associated with the same PDF, $p(m|\phi_{[l]}$). This means that $\ket{\Psi_{s}^l}$ can estimate $\phi_{[l]}$ with at least as small an estimation uncertainty as can be obtained with $\ket{\Psi_{\rho}^l}$, when all the other parameters are known and if the optimal measurement is used. Note that this measurement might not saturate the QCRB-derived bound for $E_{\Phi_{[l]}}$, and when this is the case the optimal measurement will depend on the weighting sub-matrix $W_{[ll]}$.

Return now to the actual problem of interest -- when all of the parameters are \emph{not} known. For the separable state $\ket{\Psi_s}$, all of the $\phi_{[l]}$ can be measured simultaneously to the same, or a better, precision that $\ket{\Psi_{\rho}}$ can estimate each $\phi_{[l]}$ when all of the other $\phi_{[k]}$ \emph{are} known. This is because $\ket{\Psi_s}$ is separable between duplicated sensors, and so the optimal POVM for estimating $\phi_{[l]}$ (with that state, and given $W_{[ll]}$) acts only on the $l$\textsuperscript{th} duplicate sensor. Hence, all of the measurements to optimize the estimation precision of each $\phi_{[l]}$ can be implemented in parallel. However, there is no guarantee that $\ket{\Psi_{\rho}}$ can estimate all of the $\phi_{[l]}$ simultaneously with the same estimation uncertainty that each one can be estimated with when all of the other parameters are known. 

Hence, we have shown that we can map any density operator $\rho$ to a pure state $\ket{\Psi_s}$, which is separable between duplicated sensors, and that, for some measurement, $\ket{\Psi_s}$ has an equal or lower estimation uncertainty (i.e., smaller $E_{\bs{\Phi}}$) than $\rho$, for any measurement on $\rho$. Thus, although the QCRB cannot necessarily be saturated, a separable state allows us to get as close as it is possible to saturating it.

 \subsubsection{Resource counting}
Finally, we need to compare the resources used in a given (potentially) global strategy and the local strategy obtained from this global strategy via the method above. It is at this point that a minor potential advantage of a global estimation strategy becomes apparent. In order to discuss resources it is again necessary to define the extended resource operator for the probe state $\ket{\Psi_s}$, as this should be an operator on $ \mathcal{H}_{\mathbb{S}\cup \mathbb{S}} $. As before, the obvious extension is one of the form $\hat{R}_{\text{sum}} \to \hat{R}_{\text{sum}}'$ with
\begin{equation} 
\hat{R}_{\text{sum}}'= (\hat{R}_1 +\hat{R}_1') + (\hat{R}_2 +\hat{R}_2') +\dots + (\hat{R}_{|\mathbb{S}|} +\hat{R}_{|\mathbb{S}|}'),
\end{equation}
where $\hat{R}_l'$ acts non-trivially only on the $l$\textsuperscript{th} ancillary sensor and each $\hat{R}_l'$ is a Hermitian and positive operator. We have to choose the $\hat{R}'_l$, and there is no fundamental reason to pick any particular ancillary resource operator. It is a practical question as to what is appropriate. Here we discuss two choices for these ancillary sensor resource operators that have clear physical relevance. 

Consider taking $\hat{R}_l'=\hat{R}_l$ for all $l=1,\dots,|\mathbb{S}|$. This is the relevant choice when resources used in ancillary systems or sensors should be accounted for on the same footing as the probe sensors, which undergo the unknown evolution. In this setting it may be easily confirmed that $\ket{\psi_s(\rho)}$ need contain no more than twice the amount of resources as $\rho$ -- the exact amount depends on the chosen purification, and could be considerably less than this. Hence, in this case, there is potentially a cost to using local estimation, rather than global estimation, of up to a factor of two.

The second natural choice for the resource operators is to take $\hat{R}_l'=\mathds{1}$ for all $l=1,\dots,|\mathbb{S}|$. This is the relevant choice when any properties of the ancillary systems and sensors are irrelevant from the perspective of resource counting. This is arguably the most appropriate method for counting resources when the parameters are induced by some fragile sample (relevant optical sensing examples include measurements of spin ensembles \cite{wolfgramm2013entanglement}, biological systems \cite{carlton2010fast,taylor2013biological}, atoms \cite{tey2008strong,eckert2008quantum} and single molecules \cite{pototschnig2011controlling}). In this case, it is essential to minimize the disturbance of the sample, and as any ancillary systems do not interact with the sample there is no need to minimize any property (e.g., energy) local to that part of the state. In this setting, our argument implies that a local estimation strategy is always, in principle, preferable to a global estimation strategy, as it has a smaller estimation uncertainty for the same amount of resources.

\subsubsection{Discussion}
\label{sec:non-com-dis}

It is natural to wonder whether the ancillary sensors are really essential to the argument that we have presented above. In the case where all the generators commute, we may adapt our argument in a straightforward way to confirm that, in this setting, the ancillary sensors are indeed unnecessary. This is because, as the generators commute, they are all diagonal in the same basis and the separable state obtained in Eq.~(\ref{Eq:local-state}), with respect to any one of the generators acting on that sensor, will exhibit the necessary statistics to have a block-diagonal QFIM and the same total resources. This immediately implies that the optimal estimation strategy is a local estimation strategy for any MPE problem of this sort.

However, more generally, it \emph{is} the case that the ancillary modes are essential for obtaining a minimal estimation uncertainty. This follows from previous results in the literature: it is known that entanglement with ancillas can provide an enhancement to MPE precision that is unavailable when the probes may not be entangled with any external quantum systems \cite{ballester2004estimation,fujiwara2001estimation}. Specifically, Ballester \cite{ballester2004estimation} showed this in the particular context of estimating a completely unknown general unitary (i.e., tomography) on a single $D$-dimensional Hilbert space. This is a problem with $D^2-1$ parameters and non-commuting generators. Ballester concludes that entanglement with an ancilla and a two-body measurement can increase the estimation precision by more than a factor of two.

 To be clear, our results complement, and do not contradict, those of Ballester \cite{ballester2004estimation}. Applied to this context, what we have shown is that, if we had many different and unknown unitary operators acting on different sensors, then any estimation precision enhancement gained by entangling these sensors and using global measurement could instead be obtained (and further improved upon) by using a local strategy, as long as each sensor has access to an ancillary sensor with which it may be entangled. However, the estimation uncertainty for a fixed amount of resources \emph{might} be reduced by a factor of $\frac{1}{2}$ by using a global estimation scheme (without ancillas), unless the resources used in the ancillary sensors need not be accounted for (the appropriate resource counting is a problem-specific question).

\subsection{Singular QFIMs}
\label{Sec:singular-QFIMs}
We now take a slight diversion to present a simple method for accounting for singular QFIMs and what we will call ``unwanted'' parameters. The \emph{unwanted} parameters in $\bs{\phi}$ are those elements of $\bs{\phi}$ for which the weighting matrix element is zero. That is, $\phi_k$ is unwanted if and only if $W_{kk}=0$. The terminology is natural, as for any such $\phi_k$ our estimation uncertainty function $E_{\bs{\phi}}$, which we wish to minimize, puts no weighting on the minimization of $\text{Var}(\Phi_k)$.  So far, we have considered only non-singular QFIMs and estimations where all of the elements of $\bs{\phi}$ are of interest. In this subsection we will point out how singular QFIMs and unwanted parameters affect the conclusions we have made thus far. However, the main utility in introducing these ideas is that they will be crucial in the remainder of the paper.

If an MPE problem requires the estimation of all of the elements of $\bs{\phi}$ with some non-zero weight (i.e., $W_{kk}>0$ for all $k$), then any probe state with a singular QFIM results in a failed estimation. This is because the evolved probe state contains no information about one or more of the parameters (or some linear combination of them), and so some of the parameters of interest cannot be estimated. A more interesting case is when some of the parameters are unwanted. In this situation, singular QFIMs do not necessarily result in a failed estimation. We now introduce a method for analyzing any estimation problem with unwanted parameters. This will provide the relevant QCRB-derived estimation precision for all cases when the estimation doesn't fail.

 \subsubsection{Reducing the MPE problem}

  Consider an arbitrary QFIM $\mathcal{F}(\bs{\phi})$, which might be singular, for some $d$-dimensional vector $\bs{\phi}$. We may choose the order of the elements of $\bs{\phi}$ for our convenience, so we order them such that the first $d'$ are the parameters of interest and the remaining $d-d'$ parameters are the unwanted parameters that are of no interest, where $1\leq d' \leq d$. That is, $W_{kk} > 0$ for $k \in [1,d']$ and $W_{kk} = 0$ for $k \in [d'+1,d]$. There is some $s \in [d',d]$ such that we may write
 \begin{equation} 
 \mathcal{F}(\bs{\phi}) =  \mathcal{F}(\bs{\phi}_{A}) \oplus   \mathcal{F}(\bs{\phi}_{B}) ,
  \end{equation}
where $\bs{\phi}_{A}= (\phi_1,\dots,\phi_{s})$ is an $s$-dimensional vector and $\bs{\phi}_{B} =  (\phi_{s+1},\dots,\dots,\phi_{d})$ is an $(d-s)$-dimensional vector. For example, this always holds for $s=d$, as then $\bs{\phi}_{B}$ is zero-dimensional, and so this equation is trivial. 

Now consider the $\bs{\phi}_{A}$ obtained when we minimize the value of $s \in [d',d]$ over all labelling permutations of the unwanted parameters. We call this the \emph{reduced} $\bs{\phi}$ vector and we denote it by $\tilde{\bs{\phi}}$. Therefore, $\tilde{\bs{\phi}}$ is a sub-vector of $\bs{\phi}$ containing \emph{all} of the parameters of interest and the minimal number of unwanted parameters necessary such that we may write the QFIM for $\bs{\phi}$ as a direct sum of the QFIM for $\tilde{\bs{\phi}}$ and the QFIM for the remaining unwanted parameters. 

As $\mathcal{F}(\bs{\phi})$ can be separated out in this way, it is clear that the problem of estimating $\tilde{\bs{\phi}}$ and the associated $\bs{\phi}_{B}$ are essentially decoupled and can be considered independently. 
As we are not interested in any of the elements of $\bs{\phi}_{B}$, we may discard this part of the QFIM matrix (whether it is singular or not) and consider the problem using only the QFIM $\mathcal{F}(\tilde{\bs{\phi}})$. Any remaining unwanted elements in $\tilde{\bs{\phi}}$ cannot be decoupled from the parameters of interest (at least for the probe state under consideration), and hence are so-called ``nuisance'' parameters.

 If the QFIM for the reduced vector is invertible, then the relevant QCRB is
\begin{equation} 
E_{\tilde{\bs{\Phi}}}=\frac{1}{\mu} \text{Tr}(\tilde{W}\text{Cov}(\tilde{\bs{\Phi}})) \geq \frac{1}{\mu}\text{Tr}(\tilde{W}\mathcal{F}^{-1}(\tilde{\bs{\phi}})),
\label{eq:red-qcrb}
 \end{equation}
where $\tilde{W}$ is the reduced weighting matrix calculated by reordering and reducing $W$ so that its elements match with the elements of $\tilde{\bs{\phi}}$. If the reduced QFIM is still singular, the probe state under consideration will fail at estimating the parameters of interest in $\tilde{\bs{\phi}}$, as it contains no information about at least some of the relevant parameters. 

Note that this reduction procedure respects the fact that, for probe states that \emph{do} depend on one or more unwanted zero-weighted parameters, the appropriate QCRB is obtained from the QFIM that still includes these nuisance parameters (not doing so effectively assumes that they are being held constant). Indeed, this is the entire purpose of this method. The reduction procedure can be seen as a rigorous method for removing unwanted parameters from the problem that are decoupled from those of interest and that could be causing an artificially singular QFIM.

\subsubsection{An example}

For clarity, we briefly illustrate our vector reduction procedure with an example. Consider a 4-dimensional $\bs{\phi}$ where we only wish to estimate $\phi_1$ and $\phi_2$. A possible QFIM is
 \begin{equation} 
 \mathcal{F}(\bs{\phi}) =\begin{pmatrix} 
 \mathcal{F}_{11}  &  \mathcal{F}_{12} & 0 & \mathcal{F}_{14} \\
    \mathcal{F}_{21} &\mathcal{F}_{22} & 0 & \mathcal{F}_{24} \\
0  & 0 & \mathcal{F}_{33} & 0 \\
 \mathcal{F}_{41}  & \mathcal{F}_{42} & 0 & \mathcal{F}_{44} \\
 \end{pmatrix}.
  \end{equation}
This is singular if $\mathcal{F}_{33}=0$, but the value of $\mathcal{F}_{33}$ isn't relevant to the problem of interest. If any of the non-diagonal components in the $4$\textsuperscript{th} row or column are non-zero, the reduction process results in a QFIM for $\tilde{\phi} = (\phi_1,\phi_2,\phi_4)$ (keeping the original labelling of the elements of $\bs{\phi}$) of
 \begin{equation} 
 \mathcal{F}(\tilde{\bs{\phi}}) =\begin{pmatrix} 
 \mathcal{F}_{11}  &  \mathcal{F}_{12}  & \mathcal{F}_{14} \\
    \mathcal{F}_{21} &\mathcal{F}_{22}  & \mathcal{F}_{24} \\
 \mathcal{F}_{41}  & \mathcal{F}_{42} & \mathcal{F}_{44} \\
 \end{pmatrix},
  \end{equation}
with the discarded part of the QFIM given by $ \mathcal{F}(\phi_3)=\mathcal{F}_{33}$. If $\mathcal{F}(\tilde{\bs{\phi}})$ is singular the estimation of $\phi_1$ and $\phi_2$ will fail (e.g., if all elements of this QFIM are proportional to 1), but if $\mathcal{F}(\tilde{\bs{\phi}})$ is invertible then the precision to which they can be estimated is given by the QCRB obtained from this reduced QFIM, as given by Eq.~(\ref{eq:red-qcrb}). Although this analysis may seem rather obvious, these observations will be crucial in the next section.

\subsubsection{Discussion}
 We now briefly assess how allowing zero-weighted parameters ($W_{kk} = 0$) affects our conclusions in Sections~\ref{Sec:proof} and~\ref{Sec:proof2}. The effect of zero-weighting is simple: any sensors on which only unwanted zero-weighted parameters are encoded effectively act like ancillas, and if some, but not all, of the parameters encoded into a sensor are zero-weighted the estimation problem does not change substantially from the perspective of our discussions in this section. To be clear, when zero-weighted parameters are considered it is still true that, when the aim is to estimate some subset of $\bs{\phi}$, a local estimation strategy is preferable to a global estimation strategy (modulo the resource discussions for non-commuting generators). Indeed, in order to explicitly account for zero-weighted parameters, the only changes needed to the derivations throughout this section are essentially just notational.

  Finally, we note that finite estimation uncertainty bounds can be achieved even when the probe state contains no information about one or more of the parameters to be estimated if some suitable prior knowledge is available about at least these parameters. However, the precision is then always bounded by this prior knowledge. We leave for future work an analysis of the effects of prior knowledge on the results herein, but note that prior knowledge in quantum MPE has been considered in Refs.~\cite{tsang2014multiparameter,zhang2014quantum}. Alternatively an adaptive scheme using multiple different probe states, such as that in Ref.~\cite{berry2000optimal}, can allow for an estimation when some probe states contain no information about one or more of the parameters, but we do not consider this herein.

 \section{Estimating linear functions of local parameters}\label{Sec:functions}
In the previous section we have considered the MPE problem in which the aim is to estimate the vector $\bs{\phi}$ -- where each element of $\bs{\phi}$ is locally encoded into one of the quantum sensors -- with some arbitrary importance weighting on each parameter. This is a special case of the more general problem in which the aim is to estimate some \emph{functions} of the elements of $\bs{\phi}$, rather than estimating $\bs{\phi}$ itself, and we now turn to this. 

In this more general situation, we will show that, for a range of such estimation problems, probe states that are entangled between sensors can provide a better estimation precision than any separable state. Hence, global estimation strategies \emph{are} in many cases preferable to local estimation strategies for estimating certain non-trivial functions of $\bs{\phi}$ (rather than $\bs{\phi}$ itself). The main problem that we will consider is estimating a single linear function of $\bs{\phi}$, which we turn to in Section~\ref{sec:SLF}.

\subsection{Estimating functions of vectors}
\label{sec:fovs}
To begin, we review the basics of estimating functions of an unknown vector parameter.

\subsubsection{General functions}
Consider some $d$-dimensional vector 
\begin{equation}
\bs{\theta}=(f_1(\bs{\phi}),f_2(\bs{\phi}),\dots, f_d(\bs{\phi})),
\end{equation}
 for some continuously differentiable functions, $f_k:\mathbb{R}^d \to \mathbb{R}$, with $k=1,\dots,d$, that include the functions of interest (e.g., we may only wish to estimate $\theta_1=f_1(\bs{\phi})$). The QFIM for $\bs{\theta}$, which we denote $\mathcal{F}(\bs{\theta})$, can be expressed in terms of the QFIM for $\bs{\phi}$, which we now denote $\mathcal{F}(\bs{\phi})$. Specifically, it can be shown that \cite{paris2009quantum}
\begin{equation}
\mathcal{F}(\bs{\theta}) = B^T \mathcal{F}(\bs{\phi}) B, 
\label{QFIM-J}
 \end{equation}
where $B$ is a Jacobian matrix with elements $B_{lm}=  \partial \phi_l / \partial \theta_m $. Note that, as always, the QCRB for an estimator of $\bs{\theta}$ is given by
\begin{equation}
\text{Cov}(\bs{\Theta}) \geq \frac{\mathcal{F}(\bs{\theta})^{-1}}{\mu}  ,
\label{Eq:QCRB-theta}
 \end{equation}
where $\bs{\Theta}$ denotes an estimator of $\bs{\theta}$. Our figure of merit to minimize will now be $E_{\bs{\Theta}}=\text{Tr}(W\text{Cov}(\bs{\Theta}))$ where the weighting matrix $W$ encodes which functions are of interest, and any relative weightings.
  
For any choice of $\bs{\theta}$, we can estimate any or all of the elements of $\bs{\theta}$ by first estimating $\bs{\phi}$ and then calculating $\Theta_k=f_k(\bs{\Phi})$. The associated variance for the estimator of each element in $\bs{\theta}$ can be calculated using the standard propagation of uncertainty formula, that is
\begin{equation}
\text{Var}(\Theta_k) = \sum_{l=1}^d \left(\frac{\partial f_k}{\partial \phi_l }\right)^2 \text{Var}(\Phi_l).
\label{Eq:prop-of-errors}
 \end{equation}
However, in general, an optimal estimation of $\bs{\phi}$ and taking $\Theta_k=f_k(\bs{\Phi})$ will not give a minimal uncertainty for estimating the desired $\theta_k$ functions. 

\subsubsection{Linear functions}
Herein, we will only consider a particular (but important) sub-case of the general problem of estimating functions of an unknown vector -- the case of linear functions. This encompasses a wide range of metrology problems, many of which are discussed in Sections~\ref{sec:optics} and~\ref{sec:atomic_sensing} and later in this section, such as: measuring the phase difference in a single interferometer \cite{aasi2013enhanced} or in networked interferometers \cite{knott2016local}; measuring the average or the sum of multiple parameters \cite{komar2014quantum}; or measuring a linear gradient \cite{zhang2014fitting,ng2014quantum}. In the case of linear functions each element of $\bs{\theta}$ is a linear combination (over $\mathbb{R}$) of the elements of $\bs{\phi}$, and so 
\begin{equation}
\bs{\theta}=M\bs{\phi},
\label{phiMtheta}
\end{equation}
 for some $M \in \mathbb{R}^d \times \mathbb{R}^d$. To be a valid reparameterization of the estimation problem we require that the rows of $M$ are linearly independent. 

For linear functions we have $B \in \mathbb{R}^d \times \mathbb{R}^d$, where $B$ is the Jacobian matrix introduced in Eq.~(\ref{QFIM-J}), and furthermore $\bs{\phi} = B \bs{\theta}$. Hence $B=M^{-1}$, as the inverse of $M$ is guaranteed to exist due to the linear independence of the rows of $M$. As most equations will involve $B^{-1}=M$ rather than $B=M^{-1}$, we will write all formulas in terms of $M$ throughout.
In order to facilitate convenient comparisons between estimation uncertainties for different problems, it will be useful to demand that each row of $M$ is a normalized vector, i.e., 
\begin{equation}
\sum_{l=1}^d M_{kl}^2=1,
\end{equation}
 for all $k$. This only fixes an arbitrary constant for each $\theta_k=\sum_l M_{kl}\phi_{l}$, as the estimator uncertainty for $\theta_k'=c\theta_k$, for any constant $c$, can then be calculated to be $\text{Var}(\Theta_k')=c^2\text{Var}(\Theta_k)$ \footnote{This means that the signal-to-noise ration remains the same}. 

Finally, it will be useful to note that, if $[\hat{H}_l,\hat{H}_m]=0$ for all $l$ and $m$ then, from Eqs.~(\ref{QFIM-CovH}) and~(\ref{phiMtheta}) and $U(\bs{\phi}) = \exp(-i \bs{\phi}^T \hat{\bs{H}} )$, it follows that
\begin{equation}
  \mathcal{F}_{lm}(\bs{\theta}) =  4 \big( \langle \hat{H}_l'\hat{H}_m' \rangle-  \langle \hat{H}_l'\rangle \langle \hat{H}_m' \rangle \big),
\label{Eq:QFIM-gens-theta}
   \end{equation}
 where $\bs{\hat{H}'}$ is the vector of generators for $\bs{\theta}$, which is given by 
\begin{equation} \bs{\hat{H}'} = (M^{-1})^T \bs{\hat{H}}.\end{equation}

\subsection{Optical and atomic sensing\label{sec:optics-atoms}}
In the remainder of this section it will be significantly simpler to make rigorous statements about optimal probe states and estimation strategies if we restrict ourselves to a slightly more explicit physical setting. In particular, we now introduce a fairly general set of MPE problems that cover a range of optical interferometry and atomic sensing scenarios. It is possible that this generic setup encompasses other problems of interest outside of these two important settings. One significant restriction, which we now impose for the rest of this section, is that we will only consider estimation problems in which a single parameter is encoded into each quantum sensor. Hence, the number of sensors is equal to the number of unknown parameters $d$. For example, this implies that 3-dimensional magnetic field sensing (e.g., see \cite{baumgratz2016quantum}) is \emph{not} an MPE problem encompassed by the following.

\subsubsection{Total particle number as the resource}
Let each quantum sensor have the same underlying Hilbert space, $\mathcal{H}$, and let this Hilbert space be the direct sum of a vacuum Hilbert space, a single particle Hilbert space, a two particles Hilbert space, and so on. That is, the total $d$-sensor Hilbert space is
\begin{equation}
 \mathcal{H}_{\mathbb{S}} =   \underbrace{\mathcal{H} \otimes  \mathcal{H}    \otimes\cdots  \otimes  \mathcal{H}}_{d \,\,\text{Hilbert spaces}} ,
 \end{equation}
 and the Hilbert space of each sensor is
\begin{equation}
 \mathcal{H} = \mathcal{P}_{0} \oplus  \mathcal{P}_{1} \oplus  \mathcal{P}_{2} \oplus \cdots,
\label{Hd-sum}
 \end{equation}
where $\mathcal{P}_{n}$ is the Hilbert space containing all possible states of $n$ ``particles'', with $n \in \mathbb{N}$. 

We will consider estimation problems for which the resource is simply the average total number of particles. Formally, the resource function is
\begin{equation}
 \hat{N} = \hat{N}_1 + \hat{N}_2 + \cdots + \hat{N}_d,
 \end{equation}
and $\hat{N}_k$ acts non-trivially only on the $k$\textsuperscript{th} quantum sensor with the action
\begin{equation}
 \hat{n}_{\text{gen}}= (0 \cdot \mathds{1}_0) \oplus (1 \cdot \mathds{1}_1)\oplus (2 \cdot \mathds{1}_2)  \oplus \cdots  ,
\end{equation}
where $\mathds{1}_k$ is the identity in the $k$-particles sub-space. The $\hat{n}_{\text{gen}}$ operator is simply the local particle number operator for our general construction.

To allow our results to be as general as possible, we permit ancillary sensors. By this we mean that, throughout, we will implicitly consider the enlarged Hilbert space $\mathcal{H}_{\mathbb{S}}  \to \mathcal{H}_{\mathbb{S}} \otimes \mathcal{H} \otimes \mathcal{H} \otimes \dots$, for arbitrarily many ancillary sensors. The reason that  we will largely drop the ancillary sensors from the notation is because, for \emph{almost} the entire section, they will be irrelevant.

\subsubsection{Generators with linearly-spaced eigenvalues}

We have defined the general form of the quantum sensors and the resource operator. The final object needed to define an estimation problem is the parameter generators. We will now restrict the form that these generators can take. In particular, we will consider any estimation problem in which the parameter generator that encodes $\phi_k$ into the $k$\textsuperscript{th} quantum sensor acts on this quantum sensor as
\begin{equation}
 \hat{g} = \hat{g}_{0} \oplus  \hat{g}_{1} \oplus  \hat{g}_{2}  \oplus \cdots,
\label{gen-g}
    \end{equation}
where $ \hat{g}_{n}$ is an operator acting on the $n$-particles sub-space with linearly-spaced eigenvalues, given by
\begin{equation}
  \lambda_l(n)  = \delta n + l 
\label{e-states-rel}
 \end{equation}
where $l=-sn,\dots, sn$, for some $s$ such that $2s$ is a non-negative integer, and $\delta \in \mathbb{R}$. We assume that we do not have the trivial case of $\delta= s = 0$.

Later in this section we will be interested in the sub-space containing all those states of a sensor that contain at most $n_{\max}$ particles, i.e., those states with support only on those $\mathcal{P}_n$ with $n\leq n_{\max}$. In this space of states containing $n_{\max}$ or fewer particles we denote the eigenstates of $\hat{g}$ with maximal and minimal eigenvalues by $\ket{\lambda_{\text{max}}(n_{\max})}$ and $\ket{\lambda_{\text{min}}(n_{\max})}$, respectively. The corresponding maximum and minimum eigenvalues are the maximum and minimum of $\lambda_l(n)$ over all $l=-sn,\dots, sn$ and $n \leq n_{\max}$, respectively. It may be confirmed that these eigenvalues are given by $n_{\max}\lambda_{\max}$ and $n_{\max}\lambda_{\min}$, respectively, where
\begin{align}
\lambda_{\text{min}}=  \text{min}(0,\delta-s) ,\hspace{0.5cm}
\lambda_{\text{max}} =   \text{max}(0,\delta+s).
\end{align}
Hence, for all $n_{\max} \in \mathbb{N}$, we have that
\begin{align}
\hat{g}\ket{\lambda_{\text{min}}(n_{\max})} &=n_{\max} \lambda_{\text{min}}\ket{\lambda_{\text{min}}(n_{\max})}  , \label{eqld}\\
\hat{g}\ket{\lambda_{\text{max}}(n_{\max})} & = n_{\max} \lambda_{\text{max}}\ket{\lambda_{\text{max}}(n_{\max})}  \label{eqld2}.
\end{align}

The formalism we have introduced above covers a range of important problems in optical and atomic sensing (and possibly others), although certainly not all problems of interest in either setting. To clarify how this fairly abstract construction applies in these settings, we now briefly show how it includes certain MPE problems with atomic sensing networks and multi-mode interferometry.

\subsubsection{Example: Multi-mode optical sensing}
Let each of the $d$ sensors be a single-frequency optical mode, and so $\mathcal{H}$ is the Hilbert space of a quantum harmonic oscillator. As such, $\mathcal{P}_{n}$ is fairly trivial: it is the one-dimensional Hilbert space containing the $n$-photons number state, which we denote $\ket{n}$. This is consistent with the construction of $\mathcal{H}$ as the direct sum of all the $\mathcal{P}_{n}$ spaces (see Eq.~(\ref{Hd-sum})), because $\mathcal{H}$ is spanned by $\ket{n}$  for $n\in\mathbb{N}$. Furthermore $\hat{N}$ is the sum of the standard bosonic number operators for each of the $d$ optical modes, i.e., here $\hat{n}_{\text{gen}} = \hat{n}$ where $\hat{n}\ket{n}= n\ket{n}$. 

The canonical optical estimation problem involves estimating differences between linear phase shifts. This is encoded by taking $\hat{g}= \hat{n}$, and it is easily confirmed that $\hat{n}$ obeys the conditions required of our generic $\hat{g}$ operator. In particular, we have that $\delta=1$ and $s=0$, as $\hat{n}\ket{n} = n \ket{n}$, giving $\lambda_{\text{max}}=1$ and $\lambda_{\text{min}}=0$. The eigenstate of $\hat{n}$ in the space of states containing $n_{\max}$ or fewer photons with maximal (resp., minimal) eigenvalue is $\ket{n_{\max}}$ (resp., $\ket{0}$). It is then easily confirmed that Eqs.~(\ref{eqld} -- \ref{eqld2}) hold in this case (as $\hat{n} \ket{n} = n \ket{n}$). As such, using these relations, the results that we will present later in this section can be immediately applied to multi-mode optical sensing with linear phase shifts.

\subsubsection{Example: Atomic sensing}
Let each of the $d$ sensors consist of any number of distinguishable two-level atoms, where each sensor need not contain the same number of atoms. Therefore, the total Hilbert space of each sensor is constructed as $\mathcal{H} = \oplus^{n\in \mathbb{N}}\mathcal{P}_{n}$ where $ \mathcal{P}_{n}$ is the $2^n$-dimensional Hilbert space containing all possible states of $n$ two-level atoms. E.g., 
\begin{equation} 
\mathcal{P}_{2}= \{\ket{\uparrow\uparrow}, \ket{\uparrow\downarrow}, \ket{\downarrow\uparrow},\ket{\downarrow\downarrow}\},
\end{equation}
 where $\sigma_z \ket{\uparrow}=+ \ket{\uparrow}$ and $   \sigma_z\ket{\downarrow}=-\ket{\downarrow}$ with $\sigma_z=( \begin{smallmatrix}1 & 0 \\ 0 & -1 \end{smallmatrix})$. The resource operator, $\hat{N}$, simply counts the total number of atoms in all of the sensors. Note that this formulation of a network of atomic sensors allows for probe states with an indefinite total particle number, and indefinite numbers of particles in each sensor. These states can be avoided by considering only probe states in appropriate sub-spaces. Later, we will actually consider a sub-space which \emph{does} include states with indefinite numbers of particles at each sensor, but, in the case of atomic sensing, the optimal states for almost all the estimation problems we consider will have a definite number of atoms at each site (see discussions of Section~\ref{imps}). 

One common estimation problem for ensembles of two-level atoms is measuring the strength of a one-dimensional magnetic field. This is encoded by taking the generator for the parameter mapped into each sensor to be
\begin{equation}
 \hat{J}_z = \hat{J}_{z,0} \oplus \hat{J}_{z,1}  \oplus \hat{J}_{z,2} \oplus \dots,
\label{J_zgen_def}
 \end{equation}
where $\hat{J}_{z,n} :=\frac{1}{2} \sum_{i=1}^n \sigma_{i,z}$ is the $n$-atom collective $z$-spin operator, and here $\sigma_{i,z}$ is a $\sigma_z$ operator acting on the $i$\textsuperscript{th} qubit in the sensor (we define $\hat{J}_{z,0}=\mathds{1}$). This obeys the conditions required of the generic $\hat{g}$ generator, defined in Eq.~(\ref{gen-g}). In particular, we have that $s=1/2$ and $\delta=0$, giving $\lambda_{\text{max}}=1/2$ and $\lambda_{\text{min}}=-1/2$. The eigenstate of $\hat{J}_z$ in the sub-space of states containing at most $n_{\max}$ atoms in a sensor with maximal (resp., minimal) eigenvalue is $|\uparrow\rangle^{\otimes n_{\max}}$ (resp., $|\downarrow\rangle^{\otimes n_{\max}}$). As such, the results of this section can be applied to networks of atomic-ensemble sensors whereby local one-dimensional magnetic fields are imprinted onto each sensor. Furthermore, it is simple to show that equivalent sensing problems with multi-level atoms are also encompassed by this formalism (and potentially estimation problems in many other settings, such as using NV centers to measure a variety of physical quantities \cite{schirhagl2014nitrogen}).

\subsubsection{Probe states with finite particle number}

In order to make concrete statements about \emph{optimal} estimation strategies, in this section we will largely only consider estimation strategies that use probe states with support in a finite sub-space of the total Hilbert space of the sensing network. There are a range of natural choices for how to achieve this.

One option is to consider only probe states which contain at most $n_{\max}$ particles in each sensor. That is, we consider only probe states in the sub-space of all $d$-sensor states spanned by $d$-fold tensor products of all of the eigenvectors of $\hat{n}_{\text{gen}}$ with an eigenvalue less than or equal to $n_{\max}$. In other words, each sensor may contain no more than $n_{\max}$ particles, and in total there are no more than $dn_{\max}$ particles. Formally, this sub-space is
\begin{equation}
\check{\mathcal{S}}(n_{\max}) = \bigoplus_{k_j\leq n_{\max}}\left( \mathcal{P}_{k_1} \otimes  \mathcal{P}_{k_2} \otimes \dots \otimes  \mathcal{P}_{k_{|\mathbb{S}|}}\right) ,
\end{equation}
where the direct sum is over all $k_j\leq n_{\max}$, and $\mathcal{P}_n$ is the $n$-particles sub-space. Optimizing the sensing network over states in this sub-space is asking the question: ``I can have at most $n_{\max}$ particles in each sensor. How best can I use them to estimate the parameters of interest?''

Alternatively, we could instead consider optimizing the network over only those probe states containing $N_{\max}$ or fewer particles, in total, in the entire $d$-sensor state. By this, we mean that the probe states we consider are in the sub-space of states spanned by the eigenstates of $\hat{N}$ with an eigenvalue less than or equal to $N_{\max}$ \footnote{For example, for a two-mode optical sensing problem, both $\ket{N_{\max},0}$ and $\ket{N_{\max}-1,1}$ are states in this sub-space, but $\ket{N_{\max},1}$ is not.}. Formally, the precise sub-space is
\begin{equation}
\mathcal{S}(N_{\max}) = \bigoplus_{k_j}\left( \mathcal{P}_{k_1} \otimes  \mathcal{P}_{k_2} \otimes \dots \otimes  \mathcal{P}_{k_{|\mathbb{S}|}}\right) ,
\end{equation}
where the direct sum is over those $k_j$ such that $k_1+k_2+\dots+ k_{|\mathbb{S}|} \leq N_{\max}$. Optimizing the sensing network over states in this sub-space is asking the question: ``I have at most $N_{\max}$ particles. How best can I distribute them over the sensing network to estimate the parameters of interest?''

In the rest of this section, we will largely consider the latter of these two questions. It is important to realize that there is no guarantee that conclusions drawn for this sub-space can be directly applied to the full Hilbert space, or different sub-spaces, such as $\check{\mathcal{S}}(n_{\max})$. We will make this explicitly clear later.

\subsection{Estimating a single linear function}
\label{sec:SLF}

In this subsection we consider the problem of estimating a \emph{single} scalar parameter that is a linear function of $\bs{\phi}$. This includes many problems of practical interest. For example, we may wish to determine the average or the sum of multiple parameters \cite{komar2014quantum}, the difference between two parameters (e.g., optical phase differences) or a linear gradient \cite{zhang2014fitting,ng2014quantum}. We will only consider the sub-set of such estimation problems that are within the framework of Section~\ref{sec:optics-atoms}, and we will consider optimizing the sensing network over probe states in the $\mathcal{S}(N_{\max})$ sub-space (see above). As such, these results should be applied with caution outside of this sub-space.

 Before we begin, it is important to note that (in general) the problem we are now considering is very different in nature to that of estimating $\bs{\phi}$ itself: the elements of $\bs{\phi}$ are each encoded locally into a single sensor, but a linear function of these elements is generically a very non-local property of the sensing network. Any such function can be estimated via local estimates of the locally encoded parameters at each sensor. However, the optimal protocol for estimating $\bs{\phi}$ is \emph{not} necessarily the optimal protocol for estimating a given function of $\bs{\phi}$. Keeping this in mind should make it clear that the following results complement, and do not contradict, the conclusions of Section~\ref{sec:est-phi}. 

The problem of estimating a single arbitrary linear function of $\bs{\phi}$ may be encoded by taking any $d$-dimensional $\bs{\theta}$ that is a reparameterization of $U(\bs{\phi})$ with $\theta_1=(M \bs{\phi})_1$ the parameter of interest. That is, we have
\begin{equation} 
M = \begin{pmatrix} \bs{v}^T \\ \widetilde{M} \end{pmatrix} ,
\label{M-1p}
\end{equation}
where $\theta_1 =\bs{v}^T\bs{\phi}$, and $ \widetilde{M}$ is a $(d-1) \times d$ matrix consisting of rows of normalized vectors that are all linearly independent to $\bs{v}^T$ and each other. The figure of merit that we wish to minimize is then $E_{\bs{\Theta}} = \text{Tr}(\text{Cov}(\bs{\Theta})_{11})$, which is a single element of the estimator covariance matrix. This is simply the quantity $\text{Var}(\Theta_1)$, and so we will use this more explicit notation in this section ($\bs{\Theta}$ is the vector estimator of $\bs{\theta}$, and hence $\Theta_1$ denotes the scalar estimator of $\theta_1$).

For estimation problems that are within the framework we are considering in this section (see above), it is possible to find some general bounds on the obtainable estimation uncertainty. A given probe state can be used to estimate $\theta_1$, even in many cases when $\mathcal{F}(\bs{\theta})$ is singular. In particular, a probe state will give a finite estimation uncertainty as long as after the vector ``reduction'' process, which we introduced in Section~\ref{Sec:singular-QFIMs} and which maps $\bs{\theta} \to \tilde{\bs{\theta}}$, the resultant reduced QFIM $\mathcal{F}(\tilde{\bs{\theta}})$ is invertible. Here $\tilde{\bs{\theta}}=(\theta_1,\dots)$ is an $s$ dimensional vector for some $s \in [1,d]$, which includes the parameter of interest and possibly some nuisance parameters (this is the case if $s>1$). Note that the value of $s$ depends on the particular probe state under consideration.

Consider an arbitrary pure probe state $\psi \in \mathcal{S}(N_{\max})$ that has an invertible QFIM $\mathcal{F}(\tilde{\bs{\theta}})$, where $\tilde{\bs{\theta}}$ is the reduced vector for this state. For any such $\psi$, it follows that
\begin{align}
\text{Var}(\Theta_1) 
&\geq \frac{1}{\mu} [\mathcal{F}(\tilde{\bs{\theta}})^{-1}]_{11} ,  \label{ineq-f-1}
\\&  \geq  \frac{1}{ \mu \mathcal{F}(\tilde{\bs{\theta}})_{11} }, \label{ineq-f-2}
\\&= \frac{1}{4 \mu \text{Var} (\psi,\hat{H}_{\bs{v}})},\label{ineq-f-3}
\\& \geq \frac{1}{ 4 \mu \max_{\Psi \in \mathcal{S}_{\bs{v}} }[\text{Var} (\Psi,\hat{H}_{\bs{v}})]},  \label{ineq-f-4}
\end{align}
where $\hat{H}_{\bs{v}}= \bs{v}^T\hat{\bs{H}}$, $ \mathcal{S}_{\bs{v}}$ is the sub-space of $\mathcal{S}(N_{\max})$ containing states that have an invertible reduced QFIM, and $\mu$ is the number of repeats of the experiment. These equations hold for the following reasons:

The inequality of Eq.~(\ref{ineq-f-1}) follows directly from the QCRB for the reduced $\bs{\theta}$, as given explicitly in Eq.~(\ref{eq:red-qcrb}). This inequality can always be saturated even when the QCRB for the vector $\tilde{\bs{\theta}}$ cannot be saturated, because a measurement always exists that saturates the QCRB for the estimation of any one of the elements of a vector parameter in isolation \footnote{This follows because $\mathcal{F}$ is the smallest matrix such that $\mathcal{F}\geq F$ for all measurements \cite{ballester2004estimation}. Therefore, there is a measurement with an associated $F$ such that $F_{kk} = \mathcal{F}_{kk}$ for any single $k$.}. The inequality of Eq.~(\ref{ineq-f-2}) follows from Eq.~(\ref{Eq:F>1/F}) and can be saturated only if the first row and column of the QFIM for $\bs{\theta}$ contains only zeros except on the diagonal (and hence $\tilde{\bs{\theta}}$ is a scalar, i.e., $\tilde{\bs{\theta}}=\theta_1$). 

The equality of Eq.~(\ref{ineq-f-3}) follows from Eq.~(\ref{Eq:QFIM-gens-theta}) \footnote{Where we choose the other elements of $\bs{\theta}$ such that $M^{-1}=M^{T}$. We may do this without lose of generality, as we discuss later.}. The final inequality of Eq.~(\ref{ineq-f-4}) is fairly trivial: the maximum is guaranteed to exist as by assumption we are only considering probe states in the finite-dimensional ``$N_{\max}$ particles or fewer'' sub-space $\mathcal{S}(N_{\max})$, and we only take the maximum over those states with an invertible QFIM for the reduced $\bs{\theta}$, as all other states fail at the estimation of $\theta_1$. If a probe state saturates all of the inequalities in Eqs.~(\ref{ineq-f-1} -- \ref{ineq-f-4}) then it is an optimal probe state for estimating $\theta_1$. Note, however, that we have not shown that there always exists a state that saturates all of these inequalities.

Throughout the remainder of this section we will denote the average \emph{total} number of particles in a state by $\bar{N}=\langle \hat{N} \rangle$. This quantity is distinct from $N_{\max}$, which denotes the maximum total number of particles defining the sub-space $\mathcal{S}(N_{\max})$ that we are considering in this subsection. In general $\bar{N} \neq N_{\max}$. When the probe state contains a definite total number of particles we will often make this explicit by letting $\bar{N} \to N$.

\subsubsection{Estimating sums and averages}
\label{sec:sum}
We turn now to specific problems of interest. First we consider the natural problem of estimating the parameter
  \begin{equation} \theta_{\text{sum}} = \frac{1}{\sqrt{d}} ( \phi_1 +\phi_2+ \cdots + \phi_d),
   \end{equation}
and we will see that all of the bounds in Eqs.~(\ref{ineq-f-1} -- \ref{ineq-f-4}) can be saturated in this case, using a \emph{global estimation strategy}. Up to a constant, this problem is equivalent to estimating either the average of all $d$ parameters, or the sum of all $d$ parameters. Parameter averages of this sort have been previously considered by Komar \emph{et al.} \cite{komar2014quantum}, in the specific setting of a networks of clocks, and our results include this aspect of their conclusions as a special case (we discuss the work of Komar \emph{et al.}~\cite{komar2014quantum} further in Section~\ref{sec:atomic_sensing}).

For simplicity, let us restrict ourselves to the case in which $N_{\max}$ is divisible by $d$, so that $N_{\max}=nd$ for some integer $n$. We will now show that the state with an optimal precision bound for estimating $\theta_{\text{sum}}$ is 
 \begin{equation}
  \ket{\psi_{\textsc{ghz}}(n,d)} = \frac{1}{\sqrt{2}} \left( \ket{\lambda_{\text{min}}( n) }^{\otimes d} +\ket{\lambda_{\text{max}}( n)}^{\otimes d} \right) ,
 \end{equation}
where $\ket{\lambda_{\text{min}}( n)}$ and $\ket{\lambda_{\text{max}}( n)}$ are the minimal and maximal eigenstates of the non-trivial action of each generator when restricted to the ``$n$-particles or fewer'' sub-space, as defined in Section~\ref{sec:optics-atoms}. We will refer to $  \ket{\psi_{\textsc{ghz}}(n,d)}$ as an ($N_{\max}$-particle) GHZ state, and we will largely drop the explicit $n$ and $d$ dependence from the notation for this state. 

By using Eq.~(\ref{QFIM-CovH}) and noting that $\psi_{\textsc{ghz}}$ is maximally correlated, it may be confirmed that
\begin{equation}
\mathcal{F}(\bs{\phi}) = n^2(\lambda_{\text{max}} -\lambda_{\text{min}})^2  \begin{pmatrix}
  1 & 1 & \cdots & 1 \\
  1 & 1 & \cdots & 1 \\
  \vdots  & \vdots  & \ddots & \vdots  \\
 1& 1 & \cdots & 1
 \end{pmatrix}.
\end{equation}

The precision with which $\theta_{\text{sum}}$ can be estimated should not be affected by the particular functions we choose for the remaining unwanted parameters $\theta_2$, $\theta_3,$ $\dots,$ $\theta_d$, and hence we can choose their form for our convenience (note, we are not fixing the value of these parameters). Clearly we have that $\theta_{\text{sum}} = \bs{v}^T_{\text{sum}} \bs{\phi}$ where 
\begin{equation}
 \bs{v}_{\text{sum}} = \frac{1}{\sqrt{d}}(1,1,\dots, 1)^T.
\end{equation} 
We choose to take $\theta_2$, $\theta_3,$ $\dots,$ $\theta_d$ to be all orthogonal to $\theta_{\text{sum}}$, by which we mean that  $\theta_k =\bs{v}_k^T\bs{\phi}$ with $\bs{v}_k^T \bs{v}_{\text{sum}} = 0$ for all $k=2,3,\dots,d$. Therefore $M$ (as given in Eq.~(\ref{M-1p})) is an orthogonal matrix, i.e., $M^{-1}=M^T$. With this choice for our parameterization of the unwanted parameters, we can use Eq.~(\ref{QFIM-J}) to show that
  \begin{equation}
\mathcal{F}(\bs{\theta})= dn^2 (\lambda_{\text{max}} -\lambda_{\text{min}})^2  \begin{pmatrix}
  1 & 0 & \cdots & 0 \\
  0 & 0 & \cdots & 0 \\
  \vdots  & \vdots  & \ddots & \vdots  \\
 0 & 0 & \cdots & 0
 \end{pmatrix}.
\end{equation}
Therefore, when the reduction process is applied to $\bs{\theta}$, the result is a one-parameter problem $\tilde{\bs{\theta}}=\theta_{\text{sum}}$, and $\mathcal{F}(\tilde{\bs{\theta}})=   d n^2 (\lambda_{\text{max}} -\lambda_{\text{min}})^2$. Hence we arrive at the bound
\begin{equation}
\text{Var} (\Theta_{\text{sum}}) \geq \frac{1} { \mu dn^2 (\lambda_{\text{max}} -\lambda_{\text{min}})^2}.
\label{Eq:var-ghz}
\end{equation}

To write this bound in terms of $N_{\max}$ we simply recall that we have $N_{\max}=nd$, giving
\begin{equation}
\text{Var} (\Theta_{\text{sum}}) \geq \frac{d} {\mu N_{\max}^2 (\lambda_{\text{max}} -\lambda_{\text{min}})^2}.
\label{Eq:var-ghz2}
\end{equation}
Note that this calculation still holds without taking $M$ to be orthogonal (as is to be expected -- taking $M$ to be orthogonal was not an assumption), but it is less straightforward.

In order to show that this state has the optimal precision bound via the relations of Eqs.~(\ref{ineq-f-1} -- \ref{ineq-f-4}), it only remains to show that it saturates the inequality of Eq.~(\ref{ineq-f-4}). That is, we have that
\begin{equation}
\text{Var}(\psi_{\textsc{ghz}} ,\hat{H}_{\bs{v}_{\text{sum}}})= \frac{N_{\max}^2 (\lambda_{\text{max}} -\lambda_{\text{min}})^2}{4d},
\end{equation}
 and we need to show that
\begin{equation}
\text{Var}(\psi_{\textsc{ghz}} ,\hat{H}_{\bs{v}_{\text{sum}}})=\max_{\Psi\in  \mathcal{S}_{\bs{v}_{\text{sum}}}}\left[\text{Var}(\Psi ,\hat{H}_{\bs{v}_{\text{sum}}})\right].
\label{GHZ-max-v}
\end{equation}
The proof is straightforward, but we delay this until Section~\ref{gen-rat-fun}, where we consider something slightly more general (see Eq.~(\ref{var-max----}) and the following text).

We have shown that the GHZ state is the optimal probe state for measuring parameter averages or sums -- we did this by showing that it saturates all of the bounds in Eqs.~(\ref{ineq-f-1} -- \ref{ineq-f-4}). This is perhaps not surprising. Furthermore, it may be confirmed that the GHZ state provides the optimal precision for a fixed amount of resources (we made no mention of resources in the argument above). The average number of particles in the GHZ state is either $\bar{N} = N_{\max}$ or $\bar{N}=\frac{1}{2} N_{\max}$. These values are obtained when $\lambda_{\text{max}}$ and $\lambda_{\text{min}}$ are both non-zero, or one of them is zero, respectively (this may be confirmed by reference to Section~\ref{sec:optics-atoms}, noting that in the former case the state contains a definite number of particles). Therefore, regardless of the specific details of the problem, it is guaranteed that $\text{Var} (\Theta_{\text{sum}}) \propto1/ \mu dn^2 \propto d/\mu \bar{N}^2$ and this is the Heisenberg scaling (the factor of $d$ is an artifact of normalizing the vector $\theta_{\text{sum}}$). Using more classical repeats of any state with fewer particles will result in a greater estimator uncertainty. Similar resource counting arguments apply throughout the remainder of this section and are not explicitly mentioned again.

An estimation strategy employing the $\psi_{\textsc{ghz}}$ probe state is a global estimation strategy, and as such it is often much harder to implement physically than a local estimation strategy -- whether using such a GHZ probe state (or a similar state) is at all plausible will depend on the physical problem of interest. Hence, it is also interesting to consider what the enhancement over the optimal local strategy is. The optimal local estimation strategy (when restricted to the sub-space under consideration) uses the probe state
 \begin{equation}
  \ket{\psi_{\textsc{loc}}(n,d)} = {1 \over 2^{\frac{d}{2}}} \left( \ket{\lambda_{\text{min}}( n) } +\ket{\lambda_{\text{max}}( n)}\right)^{\otimes d} .
\label{local-NO-1}
 \end{equation}
We prove this later (see Section~\ref{gen-rat-fun}, and in particular Eq.~(\ref{loc-l-b}) and the following text). Simple algebra shows that this state has the precision bound 
\begin{equation}
\text{Var} (\Theta_{\text{sum}}) \geq \frac{d^2}{ \mu N_{\max}^2  (\lambda_{\text{max}} -\lambda_{\text{min}})^2}.
\label{sep-eq-w}
\end{equation}
 By comparing this to Eq.~(\ref{Eq:var-ghz2}), it is clear that the reduction in the estimation uncertainty, obtained by using a global estimation strategy and as quantified by the estimator variance, is a multiplicative factor of $1/d$. Hence, the possible enhancement scales with the number of sensors in the network.

The derivation above is fairly general, and applies both to atomic and optical sensing. To clarify the result we provide the explicit form for this state in the case of qubits and optical modes. For two-level atoms, with the parameter encoded into each quantum sensor via the generator $\hat{J}_z$ (see Eq.~(\ref{J_zgen_def}) for definition), then $\psi_{\textsc{ghz}}$ is simply
 \begin{equation}
  \ket{\psi_{\textsc{ghz}}} = \frac{1}{\sqrt{2}} \left( \ket{\uparrow}^{\otimes N} +\ket{\downarrow}^{\otimes N} \right) ,
 \end{equation}
where $N= N_{\max}$ is the total number of atoms, and there are $N/d$ atoms per sensor. The estimation precision is $\text{Var} (\Theta_{\text{sum}}) \geq d / \mu N^2$ where $N$ is the total number of atoms in the state (as here $N_{\max}=N$).  As noted above, the discrepancy to the well-known Heisenberg limit $\text{Var}\geq 1 / \mu N^2$, for estimating the magnitude of a one-dimensional magnetic field, is due to the normalization of $\theta_{\text{sum}}$. The factor of $d$ may be removed by letting $\theta_{\text{sum}} \to \sqrt{d} \theta_{\text{sum}}$. This atomic GHZ state, and similar probe states, are discussed further in Section~\ref{sec:atomic_sensing}.

Consider the problem of measuring the sum of optical phases applied to different optical quantum sensors. In this case the $\psi_{\textsc{ghz}}$ state is
 \begin{equation}
  \ket{\psi_{\textsc{ghz}}} = \frac{1}{\sqrt{2}} \left( \ket{n,n, \dots, n } +\ket{0,0,\dots,0}\right) ,
 \end{equation}
with the precision still given by $\text{Var} (\hat{\theta}_1) \geq d / \mu N_{\max}^2$. In contrast to the atomic case, $\bar{N} \neq N_{\max}$ and the state contains an indefinite number of photons. We have that $\bar{N} = \frac{1}{2}N_{\max}$, and this is the relevant value for the resource. This state might be termed an \emph{entangled NO} (ENO) state, as it is an entangled extension of the single-mode ``NO'' state $\ket{\psi_{\textsc{no}}} \propto \ket{n}+  \ket{0} $, which has been previously considered in the optical metrology literature \cite{knott2014effect,knott2016local,tilma2010entanglement} (see also Section~\ref{sec:optics}). As estimating sums of optical phases is not (to our knowledge) a problem of practical importance, we will not discuss this optical GHZ-like state further.

\subsubsection{General equally weighted linear functions}
\label{GEWLF}
 The analysis given above can be adapted to apply to any $\theta_1 = \bs{v}^T \bs{\phi}$ where each element in $\bs{v}$ has an equal modulus, i.e., $|v_k |=1/\sqrt{d}$ for all $k=1,\dots,d$. The required change in the probe state is simply to exchange $\ket{\lambda_{\text{max}}(n)} \leftrightarrow \ket{\lambda_{\text{min}}(n)}$ for the $k$\textsuperscript{th} sensor in $ \ket{\psi_{\textsc{ghz}}}$ if $v_k = -1 /\sqrt{d}$, rather than $v_k = 1/\sqrt{d}$. Specifically, the optimal probe state is
   \begin{equation}
   \ket{\psi_{\textsc{ghz}}(\bs{v})} = \frac{1}{\sqrt{2}} \left( \ket{\psi_{\text{max}}(\bs{v})}+  \ket{\psi_{\text{min}}(\bs{v})}\right),
 \end{equation}
 where $ \ket{\psi_{\text{max}}(\bs{v})}$ and $\ket{\psi_{\text{min}}(\bs{v})}$ are given by
 \begin{align} 
 \ket{\psi_{\text{max}}(\bs{v})}&= \ket{\lambda_{m(v_1)}(n)}\cdots  \ket{\lambda_{m(v_d)}(n)},\\ 
\ket{\psi_{\text{min}}(\bs{v})} &= \ket{\lambda_{m(-v_1)}(n)}\cdots  \ket{\lambda_{m(-v_d)}(n)}, 
  \end{align}
where $m(v_k)=\text{max}$ if $v_k>0$ and $m(v_k)=\text{min}$ if $v_k<0$. This state saturates all of the bounds in Eqs. (\ref{ineq-f-1} -- \ref{ineq-f-4}) for estimating $\theta_1$, for any such $\bs{v}$. This follows in an analogous way to our proof that $\psi_{\textsc{ghz}}$ is optimal for estimating the sum of the $\phi_k$. All of the discussions of Section~\ref{sec:sum} carry over directly to this more general case, and we do not repeat them here. 

Therefore, we have shown that, when measuring \emph{any} single linear function that is an equally weighted combination of all of the scalar parameters encoded into each of the $d$ sensors, a global estimation strategy is always optimal. Furthermore, the reduction in the estimator variance, in comparison to the optimal local estimation strategy, is a factor of $1/d$, where $d$ is the number of sensors.

Before moving on, we note that a special case of this scenario recovers a minor generalization of the well-known result that a NOON state \cite{lee2002quantum} is the optimal fixed-photon-number probe state for measuring a phase difference between two modes. Specifically, the standard optical interferometry problem is a two-parameter problem (an unknown phase shift in each arm), the function of interest is $\theta_{\text{dif}} = 2^{-\frac{1}{2}}(\phi_2 - \phi_1)$, and the maximal and minimal eigenstates of the difference of two photon number operators in the sub-space of a photonic mode containing states with at most $N_{\max}$ photons are $\ket{N_{\max},0}$ and $\ket{0,N_{\max}}$, respectively. For this photonic case, we then obtain 
\begin{equation} | \psi_{\textsc{ghz}}(\bs{v}_{\text{dif}}) \rangle =\frac{1}{\sqrt{2}}( \ket{N,0} + \ket{0,N}),
\end{equation}
where $\bs{v}_{\text{dif}}= 2^{-\frac{1}{2}} (1,-1)$, and we have let $N_{\max} \to N$ as this is a fixed total photon number state. This is a standard optical NOON state. The estimation precision is then $\text{Var} (\Theta_{\text{dif}}) \geq 2 / \mu N^2$. Again, the factor of 2 discrepancy to the well know Heisenberg limit of $1/\mu N^2$ is due to the normalization of $\theta_{\text{dif}}$.

\subsubsection{Estimating a single generic rational function}
\label{gen-rat-fun}
For more general choices of the single parameter, $\theta_1$, defined by more general choices of $\bs{v}$, the situation is significantly more subtle. Here, we consider any $\bs{v}$ for which $v_k / \|\bs{v}\|_1$ is a rational number for all $k$, where $\|\bs{v}\|_1 :=\sum_k |v_k|$ denotes the 1-norm of $\bs{v}$. Moreover, we consider finding the optimal probe state in the ``$N_{\max}$ particles or fewer'' sub-space $\mathcal{S}(N_{\max})$ for any $N_{\max}$ such that $N_{\max}v_k / \|\bs{v}\|_1$ is an integer for all $k$. There is always such an $N_{\max}$, due to the rationality condition on $\bs{v}$, although the minimum such $N_{\max}$ could be arbitrarily large in general. Despite this, it seems likely that for most problems of interest $N_{\max}$ would not need to be significantly larger than $d$. 

The reason behind considering such restricted vectors and values for $N_{\max}$ is that, under the conditions above, if we allocate resources to the $k$\textsuperscript{th} sensor in proportion to $|v_k|$, then this is an integer number of particles. Note that this recovers the situation considered earlier for equally weighted functions (we assumed $N_{\max}=dn$ for some integer $n$). For notational simplicity, we also assume that $v_k \geq 0$ for all $k$. The following conclusions can easily be extended to vectors with negative elements, by following the procedure in Section~\ref{GEWLF}.

We now consider strategies for estimating $\theta_1  = \bs{v}^T \bs{\phi}$ for any such $\bs{v}$. For some vector of positive integers, $\bs{w}$ (this is entirely different from the weighting matrix, $W$, defined earlier), define the \emph{weighted} GHZ state
 \begin{equation} 
 \ket{\psi_{\textsc{w-ghz}}(\bs{w})} = \frac{1}{\sqrt{2}} \left( \ket{\psi_{\text{w-max}}(\bs{w})}+  \ket{\psi_{\text{w-min}}(\bs{w})}\right),
 \end{equation}
 where $\ket{\psi_{\text{w-max}}(\bs{w})}$ and $\ket{\psi_{\text{w-min}}(\bs{w})}$ are given by
 \begin{align} 
 \ket{\psi_{\text{w-max}}(\bs{w})}&= \ket{\lambda_{\text{max}}(w_1)}\cdots  \ket{\lambda_{\text{max}}(w_d)},\\
\ket{\psi_{\text{w-min}}(\bs{w})} &= \ket{\lambda_{\text{min}}(w_1)}\cdots  \ket{\lambda_{\text{min}}(w_d)}.
\label{dlfks}
  \end{align}
In the sub-space $\mathcal{S}(N_{\max})$ under consideration, we \emph{conjecture} that the optimal strategy for estimating $\theta_1  = \bs{v}^T \bs{\phi}$ uses the weighted GHZ state with the particular weighting $\bs{w} = N_{\max}\bs{v}/\|v\|_1$ as the probe state. We call this state a \emph{proportionally weighted} GHZ state. The basis for our conjecture is outlined below.

 It may be confirmed via Eq.~(\ref{QFIM-CovH}) that the proportionally weighted GHZ state has a QFIM for $\bs{\phi}$ of
\begin{equation} 
\mathcal{F}(\bs{\phi}) =\frac{N_{\max}^2}{\|\bs{v}\|_1^2}  (\lambda_{\text{max}} -\lambda_{\text{min}})^2  \bs{v}\bs{v}^T.
\label{Eq-F-wghz1}
 \end{equation}
As before, and without loss of generality, we may choose the unwanted parameters so that they are all orthogonal to $\theta_1$, and this implies that
\begin{equation}
M^T =\begin{pmatrix} \bs{v}, \bs{v}_2, \bs{v}_3 , \dots \end{pmatrix},
\end{equation}
where $\bs{v}^T\bs{v}_k= 0$ for $k=2,\dots,d$. From Eq.~(\ref{QFIM-J}) we have that $\mathcal{F}(\bs{\theta})=M\mathcal{F}(\bs{\phi}) M^T$, and therefore from Eq.~(\ref{Eq-F-wghz1}) it follows that
\begin{equation} 
\label{eq:reduced_QFIM_sums}
\mathcal{F}(\bs{\theta}) = \frac{N_{\max}^2}{\|\bs{v}\|_1^2}(\lambda_{\text{max}} -\lambda_{\text{min}})^2  \begin{pmatrix}
  1 & 0 & \cdots & 0 \\
  0 & 0 & \cdots & 0 \\
  \vdots  & \vdots  & \ddots & \vdots  \\
 0 & 0 & \cdots & 0
 \end{pmatrix}.
 \end{equation}
Via the reduction procedure, we then arrive at the final precision bound for estimating $\theta_1$:
\begin{equation}
\text{Var} (\Theta_{1}) \geq \frac{\|\bs{v}\|_1^2}{ \mu N_{\max}^2 (\lambda_{\text{max}} -\lambda_{\text{min}})^2}.
\label{Eq:var-wghz}
\end{equation}
This bound is saturable for the same reasons as throughout this section. Interestingly, the precision to which this proportionally weighted GHZ can estimate $\theta_1 = \bs{v}^T \bs{\phi}$ depends on the 1-norm of $\bs{v}$. In the special case of $\bs{\theta}_{\text{sum}} \propto \phi_1 +\dots+\phi_d$, Eq.~(\ref{Eq:var-wghz}) reduces to Eq.~(\ref{Eq:var-ghz2}), because $\|\bs{v}_{\text{sum}}\|_1 = \sqrt{d}$, and the proportionally weighted GHZ state is a standard GHZ state. 

The proportionally weighted GHZ state is guaranteed to be the optimal probe state for estimating $\theta_1 = \bs{v}^T \bs{\phi}$, for any $\bs{v}$ that obeys the rationality constraint outlined above, \emph{if} it saturates all of the bounds in Eq.~(\ref{ineq-f-1} -- \ref{ineq-f-4}). Because the QFIM of this state is diagonal, it is evident that it saturates the bounds in Eq.~(\ref{ineq-f-1} -- \ref{ineq-f-3}). Hence, it only remains to consider whether this state is the maximal variance state for $\hat{H}_{\bs{v}} = \bs{v}^T\bs{\hat{H}}$ in the sub-space of pure states containing $N_{\max}$ or fewer particles with an \emph{invertible} reduced QFIM, and we now turn to this.

The maximal variance of $\hat{H}_{ \bs{v}}$ over all states $\psi \in \mathcal{S}(N_{\max})$ is obtained for the state that is an equal superposition of those eigenvectors of $\hat{H}_{\bs{v}}$ in $\mathcal{S}(N_{\max})$ with minimal and maximal eigenvalues.  Let $v_{\max} = \max \{v_k\}$. By noting that the eigenvectors of $\hat{H}_{\bs{v}}$ are tensor products of the eigenvectors of $\hat{g}$ (the generating operator), and then maximising over all such eigenvectors in $\mathcal{S}(N_{\max})$, it is not difficult to confirm that the maximal and minimal eigenvalues of $\hat{H}_{ \bs{v}}$ in this sub-space are $v_{\text{max}}  N_{\max} \lambda_{\text{max}}$ and $v_{\text{max}}  N_{\max} \lambda_{\text{min}}$, respectively. Hence, for any probe state $\psi \in \mathcal{S}(N_{\max})$, we have that
\begin{align} 
\text{Var}(\psi,\hat{H}_{ \bs{v}} )
& \leq   \frac{v^2_{\text{max}}  N_{\max}^2 (\lambda_{\text{max}} -\lambda_{\text{min}})^2}{4} \label{var-max-proof4}.
\end{align}
Eqs.~(\ref{var-max-proof4}) clearly implies that
\begin{align} 
\max_{\psi \in  \mathcal{S}_{\bs{v}}} \left( \text{Var}(\psi,\hat{H}_{ \bs{v}} ) \right) \leq   \frac{v^2_{\text{max}}  N_{\max}^2 (\lambda_{\text{max}} -\lambda_{\text{min}})^2}{4} \label{var-max----}, 
\end{align}
where, as defined earlier, $ \mathcal{S}_{\bs{v}} \subset \mathcal{S}(N_{\max})$ is the sub-space of states in $\mathcal{S}(N_{\max})$ that have a reduced QFIM for $\bs{\theta}$ that is invertible. These are the states in $ \mathcal{S}(N_{\max})$ that are sensitive to $\theta_1$. If the proportionally weighted GHZ state saturates this bound then we have proven that we have found the optimal probe state. 

There is one set of circumstances under which the proportionally weighted GHZ states \emph{does} saturate this bound: when all the non-zero elements of $\bs{v}$ are equal (or have equal modulus, if we permit $\bs{v}$ to have negative elements). In this case, the parameter to estimate is the average of those $\phi_k$ with $v_k \neq0$, and the proportionally weighted GHZ state is a standard GHZ over those sensors for which $v_k\neq0$. This is a minor generalization of the case we covered in detail earlier in Section~\ref{sec:sum}.

In particular, consider the extremal case of $\bs{v}=(1,1,\dots,1)/\sqrt{d}$, and so we have $v_{\text{max}} = 1/\sqrt{d}$. Putting this into Eq.~(\ref{var-max----}) confirms Eq.~(\ref{GHZ-max-v}). This completes the proof that the GHZ state is optimal for estimating parameter averages, which we had deferred until now. The other extremal case is when only one of the $v_k$ is non-zero. In this case the proportionally weighted GHZ state is a state in which all of the particles are in a single sensor. This is the one situation where a proportionally weighted GHZ is a separable state. That such a state is optimal for this estimation problem is trivial, and it is consistent with our results in Section~\ref{sec:est-phi}.

For all other $\bs{v}$, the proportionally weighted GHZ state does \emph{not} saturate the bound of Eq.~(\ref{var-max----}). However, there is no state in $ \mathcal{S}_{\bs{v}}$ that saturates this bound. There are states in $\mathcal{S}(N_{\max})$ such that $\text{Var}(\psi,\hat{H}_{ \bs{v}})  =v^2_{\text{max}}  N_{\max}^2 (\lambda_{\text{max}} -\lambda_{\text{min}})^2/4$, but crucially these states are insensitive to $\theta_1$, and so are not in $ \mathcal{S}_{\bs{v}}$. The states for which this equality holds are those whereby all of the particles are distributed over only those sensors with a label $k$ such that $v_k = v_{\text{max}}$. For those $\bs{v}$ that are not equally weighted over the non-zero components, the states that obey this equality are completely insensitive to $\theta_1$.

In light of these observations, our conjecture that the proportionally weighted GHZ state is the optimal probe state for estimating $\theta_1=\bs{v}^T\bs{\phi}$ for general rationally proportioned $\bs{v}$ is reduced to showing that
\begin{equation}
 \max_{\psi \in  \mathcal{S}_{\bs{v}}} ( \text{Var}(\psi,\hat{H}_{ \bs{v}} ) )=  \frac{N_{\max}^2 (\lambda_{\text{max}} -\lambda_{\text{min}})^2}{4\|\bs{v}\|_1^2}.
\end{equation}
Unfortunately, we do not know how to either confirm or disprove this, for general $\bs{v}$.

Despite our inability to prove that the proportionally-weighted GHZ state is the optimal probe state for estimating $\theta_1=\bs{v}^T\bs{\phi}$, we \emph{can} prove that this state provides a better estimation precision than any local estimation strategy, for \emph{all} non-trivial $\bs{v}$ (we consider $\bs{v}$ to be trivial if only one element is non-zero). Furthermore, we can quantify the minimal precision enhancement provided by the optimal global strategy over \emph{any} local strategy. Define the weighted $d$-sensor separable state
 \begin{equation}
  \ket{\psi_{\textsc{loc}}(\bs{w})} = \frac{1}{2^{\frac{d}{2}}} \bigotimes_{k=1}^{d} \left(  \ket{\lambda_{\text{min}}(w_k) } +\ket{\lambda_{\text{max}}(w_k)} \right),
\label{local-state}
 \end{equation}
where $\bs{w}$ is a vector of integers that describes how the particles are distributed over the $d$ sensors. Note that this is a generalization of the state defined in Eq.~(\ref{local-NO-1}). 

The most general $ \ket{\psi_{\textsc{loc}}(\bs{w})} \in \mathcal{S}(N_{\max})$ that contains the maximal possible average number of particles \footnote{There is no advantage in not using all the particles available.} is given by taking $\bs{w}=N_{\max}\bs{x}/\|\bs{x}\|_1$, where $\bs{x}$ is any vector such that $N_{\max}x_k/\|\bs{x}\|_1$ is an integer for all $k$. The precision with which $\ket{\psi_{\textsc{loc}}(N_{\max}\bs{x}/\|\bs{x}\|_1)}$ can estimate $\phi_k$ is given by 
\begin{equation}
 \text{Var} (\Phi_{k}) \geq \frac{\| \bs{x} \|_1^2}{ \mu N_{\max}^2 (\lambda_{\text{max}} -\lambda_{\text{min}})^2x_k^2}.
\label{eq:loc-v-b}
\end{equation}
Using this and the propagation of uncertainty formula in Eq.~(\ref{Eq:prop-of-errors}), the precision bound for estimating $\theta_1$ is then
\begin{equation}
\text{Var} (\Theta_{1}) \geq \frac{ \| \bs{x} \|_1^2}{\mu N_{\max}^2 (\lambda_{\text{max}} -\lambda_{\text{min}})^2} \sum_{k} \left( \frac{v_k}{x_k}\right)^2 ,
\label{eq:var-loc-gen}
\end{equation}
where the summation is over those $k \in[ 1,\dots, d]$ such that $v_k \neq 0$, and we have assumed that $x_k \neq 0$ if $v_k \neq 0$, as is essential for a finite bound. This bound is saturable because the QCRB is saturable in this setting.

The most obvious choice for spreading the resources between sensors is to choose $\bs{x}=\bs{v}$, as we did to obtain the proportionally weighted GHZ state.  From Eq.~(\ref{eq:var-loc-gen}), this proportionally weighted separable state has a precision bound of
\begin{equation}
\text{Var} (\Theta_{1}) \geq \frac{ \tilde{d} \| \bs{v} \|_1^2}{\mu N_{\max}^2 (\lambda_{\text{max}} -\lambda_{\text{min}})^2} ,
\label{loc-pre-b}
\end{equation}
where $\tilde{d}$ is the number of non-zero elements in $\bs{v}$. Hence, by comparing this to Eq.~(\ref{Eq:var-ghz2}), we see that the estimation strategy using the proportionally weighted GHZ state improves on the estimation uncertainty of this proportional weighted local strategy by a multiplicative factor of $1/\tilde{d}$. This enhancement reduces as more of the elements of $\bs{v}$ become zero, and disappears if all but one of the elements of $\bs{v}$ is zero, as is expected (because then the two probe states are identical). However the most practically relevant scenario is perhaps when $v_k \neq 0$ for all $k$. In this case, the reduction in the estimation uncertainty of the proportional-weighted global estimation strategy, in comparison to this local strategy, is a factor of $1/d$.

However, the proportionally weighted local strategy is \emph{not} always the optimal local strategy. A bound on the optimal such strategy can be found by minimizing Eq.~(\ref{eq:var-loc-gen}) over all possible weighting vectors, $\bs{x}$. It can be shown that the value of $\bs{x}$ that minimizes Eq.~(\ref{eq:var-loc-gen}) is given by taking $x_k = v_k^{2/3}$ for all $k=1,\dots,d$. Putting this value for $\bs{x}$ into Eq.~(\ref{eq:var-loc-gen}) provides a lower bound on the variance of $\Theta_1$ that can be achieved with a local estimation strategy. Specifically, we have

\begin{align}
\text{Var} (\Theta_{1})
& \geq \frac{1}{\mu N_{\max}^2 (\lambda_{\text{max}} -\lambda_{\text{min}})^2} \left(\sum_{k=1}^dv_k^{2/3}\right)^{3} .
\label{loc-l-b}
\end{align}
This bound is only saturable if $N_{\max} v'_k/\| \bs{v}'\|_1 \in \mathbb{N}$ for all $k$ where $v'_k=v_k^{2/3}$, as only in these cases is the $\psi_{\textsc{loc}}$ state with this resource weighting a valid state. This will not generically be true \footnote{Even when we take account of the fact that we are considering only those $\boldsymbol{v}$ such that $v_k/ ||\boldsymbol{v}||_1$  is rational and only those $N_{\max}$ such that $N_{\max}v_k \in \mathbb{N}$ for all $k$.}. 

Regardless of whether the bound of Eq.~(\ref{loc-l-b}) is saturable, it may be used to provide a lower bound on the enhancement that the optimal global strategy provides over \emph{any} local strategy for any $\bs{v}$, as we will do below. However, before we turn to that, we note that if $\bs{v} =(1,1,\dots,1)/\sqrt{d}$ then the optimally weighted separable state \emph{is} the equally weighted separable state: the ideal weighting is $x_k \propto v_k^{2/3}$, and when the $v_k$ are all equal the ideal weighting is then an equal weighting. Hence, we have proven our earlier claim, below Eq.~(\ref{local-NO-1}), that the equally-weighted local state is the optimal local estimation strategy for estimating $\theta_{\text{sum}}$. For this case, it is simple to confirm that Eq.~(\ref{loc-l-b}) reduces to Eq.~(\ref{sep-eq-w}).

Noting that $\sum_k v_k^p \geq \|\bs{v}\|_1$ for all $p<1$ as $v_k \leq 1$, and putting this into Eq.~(\ref{loc-l-b}), for any local estimation strategy we have that
\begin{align}
\text{Var} (\Theta_{1}) & \geq \frac{\| \bs{v} \|_1^3}{\mu N_{\max}^2 (\lambda_{\text{max}} -\lambda_{\text{min}})^2}  .
\end{align}
This bound is only saturable for trivial $\bs{v}$. Despite this, the advantage of this bound is that it is simple: it implies that a global estimation strategy using the proportionally weighted GHZ state will achieve an estimator variance that is lower than any local strategy by \emph{at least} a factor of $1/\| \bs{v} \|_1 $. This is a fairly intuitive form for the enhancement factor. 

Because $\bs{v}$ is 2-norm normalized, we have that $1\leq \| \bs{v} \|_1 \leq \sqrt{d}$, with the lower bound saturated if and only if all but one of the elements of $\bs{v}$ are zero, and the upper bound saturated if and only if $\bs{v} \propto (1,1,\dots,1)$. Therefore, we can conclude that, for any $\theta = \bs{v}^T \bs{\phi}$ such that $\bs{v}$ is rationally weighted, a global estimation strategy will always provide a precision enhancement over the optimal local estimation strategy, \emph{except} for the trivial case where only one of the elements of $\bs{v}$ is non-zero. The size of the enhancement is greatest for equally weighted functions, and seems to decrease as the 1-norm of $\bs{v}$ decreases. The 1-norm of a 2-norm normalized vector can be considered a measure of how evenly weighted the components of $\bs{v}$ are.

\subsubsection{Estimating a single generic linear function}



The analysis so far has not encompassed the estimation of an arbitrary linear function of $\bs{\phi}$. The functions we have not considered are those with $\theta_1= \bs{v}^T \bs{\phi}$ where $v_k/\|\bs{v}\|_1$ is \emph{not} a rational number for at least some $k$. For example, $\theta_1 \propto \phi_1 + \pi \phi_2$ is such a linear function of $\bs{\phi}=(\phi_1,\phi_2,\dots)$, as $1/(1+\pi)$ is not rational. For functions of this sort, finding the optimal probe state in $\mathcal{S}(N_{\max})$ is an even more subtle problem than for the case of rationally weighted $\bs{v}$. For our analysis, we need to introduce an additional entangled probe state, and hence it is convenient to delay this until Section~\ref{sec:GNSlin-func}. 

We will now relate our results on estimating single linear functions to recent work in the literature. Very recently Eldredge \emph{et al.}~\cite{eldredge2016optimal} considered estimating arbitrary linear functions of $d$ parameters, whereby each parameter is encoded into a single qubit. They proposed that, instead of minimizing the estimator variance by choosing the best probe state (as has been considered here), the unitary evolution is altered instead. Translated into our general quantum sensing networks setting, this corresponds to considering the $\bs{v}$ and $\bs{\phi}$ parameterization evolution
\begin{equation}
 U_{\text{adaptive}}(\bs{\phi}, \bs{v}) = \exp \left( - i \sum_{k=1}^d \phi_k v_k \hat{H}_k \right),
 \end{equation} 
where $\bs{\phi}$ are the unknown parameters and $\bs{v}$ is a known vector defining the parameter of interest, that is $\theta_1 = \bs{v}^T \bs{\phi}$. 

This altered unitary evolution essentially transforms the problem into an estimation of $\theta= \phi'_1+\dots+ \phi'_d$, where the unitary evolution has encoded $\phi'_k=\phi_k v_k$ into the $k$\textsuperscript{th} quantum sensor, with the generator of $\phi'_k$ given by $\hat{H}_k$. Therefore, given this alternative unitary evolution, and via the arguments given earlier in this section, the optimal estimation state for this transformed problem is again the $\ket{\psi_{\textsc{ghz}}}$ state, and indeed this is one of the main result of Eldredge \emph{et al.}~\cite{eldredge2016optimal}. In some physical settings the evolution time is a highly controllable parameter, and then this estimation strategy may be possible. However, there are other settings in which an estimation strategy of this sort is not relevant. We do not consider estimation problems where the parameter-encoding unitary is adapted any further -- see Ref.~\cite{eldredge2016optimal} for further details on the interesting scheme that they propose.



\subsection{Estimating multiple linear functions of $\bs{\phi}$}
\label{m>1func}
In some MPE problems it may be necessary to estimate more than a single linear function of $\bs{\phi}$. We now consider estimation strategies for estimating the entire vector $\bs{\theta}$ for some $\bs{\theta} = M\bs{\phi}$ with $M \in \mathbb{R}^d \times \mathbb{R}^d$. Consider the particular (but important) case when reducing the uncertainty in the estimate of $\theta_k$ is equally important for all $k$, that is, $W\propto \mathds{1}$. The figure of merit to minimize is then $E_{\bs{\Theta}}\propto  \text{Tr}(\text{Cov}(\bs{\Theta}))$. Now, if the estimation problem is such that $M$ is an orthogonal matrix, then, because $M^T=M^{-1}$ and via Eq.~(\ref{QFIM-J}), the QFIM for any probe state obeys
\begin{equation*}
\text{Tr}(\mathcal{F}(\bs{\theta})^{-1}) =  \text{Tr}(M\mathcal{F}(\bs{\phi})^{-1}M^{-1}) = \text{Tr}(\mathcal{F}(\bs{\phi})^{-1}).
\end{equation*} 
The QCRB on $E_{\bs{\Theta}}$ then gives
\begin{equation} 
 E_{\bs{\Theta}}\propto\text{Tr}(\text{Cov}(\bs{\Theta})) \geq \text{Tr}(\mathcal{F}(\bs{\theta})^{-1}) =  \text{Tr}(\mathcal{F}(\bs{\phi})^{-1}) .
\label{M-dis}
\end{equation}

For \emph{any} given probe state, Eq.~(\ref{M-dis}) implies that the scalar estimation uncertainties for estimates of $\bs{\theta}$ and $\bs{\phi}$ are bounded below by the same value in both cases (the value depends on the particular state). In other words, if the QCRB is saturable, then estimating $\bs{\theta}$ and estimating $\bs{\phi}$ are entirely equivalent MPE problems, and if we find the best strategy for estimating $\bs{\phi}$ it is also the best strategy for estimating $\bs{\theta}$.

\subsubsection{General quantum sensing networks}

Consider this MPE problem for an arbitrary quantum sensing network. That is, unlike earlier parts of this section, we are \emph{not} considering only networks that fit into the framework we introduced in Section~\ref{sec:optics-atoms}. In Section~\ref{sec:est-phi} we showed that, for general quantum sensing networks, a local estimation strategy should always be preferred for estimating $\bs{\phi}$ when the parameter generators all mutually commute (and moreover, such a strategy is optimal for bounded generators). Therefore, Eq.~(\ref{M-dis}) implies that a local estimation strategy should also be preferred when estimating $\bs{\theta}=M\bs{\phi}$, as long as (1) $M$ is orthogonal, (2) estimating each of the components of $\bs{\theta}$ is equally important, i.e., $W\propto \mathds{1}$, and (3) the parameter generators are all mutually commuting (as then the QCRB is saturable). However, it is crucial for what follows to remember that this conclusion does \emph{not} necessarily hold if we consider only probe states from some sub-space of the entire Hilbert space of the sensing network (see Section~\ref{lebge}). Whether or not local estimation strategies are preferable when we consider only states from some given sub-space depends on the structure of the sub-space in question. Section~\ref{genGNSs} below provides an example of a sub-space in which a global strategy is optimal even when $M$ is orthogonal.

The simplicity of Eq.~(\ref{M-dis}) is because we considered only orthogonal $M$. However, we could be interested in estimating $\theta_k = \bs{v}_k^T \bs{\phi}$ for $k=1,\dots,d$ with $\bs{v}^T_k \bs{v}_l \neq 0$ for at least some $k\neq l$, in which case $M$ is not orthogonal. Hence, taking $M$ to be orthogonal is an assumption about the problem of interest (which it was not when the aim was to estimate a single linear function of $\bs{\phi}$). When $M$ is not orthogonal it \emph{is} possible that the ideal probe state (from the full Hilbert space) is entangled between sensors, and hence the optimal strategy can be a global estimation strategy. We show this by example in Appendix~\ref{App:refute-tim-conjecture}.

\subsubsection{Fixed total particle number sub-spaces}
\label{genGNSs}

In this section, we have largely only considered sensing networks that fit into the framework of Section~\ref{sec:optics-atoms}. Moreover, we have largely only considered optimizing these sensing networks over probe states in $\mathcal{S}(N_{\max})$, which denotes the sub-space of states containing $N_{\max}$ or fewer particles \emph{in total} for some fixed $N_{\max}$.  As we will now show, for these networks, the optimal state for estimating $\bs{\theta}= M\bs{\phi}$ for orthogonal $M$ \emph{is} entangled when we optimize over only states in this sub-space. After we have shown why this is, we will explain the implications of this.

As estimating $\bs{\phi}$ and $\bs{\theta}=M\bs{\phi}$, for $W= \mathds{1}/d$ and any orthogonal $M$, are entirely equivalent MPE problems (see above), we need only consider the simplest case of estimating $\bs{\phi}$. In this case, the QCRB for $E_{\bs{\Phi}}$ with an arbitrary sensor-separable state is simply
\begin{equation}
E_{\bs{\phi}} \geq \frac{1}{4\mu d}\sum_{i=1}^d \frac{1}{ v_i},
\label{se-bound-Nmax}
\end{equation}
where $v_i$ is the variance, with respect to the probe state, for the generator of $\phi_i$. 

We are interested in finding the optimal precision over all sensor-separable states in the sub-space $\mathcal{S}(N_{\max})$.
As earlier (see Section~\ref{sec:sum}), for simplicity assume that $n =N_{\max}/d$ is an integer. Using a similar derivation to that in Section~\ref{gen-rat-fun}, it may be confirmed that the sensor-separable state in this sub-space which minimizes $E_{\bs{\Phi}}$ is $\ket{\psi_{\textsc{loc}}(n,d)}$, where this state was defined in Eq.~(\ref{local-NO-1}). Using Eq.~(\ref{se-bound-Nmax}), the (saturable) estimation uncertainty bound for this state is easily confirmed to be \footnote{As an aside, note that this is exactly the same precision to which this state can estimate the normalized sum of the $\phi_k$ (see Eq.~(\ref{sep-eq-w})), which is as to be expected}
\begin{equation}
 E_{\bs{\Phi}} \geq \frac{d^2}{\mu N_{\max}^2(\lambda_{\max} - \lambda_{\min})^2}.
\end{equation}
The average total number of particles in $\ket{\psi_{\textsc{loc}}(n,d)}$ is proportional to $N_{\max}$ (specifically $\bar{N}=N_{\max}$ or $\bar{N}=N_{\max}/2$), and hence, in terms of the scaling of $ E_{\bs{\Phi}}$ with $\bar{N}$ and $d$, we have $ E_{\bs{\Phi}}\propto d^2/\mu\bar{N}^2$.

We now present an entangled state with a lower estimation uncertainty for this MPE problem. This state will be entangled between the $d$ sensors and a single ancillary sensor. Consider the $d+1$ sensor state defined by~\footnote{A slight generalization of this state is to including a ``balancing'' coefficient on the final part of this superposition, so that the probability of all the particles being in the ancillary sensor can be weighted to be different from the probability that the particles are in any given probe sensor. This is how the optical GNS is defined by Humphreys \emph{et al.}~\cite{humphreys2013quantum} (also, see later herein). The optimal choice for minimizing estimator variances is not the balanced case. However, for simplicity we ignore this here, as it is not particularly relevant to our argument and the improvement in the estimation precision that is obtained with the optimal weighting is only minor.}
\begin{multline}
\ket{\Psi_{\textsc{gns}}(N)} = \frac{1}{\sqrt{d}}\big( \ket{\kappa_N,0, \dots,0} +  \ket{0,\kappa_N,\dots,0} +  \dots \\ \dots +  \ket{0,0,\dots,\kappa_N} \big),
\label{eq:atom-opticsGES}
\end{multline}
where $\ket{0}$ denotes the vacuum state, that is, $\ket{0}$ is the zero particles state, and
\[
\ket{\kappa_N}=
  \begin{cases}
   \ket{\lambda_{\max}(N)}      & \quad \text{if } |\lambda_{\max}| \geq |\lambda_{\min}|,\\
    \ket{\lambda_{\min}(N)}  & \quad \text{otherwise}. \\
  \end{cases}
\]
That is, $\ket{\kappa_N}$ is the eigenvector of the parameter generator in the ``$N$ particles or fewer'' sub-space of a \emph{single} sensor that has the eigenvalue with maximal modulus. The $\ket{\Psi_{\textsc{gns}}(N)}$ state contains $N$ particles in total, and we will refer to it as a ``generalized NOON state'' (GNS), which is a direct extension of the multi-mode optical GNS state, defined by Humphreys \emph{et al.}~\cite{humphreys2013quantum}, to our more general framework.

The state $\ket{\Psi_{\textsc{gns}}(N_{\max})}$ is a definite \emph{total} particle number state, and it contains $N_{\max}$ particles. Hence, it is in $\mathcal{S}(N_{\max})$. In Appendix~\ref{QFIM-pss-inverse}, it is shown that the saturable QCRB for $E_{\bs{\theta}}$ with this state is
\begin{align}
E_{\bs{\Phi}} &\geq    \frac{ d+1 }{\mu N^2_{\max}\max\{\lambda_{\max}^2,\lambda_{\min}^2\} }.
\end{align}
Therefore, $E_{\bs{\Phi}} \propto (d+1)/\mu \bar{N}^2$. As such, for the same average number of particles, this GNS has an estimation uncertainty that is smaller than that of the \emph{optimal} local estimation strategy that uses a state from $\mathcal{S}(N_{\max})$ (see above) by a factor of $\approx 1/d$. 

In summary, we have shown that in the sub-space $\mathcal{S}(N_{\max})$ there is an entangled state which is better than any separable probe state for estimating $\bs{\phi}$, and that it has a smaller estimation uncertainty by a factor which scales with $1/d$. Before we discuss the implication of this, we make some further observations on the GNS, and return to the problem of estimating $\theta_1=\bs{v}^T\bs{\phi}$ for generic $\bs{v}$.

\subsubsection{Estimating linear functions with GNSs}
\label{sec:GNSlin-func}
We now consider the precision with which the GNS can estimate single linear functions of $\bs{\phi}$. As earlier, we consider a function $\theta_1 = \bs{v}^T \bs{\phi}$ for some 2-norm normalized, but otherwise arbitrary, $\bs{v}$. For any such $\theta_1$, we have that the GNS has a saturable estimation uncertainty for $\theta_1$ of
\begin{align}
\text{Var}(\Theta_1) \geq    \frac{ d+1 }{\mu N^2_{\max}\max\{\lambda_{\max}^2,\lambda_{\min}^2\} }.
\label{var-theta1-gns}
\end{align}
This QCRB can be obtained by using the formulae derived in Appendix~\ref{QFIM-pss-inverse}, which show that the QCRB for $\text{Var}(\Phi_i)$ with an $N_{\max}$ particle GNS is the quantity given on the right-hand-side of Eq.~(\ref{var-theta1-gns}), and then using propagation of uncertainty, and the normalization of $\bs{v}$, to show that for any sensor-symmetric state $\text{Var}(\Theta_1)$ and $\text{Var}(\Phi_i)$ have the same QCRB.

Eq.~(\ref{var-theta1-gns}) implies that the GNS has the same estimation uncertainty for estimating any linear function of $\bs{\phi}$. It is interesting to see how this estimation uncertainty for the GNS compares to the entangled states we considered earlier for linear function estimation: the GHZ state, and weighted GHZ states. In particular, in Section~\ref{gen-rat-fun} we conjectured that the proportionally weighted GHZ state (as defined below Eq.~(\ref{dlfks})) is optimal for estimating any $\theta_1 =\bs{v}^T\bs{\phi}$ for which $\bs{v}$ is proportional to a vector of rational numbers, when we considered the optimization over the sub-space $\mathcal{S}(N_{\max})$ for any $N_{\max}$ such that $N_{\max} v_k/\|\bs{v}\|_1$ is an integer for all $k$. 

In order to compare the precision with which the proportionally weighted GHZ state and the GNS can estimate such $\theta_1$, denote the QCRB for an estimator of $\theta_1$ in each case by $B(\theta_1)_{\textsc{ghz}}$ and $B(\theta_1)_{\textsc{gns}}$, respectively (i.e., these are the quantities that $\text{Var}(\Theta_1)$ are bounded below by in each case). We have that
\begin{align}
B(\theta_1)_{\textsc{gns}} &=   \frac{ (d+1 ) }{\mu N^2_{\max}\max\{\lambda_{\max}^2,\lambda_{\min}^2\} }\label{gns-ghz1} ,
\\ &\geq    \frac{ (d+1 ) }{\mu N^2_{\max} (\lambda_{\max}-\lambda_{\min})^2 } \label{gns-ghz2},
\\ & = \frac{d+1}{\| \bs{v}\|_1^2}   \text{QCRB}(\theta_1)_{\textsc{ghz}} \label{gns-ghz3},
\\ & >  B(\theta_1)_{\textsc{ghz}} \label{gns-ghz4},
\end{align}
where Eq.~(\ref{gns-ghz1}) follows immediately from Eq.~(\ref{var-theta1-gns}), Eq.~(\ref{gns-ghz3}) follows from the QCRB for $\text{Var}(\Theta_1)$ with the proportionally weighted GHZ state given in Eq.~(\ref{Eq:var-wghz}), and Eq.~(\ref{gns-ghz4}) follows because $\bs{v}$ is 2-norm normalized and so $1 \leq \| \bs{v}\|_1^2\leq d$. As in both cases the QCRB is saturable, the proportionally weighted GHZ state can provide a smaller estimation uncertainty than the GNS state for \emph{all} rationally weighted $\bs{v}$. This is consistent with our conjecture that the proportionally weighted GHZ state is the optimal probe state in $\mathcal{S}(N_{\max})$ for this estimation problem.

Unlike weighted GHZ states, the GNS can be used to estimate $\theta_1 =\bs{v}^T\bs{\phi}$ for arbitrary $\bs{v}$, and the estimation precision is independent of $\bs{v}$. As such, the GNS provides a global estimation strategy for estimating $\theta_1 =\bs{v}^T\bs{\phi}$ when $\bs{v}$ is not rationally weighted. A bound on the best possible estimation uncertainty for this problem that can be obtained with any separable state in $\mathcal{S}(N_{\max})$ is given by Eq.~(\ref{loc-l-b}). By comparing this to the QCRB for the GNS, it is easily confirmed that in some regimes for $\bs{v}$ the separable state has a lower estimation uncertainty (e.g., when only one element of $\bs{v}$ is non-zero), and in other regimes for $\bs{v}$ the optimal separable state must have a higher estimation uncertainty (e.g., if all the elements of $\bs{v}$ are equal). Due to the extreme practical difficulties in using GNSs for parameter estimation (see below), we do not fully characterize the set of linear functions for which GNSs can provide a lower estimation uncertainty, in principle, than any separable state $\mathcal{S}(N_{\max})$.




\subsubsection{Implications}
\label{imps}
When considering only probe states in $\mathcal{S}(N_{\max})$, an $N_{\max}$ particle GNS can provide a relatively small estimation uncertainty for estimating both $\bs{\phi}$ and any linear functions of $\bs{\phi}$. Moreover, for estimating a single linear function, this estimation uncertainty is in many cases (1) smaller than can be obtained with any separable state in $\mathcal{S}(N_{\max})$, and (2) only slightly larger than the uncertainty obtained with what we have conjectured is the optimal probe state, the proportionally weighted GHZ state. However, the source of the generally low estimation uncertainty for GNSs is very different to that of the GHZ states. The high estimation precision of the GNS is simply due to the particularly large generator variances in each sensor. The covariances of the GNS do not contribute to reducing the estimation uncertainty. Firstly, the ratio of the generator covariances to the variances is small, and secondly, the correlations are actually \emph{increasing} the estimation uncertainty. We will demonstrate this explicitly, in the optical sensing setting, in the following section (Section~\ref{sec:optics}). 

Moreover, there exist separable states that contain the same average number of particles, and that have the same generator variances in each sensor, as the $N_{\max}$ particle GNS. As such, these states obtain a better estimation uncertainty per resource, for estimating $\bs{\phi}$ and linear functions of $\bs{\phi}$, than the $N_{\max}$ particles GNS. However, these separable states are not in $\mathcal{S}(N_{\max})$: they are elements of the full Hilbert space. Hence, the implications of our observations on quantum sensing with GNSs depend strongly on what is the relevant question for the sensing problem at hand. The key question is: ``what is the physically relevant sub-space to optimize over?''. Picking a sub-space for mathematical convenience can lead to misleading conclusions.

The $N_{\max}$ particle GNS is a pure state in which $N_{\max}$ particles are distributed between $d+1$ sensors in such a way that all the particles are in one sensor, and there are no particles in all of the other sensors, but it is maximally uncertain which sensor the particles are all in. Hence, in many (and perhaps all) cases, the GNS is exceedingly impractical, e.g., with massive particles. As such, it might be desirable to explicitly exclude states of this sort from the analysis. In particular, we might wish to consider only states with a definite number of particles in each sensor. 

Whenever both $\lambda_{\max} \neq 0$ and $ \lambda_{\min} \neq 0$, which includes the case of magnetic field sensing with two or more level atoms, all types of GHZ state \emph{do} have a definite number of particles per sensor, and so the proofs and conjectures we have given in this section about sensing with GHZ-like states apply also in the more restricted sub-space of states with a definite number of particles in each sensor. Moreover, in this sub-space, it is easily confirmed that the optimal state for estimating $\bs{\phi}$ is a separable state. This is because we can map any state in this sub-space to a separable state in this sub-space that has the same generator variances. 
This is perhaps the most relevant analysis for most applications with massive particles: it permits entanglement between sensors, but explicitly excludes indefinite local particle number states.

When $\lambda_{\max}=0$ or $\lambda_{\min} = 0$, e.g., optical sensing of linear phase shifts, states with definite particle numbers in each mode are completely insensitive to the parameters of interest. Furthermore, in the optical setting there is little physical motivation for considering only definite total particle number states, and moreover, it is natural to consider states with a non-zero probability to contain arbitrarily many photons (e.g., coherent states, cat states, etc). As such, it is perhaps most relevant to assess optical estimation problems in the full Hilbert space. In this case, we have that separable states are ideal for estimating $\bs{\phi}$, and GNSs are not needed for high-precision estimation of $\bs{\phi}$ or linear functions of $\bs{\phi}$. However, the results of this section on estimating linear functions of $\bs{\phi}$ cannot be immediately applied to the full optical Hilbert space (for example, in this setting there is no \emph{optimal} strategy for estimating anything). We delay any further discussion of the optical MPE until Section~\ref{sec:optics}.


\subsection{Discussion}
\label{sec:fop:dis}
In this section we have considered the problem of estimating linear functions of parameters, whereby each parameter is encoded into a separate quantum sensor in a network, using a general formalism that applies to both atomic and optical sensing. We have explicitly considered two different scenarios: (1) optimizing the distribution of $N_{\max}$ particles in a $d$-sensor network to estimate a \emph{single} generic linear function of $\bs{\phi}$; (2) optimizing the network to a estimate a $d$-dimensional vector of linearly independent functions of $\bs{\phi}$. 

In the first case, we have shown that, whenever the function of interest $\theta$ is of the form $\theta = \bs{v}^T \bs{\phi}$ with $\bs{v}$ proportional to a vector of rational numbers, a probe state that is entangled between sensors will always provide a better estimation precision than the optimal separable state, \emph{except} in one special case: when all but one of the components of $\bs{v}$ are zero. The intuition behind this is the following: if all $v_k \neq 0$ then $\theta$ is a parameter describing a global process of all the sensors (and a process of some subset of them if some $v_k=0$). As such, it is natural that a globally correlated state will be the most sensitive to $\theta$. 

A global estimation strategy can provide a reduction in the estimation uncertainty, over the optimal local strategy, by at least a factor of $1/\|\bs{v}\|_1 =1/( |v_1| +|v_2| + \dots |v_d|)$ for (2-norm) normalized $\bs{v}$. Moreover, the enhancement obtained via entanglement appears to be no better than a factor of $1/d$. Indeed, for the case of equally weighted functions (i.e., $|v_k|=1/\sqrt{d}$) we have proven this to be the case. Here, the optimal estimation uncertainty is obtained by GHZ-like states and is exactly a factor of $ 1/d$ smaller than the uncertainty of the optimal local estimation strategy.

Our analysis provides a detailed answer to the question of how best to distribute $N_{\max}$ particles over $d$ sensors to estimate a given linear function. There are other closely related problems that might also be of interest, for example, we may have a fixed number of particles at each sensor, or some other constraint on the number of particles at each site, and it would be interesting to understand how to find the best probe state in this situation. Alternatively, perhaps it is possible to develop an analysis that is applicable to completely general sensing networks, and moreover does not require the restriction to finite sub-spaces of a (potentially) infinite full Hilbert space. We leave this for future work.

The second type of MPE problem we considered was the estimation of an entire  $d$-dimensional vector $\bs{\theta}$, defined by some set of $d$ linear functions of $\bs{\phi}$. For general sensing networks, we have shown that, for orthogonal functions and an equal importance weighting on each of the functions, a local estimation strategy is optimal (as long as there are no restrictions on the probe states considered). However, this conclusion does \emph{not} extend to non-orthogonal functions. A possible reason for this is the following: a set of $\theta_k=\bs{v}_k^T\bs{\phi}$ parameters can have $\bs{v}_k$ vectors that are arbitrarily close to being parallel, whilst still being linearly independent. As we have seen, when we wish to estimate a single linear function (as described by a single vector), entanglement between sensors can provide enhancements in the estimation precision. Hence, it is perhaps unsurprising that, as the estimation problem becomes closer in nature to the estimation of a single parameter, global estimation strategies are preferable to local estimation strategies. It would be interesting to see if there is some relationship between the amount of entanglement in the ideal probe state, and the degree to which the $\bs{v}_k$ vectors are clustered around one or more directions in $\mathbb{R}^d$.

We have covered a significant subset of those estimation problems in which the aim is to estimate some linear functions of locally encoded parameters. However, there are a range of cases that we have not addressed. We have already mentioned the limitation of our results in this section to particular sub-spaces of the full Hilbert space, and we refer back to Section~\ref{m>1func} for detailed discussions of this subtlety. One further limitation is that we have not considered any MPE problem in which we wish to estimate $d'$ linearly independent functions of $\bs{\phi}$ with $1 <d'< d$. For brevity, we relegate discussions of MPE problems of this sort to Appendix~\ref{app-m>1}, where we again consider optimizing the distribution of $N_{\max}$ particles over the network.

\section{Optical multi-parameter estimation}
\label{sec:optics}
Throughout this paper we have considered how ``quantum sensing networks'' can be optimized to obtain the best possible sensing precision for a range of generic MPE problems of practical interest. We now demonstrate how the general results we have derived can be applied to various problems in multi-mode optical interferometry. Although we will rely on results that we have derived earlier in this paper, this section has been written to be fairly self-contained, with readers interested only in optical sensing in mind.

\begin{figure}
\includegraphics[width=7.0cm]{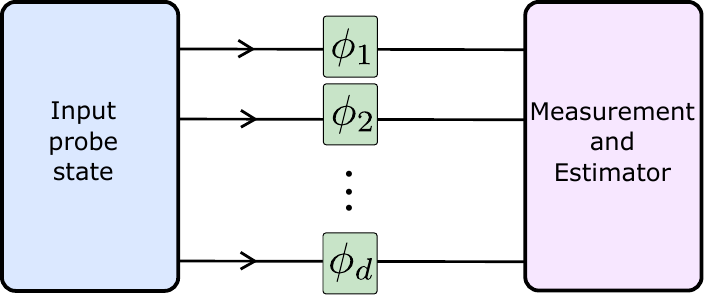}
\caption{{\bf Many-mode optical sensing:} $d$ unknown phases are imprinted onto $d$ optical modes by linear phase shifts. In the main text we consider how to choose the input state, and the measurement, in order to implement low-uncertainty estimates of some given linear functions of these unknown phase shifts, $\bs{\phi} = (\phi_1,\dots,\phi_d)$. Problems of this sort have been considered by a range of authors, e.g., see Refs.~\cite{humphreys2013quantum,liu2014quantum,yue2014quantum,ciampini2015quantum,knott2016local,gagatsos2016gaussian}. This generic setup encompasses a model of quantum enhanced imaging \cite{humphreys2013quantum} and networks of two-mode interferometers \cite{knott2016local}.}
\label{fig:MPE_general_noref}
\end{figure}

We will consider optical estimation problems in which the unknown parameters are linear phase shifts. More specifically, consider $d$ optical modes where the $k$\textsuperscript{th} optical mode undergoes the unitary evolution $\hat{U}(\phi_k) = \exp(-i \phi_k \hat{n}_k)$. Here, $\hat{n}_k$ denotes the bosonic number operator on mode $k$, and we will use $\ket{n}$ to denote the $n$-photons eigenstate. The total evolution of all $d$ modes is
\begin{equation}
\hat{U}(\bs{\phi}) = \exp(-i (\phi_1 \hat{n}_1 +\dots +  \phi_d \hat{n}_d)).
\end{equation}
This general setup is demonstrated in Figure~\ref{fig:MPE_general_noref}. 

We will consider how to choose the input state, and the measurement of the $\bs{\phi}$-dependent output state, in order to implement low-uncertainty estimates of some given linear functions of these unknown phase shifts. Estimation problems of this sort have been considered in Refs.~\cite{humphreys2013quantum,liu2014quantum,yue2014quantum,ciampini2015quantum,knott2016local,gagatsos2016gaussian}. We will be particularly interested in the relationship between mode-entanglement and estimation precision.

The aim, in all the optical estimation problems we consider, is to obtain a small estimation uncertainty as a function of the average \emph{total} photon number. That is, we consider the expectation value of
\begin{equation}
\hat{N} = \hat{n}_1 + \hat{n}_2 + \dots + \hat{n}_d,
\end{equation}
to be the amount of resources contained in a probe state. As we will see, there are certain complications in estimation problems of this sort that are rooted in the unbounded nature of the bosonic number operator and the physical possibility for indefinite photon number states. These have been pointed out in Ref.~\cite{knott2016local} for optical MPE problems, and are now well-known in the standard two-mode interferometry literature \cite{rivas2012sub,hall2012universality,giovannetti2012sub}. We delay any further discussion of these issues until later.

 It will be useful to define the $d$-mode entangled state
\begin{multline}
\ket{\Psi_g(\psi)} = \mathcal{N}_g \big( \ket{\psi,0, \dots,0,0} + \ket{0,\psi,\dots,0,0} + \dots \\ + \ket{0,0,\dots,\psi,0}  +  \ket{0,0,\dots,0,\psi} \big),
\label{eq:GES}
\end{multline}
where $ \psi $ is any single-mode normalised state and $\mathcal{N}_g$ is a normalization factor:
\begin{equation}
\mathcal{N}_g= \left(d+d(d-1)|\langle 0 | \psi \rangle |^2\right)^{-\frac{1}{2}}.
\end{equation}
Two states of this sort have been considered in the literature, and will be important here: a (balanced) generalized NOON state (GNS), which is given by $\ket{\psi}=\ket{N}$ \cite{humphreys2013quantum,yue2014quantum,knott2016local} and which is the optical version of the more general state introduced in Section~\ref{genGNSs}, and a generalised entangled coherent state (GECS), which is given by $\ket{\psi}=\ket{\alpha}$  \cite{liu2014quantum,knott2016local}, where $\ket{\alpha}$ is a coherent state. That is, $\hat{a}\ket{\alpha}=\alpha\ket{\alpha}$ where $\hat{a}$ is the bosonic annihilation operator. We could also consider the ``generalized'' entangled state given by any other choice for $\psi$. For example, we could take $\psi$ to be a squeezed vacuum state, which extends the two-mode squeezed entangled state introduced in Ref.~\cite{knott2016practical} (see also Ref.~\cite{lee2015quantum}) to the many-mode domain.




\subsection{Quantum enhanced imaging}
\label{sec:imaging}

We now consider the MPE problem of \emph{quantum enhanced imaging}, which was introduced by Humphreys \emph{et al.}~\cite{humphreys2013quantum}. In this estimation problem there are $d$ linear phase shifts, and we wish to estimate the differences between each of the first $d-1$ phase shifts and the final phase shift, which acts as a phase reference. That is, the aim is to estimate the $d'=d-1$ parameters $\theta_k = \phi_k -\phi_d$ for $k=1,\dots,d'$. Note that we have not normalized the $\theta_k$ (i.e., these would be the parameters $\theta_k \to \theta_k/\sqrt{2}$), in contrast to the convention taken throughout the rest of this paper, for consistency with Refs.~\cite{humphreys2013quantum,liu2014quantum,yue2014quantum,knott2016local,gagatsos2016gaussian}. Moreover, note that our $d'$ is equivalent to $d$ in Refs.~\cite{humphreys2013quantum,liu2014quantum,yue2014quantum,knott2016local,gagatsos2016gaussian}, as we have chosen to use $d$ to denote the total number of independent parameters, as throughout this paper.

In general, some of the $\theta_k$ may be of more importance than others, and so we assign a weighting to the importance of reducing the variance of each $\Theta_k$ ($\Theta_k$ denotes the estimate of $\theta_k$). Hence, the aim is to find an estimation strategy with a small scalar uncertainty, defined by 
\begin{equation}
E_{\bs{\Theta}} = \sum_{i=1}^{d'} W_{ii}\text{Var}(\Theta_i),
\end{equation}
 for some given $d'\times d'$ diagonal weighting matrix $W$, with $W_{ii}\geq0$ for all $i$. This is a particular estimation problem for the general optical setup presented in Figure~\ref{fig:MPE_general_noref}, and this MPE problem has been considered, for the case of $W\propto \mathds{1}$, in Refs.~\cite{humphreys2013quantum,liu2014quantum,yue2014quantum,knott2016local,gagatsos2016gaussian}.

In order to put our results into context, and to show precisely how they relate to previous work, we first review the most relevant literature. Refs.~\cite{humphreys2013quantum,liu2014quantum,yue2014quantum} compare the precision obtainable in this MPE problem with \emph{simultaneous estimation} (SE) and \emph{individual estimation} (IE) strategies (IE is also sometimes termed \emph{independent} estimation \cite{liu2014quantum}). SE simply refers to an estimation strategy in which the schematic of Figure~\ref{fig:MPE_general_noref} is used to estimate the $d'$ phase differences simultaneously with any input state and any measurement. Exactly what IE refers to is less clear. It appears to refer to either estimating the $\theta_k$ by implementing the estimations for each $k$ in turn, or, to  estimating (possibly in parallel) the $d'$ quantities $\theta_k' = \phi_k - \phi_{k}^{\text{ref}}$, where $ \phi_{k}^{\text{ref}}$ is the phase of a reference mode for the $k$\textsuperscript{th} probe mode. That is, each probe mode has a separate phase reference. From an abstract point of view, these two notions of IE are roughly equivalent, although they are clearly different experimental procedures.

The key claims of Refs.~\cite{humphreys2013quantum,liu2014quantum,yue2014quantum} are that: (1) A SE strategy can provide a better estimation precision for this MPE problem than is possible with any IE strategy; (2) The improvement in the estimation precision using a SE strategy, over the optimal precision with an IE strategy, scales with $\mathcal{O}(d)$ (in the lossless case, which is all we consider here). For example, Humphreys \emph{et al.}~\cite{humphreys2013quantum} (see abstract) say: 
\vspace{0.1cm}

``\emph{We identify quantum probe states that provide an enhancement compared to the best quantum scheme for the estimation of each individual phase separately...}'' 

\vspace{0.1cm}
\noindent Similarly, Yue \emph{et al.}~\cite{liu2014quantum} (see abstract) say: 

\vspace{0.1cm}
``\emph{Our results show that simultaneous estimation (SE) of multiple phases is always better than individual estimation (IE) of each phase even in noisy environment.}''

\vspace{0.1cm}
A SE strategy that uses a mode-separable state and a mode-separable measurement can also be considered to be an IE strategy (as IE is essentially a subset of SE \cite{yue2014quantum}). This is because then the estimation of $\phi_k$ is independent of the estimation of $\phi_l$ for $l \neq k$, and as such, the estimations can all be implemented in sequence if so desired. Therefore, an important question for assessing the comparative merits of SE and IE is: Can entanglement between modes enhance the estimation precision? A closely related question, in the language used earlier in this paper (see Section~\ref{sec:global-local}), is: Do global estimation strategies attain a higher precision than local estimation strategies? Moreover, if entanglement can enhance the estimation precision, can it provide an $\mathcal{O}(d)$ enhancements in the estimation precision?

The results we have derived on general quantum sensing networks can be used to carefully analyze the claims of Refs.~\cite{humphreys2013quantum,liu2014quantum,yue2014quantum}. Our conclusions will depend on certain subtle details of the estimation problem that so far haven't been specified, namely: 
\begin{enumerate}
\item What probe states are permissible?
\item Are external reference beams available?
\item Do we know the reference phase? 
\end{enumerate}
If, in principle, arbitrary probe states and measurements are available (which implicitly allows for external reference beams -- see later), then we will show that: if we already know the value of the reference phase, then a SE strategy that uses entangled states attains a worse precision than an IE using separable states and containing the same number of photons. If we do not know the reference phase, then our results suggest that entanglement can -- at most -- provide a reduction in the estimation uncertainty of a small constant multiplicative factor $\approx 1/4$. If only probe states from sub-spaces are permissible then these conclusions can be dramatically altered, and this depends entirely on what sub-space of states is considered. We will explicitly discuss certain restricted state spaces and measurements, but, as there are any number of ways to restrain the states and measurements, we will leave many cases unconsidered.

\subsubsection{A known reference phase}
\label{knownphid}

In Humphreys \emph{et al.}~\cite{humphreys2013quantum}, and the extensions of this work in Refs.~\cite{liu2014quantum,knott2016local}, the phase of the reference mode is taken to be known (which is not an assumption in some cases, as we will make clear later). This is equivalent to setting $\phi_d=0$, or taking $\phi_d$ to be any other constant. As such, the aim is to estimate $\theta_k = \phi_k$ for $k=1,\dots,d'$. We will discuss the physical meaning of taking $\phi_d$ to be constant below. 

If we assume that all probe states and measurements are, in principle, permissible, then this estimation problem is now a particularly simple case of MPE in our general framework of quantum sensing networks (see Section~\ref{sec:basic_formalism}). Here, each of the first $d'$ optical modes can be considered to be a single sensor, each sensor has a single linear phase shift encoded into it, and we wish to estimate these linear phase shifts (as $\phi_d=0$, we are effectively no longer considering phase differences). As such, our results in Section~\ref{Sec:proof} \emph{prove} that for \emph{any} mode-entangled probe state, we may find a mode-separable probe state that contains the same number of photons on average, and that has a strictly smaller estimation uncertainty (i.e., a smaller $E_{\bs{\Theta}}$). This holds for any weighting matrix $W$. Moreover, the results of Section~\ref{Sec:proof} also prove that the optimal measurement is a set of local measurements of each optical mode. 

Therefore, we have proven that SE does \emph{not} attain a higher precision than IE for this MPE problem when: (1) the phase of the reference mode is known, and (2) there are no specific constraints on the available probe states or measurements (which hence allows for external reference beams: see later). This contrasts with the message of Refs.~\cite{humphreys2013quantum,liu2014quantum,yue2014quantum}, but importantly note that Humphreys \emph{et al.}~\cite{humphreys2013quantum} do \emph{not} consider all possible probe states -- they only consider fixed total particle number states. We return to this point later.

\subsubsection{Mode-symmetric probe states}
Although the proof given above is definitive, it is fairly abstract. Hence, in order to illustrate this more explicitly, we now compare precision bounds for certain mode-entangled and mode-separable states. Note that the following analysis has been presented by ourselves and our collaborators in Ref.~\cite{knott2016local}, in a similar form to this (with slightly weaker conclusions). For simplicity, assume $W= \mathds{1}/d'$ (i.e., all the phases are of equal importance). Hence, $E_{\bs{\Theta}} = \text{avg}_i[\text{Var}(\Theta_i)]$. The relevant QCRB is now 
\begin{equation}
E_{\bs{\Theta}} \geq \frac{ \text{Tr}[ \mathcal{F}(\bs{\theta})^{-1}]}{\mu d'},
\end{equation}
 where $\mathcal{F}(\bs{\theta})$ is the $d' \times d'$ QFIM for $\bs{\theta}$ and, as always, $\mu$ is the number of classical repeats of the experiment. This bound is saturable, as the phase generators commute (see Section~\ref{sec:uni-est}). 

Up to a factor of 4, the diagonal elements of $\mathcal{F}(\bs{\theta})$ are the photon number variances of the probe modes, and the off-diagonal elements are the photon number covariances between probe modes (this follows from Eq.~(\ref{QFIM-CovH})). Hence, for any pure state that is symmetric with respect to the $d'$ probe modes, the QFIM for $\bs{\phi}$ is given by $\mathcal{F} = 4\left( (v- c)  \mathds{1}+ c  \mathcal{I} \right)$, where $\mathcal{I}$ is the $d'\times d'$ matrix of all ones, $c = \langle \hat{n}_k\hat{n}_l \rangle - \langle \hat{n}_k \rangle\langle \hat{n}_l \rangle$ for any $k\neq l$ with $k,l \in [1,\dots,d']$, and $v = \langle \hat{n}_k^2 \rangle - \langle \hat{n}_k \rangle^2$ for any $k \in [1,\dots,d']$.

 Let $\mathcal{J} =c/v$, which is a measure of the two-mode correlations \cite{sahota2015quantum} between probe modes, and let
\begin{equation} 
g(\Psi) = \frac{1+(d'-2)\mathcal{J}}{(1-\mathcal{J})(1+(d'-1)\mathcal{J})}, 
\end{equation}
which depends only on the number of modes and the two-mode correlations in the probe mode symmetric $d$-mode state $\Psi$. In Appendix~\ref{QFIM-pss-inverse} we show that, for any pure state that is symmetric with respect to the $d'$ probe modes, and which has an invertible QFIM for $\bs{\theta}$, the QCRB for $E_{\bs{\Theta}} $ is given by
\begin{align}
E_{\bs{\Theta}} &\geq   \frac{ g(\Psi) }{4 \mu v }.
\label{Egpsi}
\end{align}

Any entangled state can be mapped onto a separable pure state with identical single-mode photon number statistics, and hence an identical $v$. As such, the only way in which entanglement can aid the estimation precision is if we can have $g(\Psi)<1$. The argument we have given above (based on Section~\ref{Sec:proof}) implies that SE is not preferable to IE in general, and so it guarantees that there is no probe mode symmetric state with $g(\Psi)<1$ \footnote{Note, this formula for $E_{\bs{\Theta}}$ is only valid for mode-symmetric states, which cannot have an arbitrary $\mathcal{J}$ in $[-1,1]$.}. Hence, we do not explicitly prove this. Instead, we now illustrate this by example.

\subsubsection{Example: generalized NOON states}
\label{sec:gns}

Consider the probe state proposed by Humphreys \emph{et al.}~\cite{humphreys2013quantum}: the $d$-mode generalized noon state (GNS)
\begin{multline*}
\label{eq:unbal_GNS}
\ket{\Psi_{\textsc{gns}}} =\mathcal{N}\big( \ket{N,0, \dots,0,0} + \ket{0,N,\dots,0,0} + \dots \\ + \ket{0,0,\dots,N,0}  + \gamma\ket{0,0,\dots,0,N} \big),
\end{multline*}
where  $\mathcal{N} = 1/\sqrt{d'+\gamma^2}$, and $\gamma \geq 0$ is a ``balancing'' parameter. Taking $\gamma=1$ gives the balanced GNS, which we already introduced earlier (see Eq.~(\ref{eq:GES})). Although this is not the optimal value for $\gamma$ (for minimizing $E_{\bs{\Theta}}$), Humphreys \emph{et al.}~\cite{humphreys2013quantum} point out that the $\mathcal{O}(d')$ scaling characteristics for the estimation precision of this state, which they observe therein, are the same for the balanced ($\gamma=1$) and optimal ($\gamma=d'^{1/4}$) state. Hence, we consider the balanced GNS case for simplicity \footnote{We have presented a similar analysis for the optimal GNS in Ref.~\cite{knott2016local}.}.

The balanced GNS has a single-mode photon number variance of $v=d'N^2/(d'+1)^2$ and a correlation parameter $\mathcal{J}=-1/d'$. As such, for this state we have 
\begin{equation}
g(\Psi_{\textsc{gns}}) = \frac{2d'}{d'+1} \geq 1.
\end{equation}
Moreover, $g(\Psi_{\textsc{gns}}) \to 2$ as $d' \to \infty$. Hence, the quantum correlations between modes in the GNS actually \emph{increase} the estimation uncertainty -- by a multiplicative factor, between 1 and 2, that depends on $d'$. The saturable QCRB is $E_{\bs{\Theta}}^{\textsc{gns}} \geq (d'+1)/( 2\mu N^2 )$.
For $d'>1$, we have that $E_{\bs{\Theta}}^{\textsc{gns}}$ is strictly greater than $1/4\mu v$. That is, $E_{\bs{\Theta}}^{\textsc{gns}}>1/4\mu v$.

There is a separable state, containing the same average number of photons, and the same photon number variance $v$, which saturates the phase precision bound of $E_{\bs{\Theta}} \geq 1/4\mu v$. Specifically, consider the $d$-mode \emph{unbalanced NO state} (UNS)
\begin{equation}
\label{eq:UNS}
\ket{\Psi_{\textsc{uns}}} = d^{-\frac{d}{2}}\left( \ket{N} + \sqrt{d-1} \ket{0} \right)^{\otimes d}.
\end{equation}
By construction, this is a $d$-mode mode-symmetric pure state with a photon number variance of $v=d'N^2/(d'+1)^2$, and a photon number covariance of $c=0$ (so $\mathcal{J}=0$), meaning that $g(\Psi_{\textsc{uno}})=1$. Moreover, this state contains exactly the same average number of photons as the balanced GNS: we have $\langle \hat{N} \rangle = N$. 

The saturable QCRB for the UNS is $E_{\bs{\Theta}}^{\textsc{uns}} \geq (d'+1)^2/ (4\mu d' N^2) $, and so $E_{\bs{\Theta}}^{\textsc{gns}}>E_{\bs{\Theta}}^{\textsc{uno}}$ for $d>1$. To maintain the same estimation uncertainty for different numbers of modes, we must scale $N$ with $d'$. In particular, taking $N=\sqrt{d'} N_{\textsc{f}}$ for some constant $N_{\textsc{f}}$, gives the almost $d$-independent bounds
\begin{align}
E_{\bs{\Theta}}^{\textsc{gns}} \geq \frac{k}{2\mu N_{\textsc{f}}^2}, \hspace{0.8cm} E_{\bs{\Theta}}^{\textsc{uns}}\geq \frac{k^2}{4\mu N_{\textsc{f}}^2}\label{eq:uns-qcrb},
\end{align}
 where $k =(d'+1)/d' \approx 1$ for $d' \gg 1$. This makes the minor precision enhancement, over a GNS, obtained using the UNS very clear. Hence, we may map the SE with a GNS to an IE with an UNS with a better estimation precision, and identical single-mode photon statistics.

Essentially this same explicit derivation holds for all SE strategies that use mode-symmetric states (and the general argument of Section~\ref{Sec:proof} applies to \emph{all} SE strategies). For example, we can compare a generic ``generalized entangled state'' $\ket{\Psi_g(\psi)}$ for arbitrary $\psi$ (this state is defined in Eq.~(\ref{eq:GES})), to an equivalent $d$-mode mode-separable state using exactly the same method. The only difference is that the algebra is more cumbersome.

\subsubsection{The origin of $\mathcal{O}(d)$ enhancement factors}

As we have already noted, one of the key claims of Refs.~\cite{humphreys2013quantum,liu2014quantum,yue2014quantum} is that SE provides an $\mathcal{O}(d)$ precision enhancement over the optimal IE strategy. We have shown that this is not the case when the reference phase is known, and there are no explicit restrictions on the probe states or measurements. The origin of the idea that there is an estimation enhancement for SE, and that it scales with $d$, is that SE is better than certain IE strategies by such a factor.

Consider estimating each phase with a standard $n$-photon NOON state
\begin{equation}
\ket{\Psi_{\textsc{noon}}} = \frac{1}{\sqrt{2}}(\ket{n,0} + \ket{0,n}).
\end{equation}
 These estimations are implemented either in sequence, in which case the NOON state is entangled between the relevant probe mode and the single reference mode, or in parallel, in which case each NOON state is entangled between the relevant probe mode and one of $d'$ dedicated phase reference modes. In either case, the phase precision for estimating $\theta_i$ is the well-known Heisenberg limit $\text{Var}(\theta_i) \geq 1/\mu n^2$ (this can be obtained from the precision bound for the GNS with $d'=1$). For $d'$ NOON states to contain the same number of photons as a GNS (or a $d$-mode UNO state) then we must have $n= N/d'$, where, for simplicity, we assume that $N/d'$ is an integer. As such, we arrive at the precision bound $E_{\bs{\theta}}^{\textsc{noon}} \geq d'^2/\mu N^2$. 

In order to compare the NOON state precision bound to that for the GNS and the UNS, we need to consider the estimation uncertainty for the same total number of photons. Hence, to compare $E_{\bs{\theta}}^{\textsc{noon}}$ to the (almost) $d$-independent bounds in Eq.~(\ref{eq:uns-qcrb}), we should allow the total number of photons in the $d'$ NOON states to scale with $\sqrt{d'}$. Specifically, we should take $N=\sqrt{d'} N_F$ for some fixed constant $N_F$. Upon doing this, we obtain
 \begin{equation}
E_{\bs{\Theta}}^{\textsc{noon}} \geq \frac{d'}{\mu N_F^2}.\\
\end{equation}

It is clear that the saturable estimation uncertainty bounds for the UNS and the GNS are smaller than the bound obtained with $d'$ NOON states, by a factor of $\approx d'$ (note, $k\approx 1$ in Eq.~(\ref{eq:uns-qcrb})). By comparing the GNS to NOON states \cite{humphreys2013quantum,yue2014quantum},  or comparing generalized entangled coherent states (GECS) to ECS \cite{liu2014quantum}, this is essentially what was shown in Refs.~\cite{humphreys2013quantum,liu2014quantum,yue2014quantum}. Moreover, it is the main basis for their claims that SE can provide $\mathcal{O}(d)$ enhancements over the best possible IE strategy. 

However, estimating the phases with many standard NOON states is \emph{not} the best IE strategy, as we have shown. There are probe states with higher photon number variances than the NOON state, such as the GNS, the UNS, and many other states \cite{knott2016local,knott2016practical,lee2015quantum,rivas2012sub,knott2016search,knott2014effect,sahota2015quantum}. These states exhibit an enhancement over the estimation precision obtained with many standard NOON states (i.e., the standard ``Heisenberg limit''), but this does not mean the enhancement can be ascribed to mode-entanglement or simultaneous estimation. The precision enhancements are obtained if the probe state has higher photon number variances for the same total number of photons, which can be obtained with or without mode-entanglement, and with IE as well as SE. 



\subsubsection{Restrictions on the probe states}
The arguments we have given so far in this section assume that we do not \emph{a priori} discount the possibility of certain probe states (or measurements). One possible restriction is to consider only probe states containing a definite \emph{total} number of photons. The GNS contains a definite number of photons. However, the separable probe states that contain \emph{on average} the same number of total photons, and that have a lower estimation uncertainty than the GNS, are \emph{not} definite photon number states. 

If we restrict the analysis to only definite photon number states it is not clear that there is any state containing $N$ photons with a smaller estimation uncertainty than the GNS, and the estimation uncertainty of the GNS is a factor of $\approx 1/d$ smaller than that obtained with the optimal IE -- which here uses multiple NOON states each containing $\approx N/d$ photons (see above). As such, under this restriction, the claim that SE is better than IE \cite{humphreys2013quantum,liu2014quantum,yue2014quantum} is indeed true. Moreover, it is under these circumstances that Humphreys \emph{et al.}~\cite{humphreys2013quantum} made this claim. 

The critical question is then: is it physically relevant to consider only fixed total particle number probe state? We would argue that, for optical sensing, the answer to this is no. This is because indefinite photon number states are the norm in optics (e.g., coherent states). It is also possible to consider optimizing over probe states from some other, perhaps more physically well-motivated, sub-space. For example, Gagatsos \emph{et al.}~\cite{gagatsos2016gaussian} consider only Gaussian probe states. Interestingly, they conclude that SE is of limited benefit under these conditions.

Returning to the setting of fixed total particle number states, we note that in this case taking $\phi_d=0$ is \emph{not} an assumption. This is because fixed total particle number states are completely insensitive to the total phase $\phi_1+\dots+\phi_d$. Equivalently, they are completely insensitive to $\phi_d$ and it may therefore be set to a constant. However, more generally, taking $\phi_d=0$ is an assumption (the meaning of which we explain later). Hence, in the following section we extend our analysis to the case where $\phi_d$ is unknown.

\subsubsection{An unknown reference phase}
Consider now the case when $\phi_d$ is unknown. Therefore, the parameters to be estimated are $\theta_k = \phi_k -\phi_d$ for $k = 1 , ..., d'$ with both $\phi_k$ and $\phi_d$ unknown. This is properly encoded by taking a $d$-dimensional vector $\bs{\theta}=(\theta_1,\theta_2,\dots,\theta_d)$ where $\theta_d$ is some arbitrary linear combination of the $\phi_k$ that is linearly independent of all the other $\theta_k$. We are then interested in minimizing $E_{\bs{\Theta}} =\sum_i W_{ii} \text{Var}(\Theta_i)$ where $W_{dd}=0$ and the other $W_{ii} \geq 0$ are arbitrary (but sum to unity), and define the importance of optimizing the estimate of the other $\theta_k$. 

By reference to Section~\ref{sec:fovs}, it is clear that, whenever $\mathcal{F}(\bs{\phi})$ is invertible, to obtain the relevant QCRB we must consider the $d\times d$ inverse QFIM for $\bs{\theta}$ given by $\mathcal{F}^{-1}(\bs{\theta}) = M \mathcal{F}^{-1}(\bs{\phi}) M^{T}$
where $M$ is the Jacobian matrix
\begin{equation}
M =\frac{1}{\sqrt{2}}\begin{pmatrix}
  1 &   0  & \cdots &0 & -1 \\
   0&  1 & \cdots & 0 & -1 \\
  \vdots  & \vdots  & \ddots & \vdots  \\
 0 & 0 & \cdots &  1 & -1 \\
y_1 & y_2   &\cdots &  y_{d-1} & y_d
 \end{pmatrix}.\end{equation}
The values for the $y_k$ in the last row of $M$ are arbitrary, except that the last row should be linearly independent from all other rows (e.g., set $y_k=1$ for all $k$) \footnote{Perhaps the most convenient choice is $v_k= 1/\sqrt{d}$, as this is orthogonal to all other rows, and moreover, if $M \to M/\sqrt{2}$ then all the rows are normalized, which is the convention we used earlier in this paper to avoid trivial differences between $M$ matrices.}. We will only consider the case where the QFIM is invertible. When the probe state contains a fixed total number of photons the full QFIM is not invertable. However, this case has already been completely analyzed, as then taking $\phi_d=0$ is not an assumption (see above).


For simplicity, consider only the case of probe states which are symmetric with respect to the exchange of any of the first $d-1$ modes. As such, the QFIM for $\bs{\phi}$ is given by
\begin{equation} \mathcal{F}(\bs{\phi}) =\begin{pmatrix}
  v &   c& \cdots &c  & c' \\
   c& v& \cdots & c& c' \\
  \vdots  & \vdots  & \ddots & \vdots  \\
 c & c& \cdots &  v&  c' \\
c'& c' &\cdots &c'& v'
 \end{pmatrix} ,\end{equation}
where $v$ ($v'$) is the photon number variance of any of the first $d-1$ modes (the reference mode), $c$ is the photon number covariance between any pair of the first $d-1$ modes, and $c'$ is the photon number covariance between the reference mode and any of the first $d-1$ modes.
Let $\mathcal{J}=c/v$ and $\mathcal{J}'=c'/\sqrt{vv'}$, which quantify the correlations between any pair of probe modes, and the correlations between a probe mode and the reference mode, respectively. 

The assumption of probe mode symmetric states is only well-motivated if all of the parameters are of equal importance, i.e., $W_{ii}= 1/d'$ for $i=1,\dots,d'$, and so we now assume this. With an explicit calculation, it may be shown that
\begin{equation}
E_{\bs{\Theta}} \geq  \frac{1}{\mu \alpha} \left( \frac{\beta}{2v}+ \frac{\mathcal{J}'}{\sqrt{vv'}} +\frac{\gamma}{2v'} \right) ,
\label{QCRBthetaphid}
\end{equation}
where $\alpha$, $\beta$ and $\gamma$ are functions of $\mathcal{J}$, $\mathcal{J}'$, and $d'$, given by $\alpha=\delta(\mathcal{J}',d')$, $\beta=\delta(\mathcal{J}',d'-1)/(1-\mathcal{J})$, and $\gamma=\delta(0,d')$, where
\begin{equation} 
\delta(a,b) = 1+\mathcal{J}(b-1) -a^2 b .
\end{equation}

If we consider only separable input states (meaning that $\mathcal{J}=\mathcal{J}'=0$), then the estimation precision is much simpler. It is easily confirmed that in this case $\alpha=\beta=\gamma=1$, and so
\begin{equation}
E_{\bs{\Theta}} \geq   \frac{v+v'}{2\mu vv'} .
\label{sepphi}
\end{equation}

To assess the role of entanglement in obtaining low uncertainty estimates of $\bs{\theta}$, the most obvious first step would be to bound the maximal possible enhancement that entanglement can provide over any separable state, for a fixed average total number of photons. However, there are separable states which have arbitrarily small and saturable QCRBs (see discussion below), and so to provide a meaningful comparison we need to do something slightly more subtle than this. In particular, we will compare an arbitrary probe mode symmetric entangled state $\Psi$ with a separable state that has the same single mode photon statistics in each of the probe modes as $\Psi$.

More specifically, consider an arbitrary probe mode symmetric state of the $d$ optical modes $\Psi$, that has an invertible QFIM for $\bs{\theta}$, with corresponding values for $v$, $v'$, $c$ and $c'$ (as defined above), and denote the average total number of photons in $\Psi$ by $\bar{N}_{\Psi}$. Denoting the estimator uncertainty obtained with this state by $E_{\bs{\theta}}^{\Psi}$, we have that
\begin{equation}
E_{\bs{\Theta}}^{\Psi} \geq  \frac{1}{\mu \alpha} \left( \frac{\beta}{2v}+ \frac{\mathcal{J}'}{\sqrt{vv'}} +\frac{\gamma}{2v'} \right) \geq \frac{1}{4\mu v} ,
\label{Psi-QCRB-the}
\end{equation}
where the first inequality is simply the saturable QCRB of Eq.~(\ref{QCRBthetaphid}), and the second inequality follows because the estimator uncertainty can only be reduced if we know $\phi_d$, and if we know $\phi_d$ the best possible estimation precision for a state with probe mode variances of $v$ is bounded by $1/(4\mu v)$. This follows from our argument in Section~\ref{knownphid}, where we showed that a separable state is optimal. This is perhaps most explicitly clear in Eq.~(\ref{Egpsi}) and the following discussion.

There exists a separable state $\Psi_{s'}$ with the same single-mode photon number statistics, and hence the same photon number variances and average total number of particles, as $\Psi$. In particular, this state can be found using the mapping of Eq.~(\ref{eq:psi-s}), or see Ref.~\cite{knott2016local}. Hence, by reference to Eq.~(\ref{sepphi}) we see that, for the same average total number of photons, this state has the QCRB $E_{\bs{\Theta}}^{\Psi_{s'}} \geq  (v+v')/(2\mu vv')$.

 It is not particularly clear how to compare the QCRB for $\Psi_{s'}$ to the estimation uncertainty with $\Psi$. Hence, we slightly alter $\Psi_{s'}$. Specifically, we replace the state in the reference mode by the same state that is in all of the probe modes. Denoting this total $d$-mode state by $\Psi_{s}$, from Eq.~(\ref{sepphi}) we have that $E_{\bs{\Theta}}^{\Psi_{s}} \geq  1/(\check{\mu} v)$ where $\check{\mu}$ is the number of independent repeats of this protocol.  $\Psi_{s}$ is now not guaranteed to contain the same average total number of photons as $\Psi$, and so we cannot simply set $\check{\mu}=\mu$ and compare bounds directly. The average total number of photons in $\Psi_{s}$, denoted $\bar{N}_{\Psi_{s}}$, satisfies $\bar{N}_{\Psi_{s}}\leq \frac{d+1}{d}\bar{N}_{\Psi}$ (which is saturated only when $\Psi$ has no photons in the reference mode, which cannot be the case here as $\Psi$ has an invertible QFIM). Hence to compare bounds on the estimator uncertainty for $\Psi$ and $\Psi_{s}$ with $\check{\mu}\bar{N}_{\Psi_s} \leq \mu \bar{N}_{\Psi}$ we may set $\check{\mu} = \frac{d}{d+1}\mu$. As such we obtain the bound
\begin{equation}
E^{\Psi_s}_{\bs{\Theta}} \geq \frac{d+1}{d\mu v},
\label{eq;locls}
\end{equation}
where the resources used for a given $\mu$ are no greater than those used in the QCRB of Eq.~(\ref{Psi-QCRB-the}) for $\Psi$, for the same $\mu$.

By comparing Eq.~(\ref{Psi-QCRB-the}) and~(\ref{eq;locls}), we see that for any state $\Psi$ there exists a separable state $\Psi_s$ with a QCRB on the estimation uncertainty that, for the same total number of resources used ($\mu\bar{N}$), is no more than a factor of $4(d+1)/d \approx 4$ larger than that for $\Psi$. As such, entanglement can, at best, provide a reduction in the estimation uncertainty by a multiplicative factor of $\approx 1/4$. Moreover, note that our argument is not particularly elegant, and so we suspect that the constant factor reduction in the estimation uncertainty that entanglement can provide is probably $>d/4(d+1)$. In particular, when $d=2$, the entanglement enhancement factor is at most $1/2$ (see later). However, note that for $d=2$ there exists a state (the NOON state) that demonstrates a $1/2$ enhancement, in comparison to the equivalent separable state (a NO state in each mode). Hence, it \emph{is} definitely the case that entanglement can provide a minor reduction in the estimation uncertainty for at least some values of $d$. It would interesting to see if it is possible to derive a tighter bound than the one provided here, and to extend the analysis to include arbitrary states, rather than only probe mode symmetric states

In summary: when the reference phase is known, we have proven that entanglement is detrimental to the phase precision, unless only states for certain sub-spaces are permissible. When the reference phase is not known, then entanglement appears to only provide, at best, a small constant precision enhancement (for any $d$). We have proven this for mode-symmetric probe states -- a more general treatment is left to future work. Again, note that this conclusion will also not necessarily hold if only states from certain sub-spaces are permissible (e.g., fixed total photon number states).

\subsubsection{External reference beams and a constant phase reference}

As throughout this paper, our analysis in this section has been based exclusively on the QCRB and the QFIM. Hence, for the precision bounds we derived from the QFIM to be guaranteed to be saturable (in the large $\mu$ limit), it is essential that we should not be explicitly discounting the possibility of certain measurements. Otherwise, the measurement which saturates the bound may not be possible, and the precision bounds could be over optimistic (e.g., a finite estimation uncertainty might not be possible even if the QCRB is small). 

In particular, if the probe state contains an indefinite number of photons, then the optimal measurement may implicitly require an external reference beam that is phase-locked with the input state \cite{jarzyna2012quantum}. For example, a single-mode coherent state that has undergone a phase shift $\phi$ is given by $\ket{\alpha e^{-i\phi}}$. This state has a non-zero QFI, but it is impossible to perform a measurement on this state to determine $\phi$ without the use of another state that is phase-locked with it -- this additional state is termed the ``external reference'', or a local oscillator (LO). The simplest measurement is to mix $\ket{\alpha e^{-i\phi}}$ with another coherent state $\ket{\beta}$ at a beam splitter, and then count the number of photons at the output. 

As such, it is clear that the QFIM implicitly assumes that such a phase reference is possible. With this in mind, we see that in our analysis in this section (and in many papers on optical quantum metrology) we are effectively assuming that: (1) external reference beams are available, and (2) any photons in these reference beams should not be included in the resource counting.

In many physical scenarios, assuming that an external phase reference is available, and that any photons in this phase reference should not count towards the resources used in the estimation, is the most appropriate theoretical analysis. For example, if the sample being imaged is fragile, such as in Ref.~\cite{wolfgramm2013entanglement,carlton2010fast,taylor2013biological,tey2008strong,eckert2008quantum,pototschnig2011controlling}, then the most relevant resource counting should not include any LO. Moreover, if the LO is only a coherent state or another classical-like state, it is not challenging or expensive to produce (at least compared to entangled, squeezed or superpoissonian states that are ideally used to probe the sample).

We now turn to discussing the meaning of setting $\phi_d=0$, or instead taking $\phi_d$ to be an unknown parameter, and when each analysis is appropriate. Setting $\phi_d=0$, and implicitly allowing for external reference beams, can be understood to encode an estimation problem in which (a) the parameters of interest are effectively absolute phases (see below), and (2) it is possible to entangle probe modes with a phase reference (mode $d$), but that photons in that mode should be counted as contributing towards the total resources used (which is perhaps a sensible way to count resources, as creating entangled states is hard). Hence, in this analysis, if we do not wish to entangle any probe modes with the reference mode, mode $d$ can simply be discarded entirely, which is why the QFIM does not explicitly depend on any property of mode $d$.

In many physically motivated estimation problems that can be termed ``imaging'', e.g., phase contrast imaging, the aim is to estimate some property of a sample by estimating a collection of phases ($\phi_k$) that the sample imprints onto a light beam. An experiment may achieve this by measuring the difference between $\phi_k$ and some reference phase, $\phi_{\text{ref}}$, but essentially it is the $\phi_k$ that we are interested in, not $\phi_k-\phi_{\text{ref}}$. To successfully determine $\phi_k$ we therefore must know what $\phi_{\text{ref}}$ is already, e.g., by first calibrating the experiment. If the experiment is correctly calibrated, then we may set $\phi_{\text{ref}}=0$, and hence in the notation of this section we set $\phi_d=0$. On the other hand, if $\phi_d$ is taken to be unknown, we are encoding the notion that we are explicitly interested in the $d-1$ phase differences to this mode. Experiments of this sort cannot determine the $\phi_k$.


Finally, in some cases, either (i) any resources in external reference beams should be accounted for, or (ii) no external reference beams are possible. One way to analyze an estimation problem of this sort, whilst still using the QFIM and the QCRB, is the following: Explicitly include any permissible references beams in the analysis (and count photons in them towards the resource, if necessary), and consider only input states of the combined probe modes and reference modes which are insensitive to the global phase of all of the modes. This can be achieved by integrating over this global phase: see Ref.~\cite{jarzyna2012quantum} for details. We leave an analysis of MPE problems of this sort for future work.

\subsection{Networked interferometers}
\label{sec:parallel_intf}
Two-mode interferometry is important for a broad range of sensing problems, and hence a network of two-mode interferometers is a natural multi-mode optical sensing problem -- this was recently studied in Ref.~\cite{knott2016local}. Again, this is a special case of the scheme in Figure~\ref{fig:MPE_general_noref} in which $d$ optical modes undergo linear phase shifts. In particular, in this case the aim is to estimate $d/2$ parameters
\begin{equation}
\theta_k = \phi_{2k} - \phi_{2k-1}  ,
\end{equation}
for $k=1,\dots d/2$ (and even $d$). That is, we have chosen the first and second mode to form the two arms of the first interferometer, and so on. The scalar estimation uncertainty relevant for this problem is then $E_{\bs{\Theta}} = \sum_{i} W_{ii} \text{Var}(\Theta_i)$, where $W_{ii} \geq 0$ are $d/2$ arbitrary weightings defining the importance of each phase.

The analysis given in Ref.~\cite{knott2016local} considered only probe states that are interferometer-symmetric and symmetric with respect to the two modes of each interferometer. That is, those states that are invariant under exchanging any pair of interferometers in the network and under exchanging the two modes within each interferometer. Moreover, it was assumed that minimizing the uncertainty in the estimates of each of the phase differences is equally important ($W_{ii} = 2/d$ for all $i$). Under these conditions, in Ref.~\cite{knott2016local} it was shown that entanglement \emph{between interferometers} cannot enhance the estimation precision of each of the phase differences. 

The results we have derived herein for general quantum sensing networks can be immediately applied to this problem in its full generality. Our results (see Section~\ref{Sec:proof}) imply that, without any restriction on the probe states or the relative importance of the parameters, entanglement between interferometers can never improve the estimation precision. That is, any probe state which is entangled between interferometers can always be mapped onto an equivalent separable state, which has identical single-interferometer photon statistics, and that has an equal or better estimation precision (smaller $E_{\bs{\Theta}}$). Moreover, we proved that the optimal measurement is an independent measurement on each of the interferometers. The explicit method for doing this was provided in Section~\ref{Sec:proof}, and demonstrated for optical states in Section~\ref{sec:gns}. Hence, again IE attains a higher precision than SE. Note however that the same caveats apply as throughout this paper: if only probe states from some sub-space are allowed, this conclusion may not hold, and this will depend on the structure of the sub-space under consideration.

As in the ``quantum enhanced imaging'' problem, the key to high-precision estimation is a probe state with a high photon number variance in each mode (see below). Again, this can be achieved with SE using ``generalized'' entangled states, such as the GNS or the GECS. As before, $d$-mode-entangled states can provide significant precision gains in comparison to, say, parallel NOON state interferometry. However, the source of the precision enhancements is in utilizing a probe state with a photon number variance which is greater than that of the NOON state, rather than the entanglement. The same precision can be obtained with IE using mode separable states that have photon number variances greater than that of the NOON state \cite{knott2016local,knott2016practical,lee2015quantum,rivas2012sub,knott2016search,knott2014effect,sahota2015quantum}, such as the UNS (see Eq.~(\ref{eq:UNS})).

\subsubsection{Entanglement within an interferometer is useful}

It is important to realize that our results do \emph{not} show that all entanglement is detrimental to the estimation precision: entanglement \emph{within} an interferometer may enhance the estimation precision. It \emph{is} the case that entanglement between the arms of each interferometer provides estimation precision enhancements. In particular, for any symmetric state of a single two-mode interferometer, the saturable \cite{knott2016local} QCRB for the estimator variance of $\theta_k$ is given by
\begin{equation}
\text{Var}(\Theta_k) \geq \frac{1}{2v(1-\mathcal{J})},
\end{equation}
where $v$ is the photon number variance in either arm, and $\mathcal{J}=c/v \in [-1,1]$ quantifies the correlations between the arms \cite{knott2016local,knott2016practical,sahota2015quantum}.

Hence, a symmetric state with a variance of $v$ can be mapped to an entangled state with a better estimation precision, but exactly the same single-mode characteristics, simply by anti-correlating the two single-mode photon number probability distributions for the two-mode state, i.e., by adding some non-zero anti-correlation into the probability distribution. For example, a NO state $\ket{\psi_{\textsc{no}}} \propto \ket{N} + \ket{0}$ \cite{knott2014effect} in each mode (which is obviously separable) can be mapped to a NOON state, and an optical cat state \cite{ralph2002coherent,munro2001weak,knott2016practical} in each mode can be mapped to a state that is similar to an entangled coherent state (ECS) \cite{sanders1992entangled,gerry2002nonlinear,knott2014attaining}. However, the reduction in the estimation uncertainty that can obtained via entanglement is -- at most -- only a factor of $1/2$  \cite{knott2016practical,sahota2015quantum}. 

Interestingly, this conclusion is consistent with our broad message in Section~\ref{GEWLF} (although the results there only strictly apply in a particular sub-space). This is because there we showed that, in a particular sub-space, the estimation of balanced linear functions of $d$ parameters encoded into $d$ sensors can be enhanced by a factor of $1/d$ using an entangled probe, and here $d=2$. Note that, as pointed out in Ref.~\cite{jarzyna2012quantum}, when the phase imprinted into each arm of the interferometer is unknown, it essential to model a standard interferometer as a two-parameter estimation problem where we wish to only estimate a single function of these parameters, in order to always obtain the correct QCRB-derived estimation precision.

Finally, note that if we wished to estimate both the difference and the sum of the phases in each interferometer, then entanglement within each interferometer would not provide any enhancement in the estimation precision. This is implied by the discussion of Section~\ref{sec:fop:dis}, and in this case we always attain a higher precision when using a completely mode-separable input state, and an independent measurement of each mode is sufficient to saturate the QCRB.

\subsection{Discussion}

In this section we have analyzed two optical MPE problems: quantum enhanced imaging \cite{humphreys2013quantum,liu2014quantum,yue2014quantum,ciampini2015quantum,knott2016local,gagatsos2016gaussian} and networked interferometers \cite{knott2016local}. In all cases, the critical resource for quantum-enhanced parameter estimation is the photon number variance in each mode: ideally, this should be as large as possible. If only restricted probe states are permissible for some reason, e.g., the analysis assumes fixed total photon number states, then the ideal probe state might be highly entangled. But, it is \emph{not} entanglement -- or ``simultaneous estimation'' -- that is the critical resource. It is simply that, in some special circumstances, the only way to obtain states with high photon number variances is via entanglement. 

However, optimizing over fixed total photon number states seems poorly motivated from a physical perspective, given that indefinite photon number states are the norm in optics. Moreover, if we consider either MPE problem in the full Hilbert space, then the priority, in order to obtain low uncertainty estimations, is maximizing the photon number variances in each mode. Interestingly, there are practical methods for creating single-mode states with photon number variances that are higher than that of NOON states \cite{knott2016practical,knott2016search,sahota2015quantum}, in particular see Ref.~\cite{knott2016search}.

The analysis we have presented is based entirely on the QFIM and the QCRB, and there are some important shortcomings to using only these metrics to judge the performance of a state for parameter estimation. In particular, because of the unbounded nature of the photonic number operator, it is possible to find probe states with arbitrarily large photon number variances, and hence arbitrarily small, but saturable, QCRBs, for a fixed total \emph{average} number of photons \cite{rivas2012sub,zhang2013unbounded,hall2012universality,giovannetti2012sub}. However, it is known that it is not possible to obtain an arbitrarily high precision-per-resource in practice \cite{hall2012universality,giovannetti2012sub}, as the number of experimental repeats needed to come close to saturating the QCRB, and the experimenters prior knowledge, need to be taken into account. To rigorously analyze the merits of different states, beyond the framework of the QCRB and the QFIM, it is necessary to perform a statistical analysis of the estimation protocol, e.g., based on Bayesian inference. Using such an analysis, it would be interesting to compare the estimation uncertainties obtained with GNSs, or similar highly entangled states, to those of mode-separable states. We conjecture that there are mode-separable states which perform at least as well as GNSs for both problems considered herein, under such an analysis.

Finally, there are a range of other optical MPE estimation problems that we have not covered here which might be of practical interest. In the case of $d$ linear phase shifts, we might be interested in average phase differences, or the gradient of phase differences. Our results on estimating linear functions of many parameters (see Section~\ref{Sec:functions}) can be applied to situations such as this, but further work is required to be able to rigorously use them when considering probe states from the full Hilbert space of the optical network. Beyond linear phase shifts, there are a range of other optical estimation problems that have been investigated in the literature, including estimating non-linear phase shifts \cite{liu2014quantum,tilma2010entanglement,joo2012quantum} and coherent displacements \cite{duivenvoorden2017single,genoni2013optimal}. Moreover, there a range of metrology problems of interest in more general bosonic systems \cite{volkoff2016optimal,woolley2008nonlinear}. It is likely that many of our results can be applied to networked sensing in these settings. 

\section{Networked atomic sensing}
\label{sec:atomic_sensing}
We will now demonstrate how the results derived throughout this paper apply to the problem of optimizing a sensing network of 2-level atoms (i.e., qubits). Although we will rely on results that we have derived earlier in this paper, this section has been written to be fairly self-contained, with readers interested only in atomic sensing in mind. Hence, we will begin by defining some notation. Let
\begin{equation}
 \sigma_x=\begin{pmatrix} 0 & 1 \\ 1 & 0 \end{pmatrix}, \hspace{0.3cm}  \sigma_y=\begin{pmatrix} 0 & -i \\ i & 0 \end{pmatrix} , \hspace{0.3cm} \sigma_z=\begin{pmatrix} 1 & 0 \\ 0 & -1 \end{pmatrix} ,
\end{equation}
and let $\ket{\uparrow}$ and $\ket{\downarrow}$ be the $+1$ and $-1$ eigenstates of $\sigma_z$, respectively. Furthermore, using standard notation, define the $n$-atom ensemble spin operators by
\begin{equation}
 \hat{J}_q := \frac{1}{2} \sum_{i=1}^n \sigma_{q,i} ,
\end{equation}
for $q=x,y,z$, where $\sigma_{q,i}$ is the $\sigma_{q}$ operator acting on qubit $i$. 

We will consider sensing problems of the following sort: we have $N$ qubits, which are partitioned into $K$ ``sensors'', with the $k$\textsuperscript{th} sensor containing $n_k$ qubits (and hence $\sum_k n_k = N$).  Each sensor has some parameters unitarily encoded into it, and we wish to estimate these parameters, or some functions of these parameters. We will consider estimation problems with a unitary evolution of
\begin{equation}
\hat{U}_k(\phi_k) = \exp(-i (\phi_{x,k} \hat{J}_x+\phi_{y,k} \hat{J}_{y}+ \phi_{z,k} \hat{J}_{z})),
\label{mag-field}
\end{equation}
at sensor $k$, where these collective spin operators act on the ensemble of qubits in this sensor.  Often we will consider the special case where $\phi_{x,k}= \phi_{y,k}=0$.

A sensor evolving in this way can be used as a model for a range of applications: NV centres measuring magnetic fields, electric fields, temperature, etc \cite{rondin2014magnetometry,schirhagl2014nitrogen}; Ramsey interferometry \cite{gross2010nonlinear}; frequency standards \cite{komar2014quantum} (and see later); and many more \cite{huelga1997improvement,tanaka2014robust}. A network of such sensors encompasses the notion of mapping out the spatial profile of a electric or magnetic field at discrete sites, and many related problems.

Herein, we will consider the total number of atoms used in the estimation to be the ``resource'', so that we wish to optimize the estimation precision for a given $N$. The evolution time $t$ can also be considered to be a resource in atomic sensing, but by fixing each qubit to evolve for an identical time we may consider the number of qubits as the resource of interest. For example, if measuring a one-dimensional magnetic field using a single qubit, the field is imprinted via the unitary operator $\hat{U}=\exp(-i\omega t \sigma_z/2)$. If we fix the evolution time $t$, then we can set $\phi=\omega t$. This then reduces to the standard phase estimation problem, as measuring $\phi$ is equivalent to measuring $\omega$, and the only resource to be accounted for is the number of atoms used \cite{huelga1997improvement}.

We will be interested in finding the optimal probe states for a range of estimation problems of this sort. However, we will not consider arbitrary probe states. Specifically, we will assume that we have $N$ atoms, and that we wish to find the optimal way to distribute them over the sensors to maximize the estimation precision, but under the restriction that there is a definite number of atoms in each sensor. This is a physically well-motivated assumption.

\subsection{Estimating local parameters}
In order to fully define the estimation problem of interest, we need to specify what we want to estimate. To begin, we consider estimation problems in which the aim is to estimate all of the unknown parameters that are encoded into the sensors (i.e., we are not only trying to estimate, say, the average of all of the parameters). It is important to realize that the optimal protocol for this estimation problem (for fixed $N$) is \emph{not} generically the optimal method for estimating some function of the unknown parameters, e.g., the average field strength.

Consider the problem of mapping out a one-dimensional magnetic field, or any equivalent estimation problem. Specifically, we consider the unitary evolution $\hat{U}_k(\phi) = \exp(-i \phi_k \hat{J}_z)$ at sensor $k$. The results that we have derived herein for general quantum sensing networks (see Section~\ref{Sec:proof}) show that the optimal probe state for this problem is a local $N/K$ qubit GHZ state at each sensor (assuming $N/K$ is an integer), where an $n$ qubit GHZ state is defined by
\begin{equation} 
\ket{\psi_{\textsc{ghz}}(n)} = \frac{1}{\sqrt{2}} \left( \ket{\downarrow}^{\otimes n} + \ket{\uparrow}^{\otimes n}\right).
\end{equation}
Entanglement between sensors does not enhance the estimation precision. 

In the above, we have implicitly assumed that measuring the field strength at any given sensor is as important as measuring it at any other sensors, i.e., we have an equal incentive to minimize the estimator variance for all of the $\phi_k$. However, our results of Section~\ref{Sec:proof} show that this assumption may be relaxed, and the only consequence of this is that we should no longer necessarily allocate the same number of atoms to each sensor. However, in all cases, the atoms at each site should ideally be in a local GHZ state.

Consider now the more general problem of mapping out a three-dimensional magnetic field. Specifically, we consider the general unitary evolution given in Eq.~(\ref{mag-field}) at each sensor. In this more general setting, because the parameter generators do not all mutually commute, it is possible that entanglement between the sensors might reduce the estimation uncertainty in some cases \cite{ballester2004estimation,fujiwara2001estimation}. However, our results in Section~\ref{Sec:proof2}  show that, if some of the atoms in each sensor can be designated as ``ancillas'', meaning that they do not undergo the unknown evolution, then entanglement between sensors can, at most, provide a multiplicative factor of $1/2$ reduction in the estimation uncertainty (as quantified by the sum of the variances of the estimators of all $3K$ unknown parameters). Moreover, note that atoms need not be used as these ancillas -- they need only be some quantum systems that can be phase-locked and entangled with the sensors.

For estimating a multi-dimensional field, it is again the case that entanglement \emph{within} a sensor is a useful resource for obtaining a quantum-enhanced estimation precision. However, too much entanglement can be detrimental to the estimation precision \cite{baumgratz2016quantum}. The reader is referred to the work of Baumgratz and Datta \cite{baumgratz2016quantum} for further information on how to optimize a single sensor for multi-dimensional field estimation.

\subsection{Estimating global parameters}
We now consider atomic sensing estimation problems in which the aim is to estimate some global property of a sensing network. Perhaps the most obvious quantity of interest is the average field strength. In the case of a one-dimensional magnetic field, our results for general quantum sensing networks (see Section~\ref{Sec:functions}) prove that the optimal probe state is a \emph{global} GHZ state of the $K$ sensors. Specifically, this is the state $\ket{\psi_{\textsc{ghz}}(N)} \propto \ket{\downarrow}^{\otimes N} + \ket{\uparrow}^{\otimes N}$, where $N/K$ of these qubits are in each of the sensors.

This is consistent with Ref.~\cite{komar2014quantum}, which considers essentially the same problem in the context of a ``network of clocks''. Komar \emph{et al.}~\cite{komar2014quantum} propose a network consisting of $N/K$ geographically remote clocks, each containing $K$ atoms. They then show that a global GHZ state is the optimal state for estimating the ``centre of mass frequency'' of all of the atoms, and this is mathematically equivalent to our result.

The observation that a global GHZ state is optimal for estimating the average strength of a one-dimensional field can also provide an interesting insight into a basic result in quantum-enhanced metrology. Consider the case in which each sensor is a single qubit (i.e., $K=N$), and so the total unitary evolution is
\begin{equation}
\hat{U}(\bs{\phi}) = \exp\left(-\frac{i}{2} (\phi_1 \sigma_{z,1} + \dots + \phi_N \sigma_{z,N})\right).
\end{equation}
The $N$-qubit GHZ state is sensitive only to the average of the $\phi_{k}$, and hence when acting on a GHZ state we have that $\hat{U}(\bs{\phi})$ is indistinguishable from the unitary $\hat{U}(\bar{\phi}) = \exp(-i \bar{\phi} \hat{J}_{z})$ where $\bar{\phi}$ is the average of the $\phi_{k}$. As such, rather than considering GHZ states to be useful for estimating the strength of a uniform field on many atoms, it is more natural to consider them to be useful for estimating the average field strength of a potentially non-uniform one-dimensional field on many atoms.

Returning to the setting of $K$ sensors and $N$ atoms to be distributed over these sensors, more generally we might be interested in estimating some linear function of the $\phi_k$ which is not simply the average. In this case, entanglement between sensors is generically still useful for enhancing the estimation precision. The details of precisely what probe state is optimal are more complicated in this case: in many cases, we have conjectured that GHZ states with different numbers of particles in each sensor are the optimal probe states. Moreover, in a large number of cases entanglement between sensors \emph{is} provably useful for minimizing the estimation uncertainty. We refer the reader back to Section~\ref{Sec:functions} for full details on this (see also Ref.~\cite{eldredge2016optimal} for similar work).

Finally, we note two problems of potential practical interest that our results do not address: (1) optimal strategies for estimating functions of multi-dimensional fields, and (2) optimizing the sensing network over states with a pre-determined fixed number of atoms in each sensor. We suggest that addressing either of these problems would be interesting future work.

\section{Conclusions}
In this paper we have introduced a general model for a network of quantum sensors, where each sensor is some arbitrary quantum system into which unknown parameters are encoded via a local unitary evolution. Using this model we have derived a number of results that shed light on the question: can entanglement between the sensors enhance the precision with which the unknown parameters can be estimated? We first studied this question for a generic estimation problem within our framework where each sensor in the network is used to measure a single parameter. For any such problem, we showed that correlations between sensors reduce the estimation precision, and a state that is separable between sensors is preferred. Furthermore, we showed that both pure states and separable measurements are optimal for this estimation problem. 

These conclusions are intuitive: when the aim is to measure locally encoded parameters, there is no obvious reason why global states or measurements, exhibiting entanglement between the sensors, should be expected to improve the estimation precision. With this in mind, our results clarify claims in the literature on the fundamental advantages of ``simultaneous estimation'' in optical multi-parameter estimation problems \cite{humphreys2013quantum,liu2014quantum,yue2014quantum} .

Next, we considered the more general estimation problem whereby each sensor may be used to measure more than one parameter, i.e., each sensor is measuring a vector of parameters. In this case, whether correlations can enhance the estimation precision depends on the properties of the parameter generators. In particular, if all generators commute, then entanglement between sensors is still detrimental to the estimation precision. In contrast, if the generators do not all commute (e.g., when measuring a vector field at each sensor) then entanglement between sensors \emph{may} in some cases give a small constant reduction in the estimation uncertainty. 

However, any advantage obtained from entangling the sensors entirely disappears if each sensor has an ancillary system (i.e., some particles which do not undergo the unknown evolution) to aid the measurement, and when no property of these ancillas (e.g., the number of particles used) counts towards the total ``resources'' used in the estimation. This analysis is likely to be the most relevant in any practical setting in which the reason for limiting some ``resource'' used in the estimation is to avoid damaging a sample whose properties are being probed~\cite{wolfgramm2013entanglement,carlton2010fast,taylor2013biological,tey2008strong,eckert2008quantum,pototschnig2011controlling}.

Estimating one or more global functions of the unknown parameters (e.g., the average) is generally a fundamentally different problem to estimating the local encoded parameters themselves. Using a framework that is not completely general, but is suitable for analyzing a range of networked atomic and optical sensing estimation problems, we analyzed a generic estimation problem in which one parameter is encoded into each sensor, and where the aim is to measure some \emph{linear function} of all the parameters. In this case, we showed that for almost all linear functions entanglement can give precision advantages. The degree to which entanglement can reduce the estimation uncertainty depends on the details of the linear function of interest, and can scale with the number of parameters.


The main message of this paper is that ``simultaneous estimation'' and entanglement are not always useful in quantum-enhanced multi-parameter estimation. In some cases entanglement can be an important resource for minimizing estimation uncertainty, but in many cases entanglement can actually be detrimental to the estimation precision. The utility of entanglement in quantum metrology depends strongly on whether the parameters of interest are local or global properties of a set of systems. Our general model of networked quantum sensors, used to derived these results, can provide a rigorous framework to further illuminate the role of quantum correlations in both theoretical multi-parameter estimation research, and in technologies utilizing networked quantum sensing and metrology.

\section*{Acknowledgements}
We thank Jes{\'u}s Rubio for helpful discussions. This work was partly funded by the UK EPSRC through the Quantum Technology Hub: Networked Quantum Information Technology (grant reference EP/M013243/1). Sandia National Laboratories is a multi-mission laboratory managed and operated by Sandia Corporation, a wholly owned subsidiary of Lockheed Martin Corporation, for the U.S. Department of Energy's National Nuclear Security Administration under contract DE-AC04-94AL85000.
\section*{References}


\bibliographystyle{apsrev}

\bibliography{MyLib_Thesis_multiparameter_new}

\appendix
\section{}\label{App:Finequality}
In this appendix it is shown that for any finite-dimensional $d \times d$ real, symmetric and positive definite matrix, $A$, then 
\begin{equation}
 [A^{-1}]_{kk} \geq \frac{1}{A_{kk}},
 \label{Eq-AA}
\end{equation}
for all $k$. Furthermore, the equality holds if and only if the $k$\textsuperscript{th} column and $k$\textsuperscript{th} row only have a non-zero entry on the diagonal. Hence, the equality holds for all $k$ if and only if $A$ is diagonal. As the QFIM, $\mathcal{F}$, is real, symmetric and positive semi-definite, this equation applies to the QFIM whenever it is invertible (and therefore positive definite). This is what is stated in Eq.~(\ref{Eq:F>1/F}) of the main text. An almost identical result for the \emph{classical} FIM has been shown by \cite{bobrovsky1987some} and more recently by Ciampini \emph{et al.}~\cite{ciampini2015quantum}, with the derivation given below very similar to the latter proof for the FIM.
 
\emph{Proof:} In the following we will need the Cauchy-Schwarz inequality, which states that
 \begin{equation}
\left(\sum_{i=1}^{n} u_iv_i\right)^2 \leq \sum_{i=1}^{n} u_i^2\sum_{i=1}^{n} v_i^2,
\end{equation}
 for real $u_i$ and $v_i$, $i=1,\dots,n$. The equality holds only when $u_i = cv_i$ for all $i$ and some constant $c \in\mathbb{R}$ \cite{meyer2000matrix}. Any real and symmetric matrix $A$ is diagonalizable, and therefore we may write the $n$\textsuperscript{th} power of $A$, for all $n\geq0$, as $A^{n}= V  D^{n} V^T$,
 where $V$ is an orthogonal matrix (i.e., $VV^{T}=V^{T}V=\mathds{1}$) and $D^n$ is a diagonal matrix whose elements are $D^n_{kk} = \lambda_k^n$ for real $\lambda_k$, which are the eigenvalues of $A$. Furthermore, because $A$ is assumed to be \emph{positive definite}, then $\lambda_k > 0$ for all $k$. It then follows that
\begin{equation}\mathds{1} = \sqrt{A}\sqrt{A^{-1}}.
\end{equation}

Therefore, $[\sqrt{A}\sqrt{A^{-1}}]_{kk}^n=1$ for any $k=1,\dots d$ and integer $n$. Therefore,
\begin{align}
1&= [\sqrt{A}\sqrt{A^{-1}}]_{kk}^2 \\
&= \bigg(\sum_{j=1}^d [\sqrt{A}]_{kj}[\sqrt{A^{-1}}]_{jk} \bigg)^2, \label{p--b}\\
&\leq \sum_{j=1}^d[\sqrt{A}]_{kj}^2 \sum_{j=1}^d[\sqrt{A^{-1}}]_{jk}^2,\label{p--c}\\
&= \sum_{j=1}^d[\sqrt{A}]_{kj}[\sqrt{A}]_{jk} \label{p--d} \sum_{j=1}^d[\sqrt{A^{-1}}]_{kj}[\sqrt{A^{-1}}]_{jk},\\
&= A_{kk} [A^{-1}]_{kk},\label{p--e}
\end{align}
where the inequality of Eqs.~(\ref{p--b} -- \ref{p--c}) is implied by the Cauchy-Schwarz inequality and the equality of Eqs.~(\ref{p--c} -- \ref{p--d}) is because $A$ is symmetric and hence $[\sqrt{A^{\pm 1}}]_{kj}=[\sqrt{A^{
\pm1}}]_{jk}$. 

Eq.~(\ref{p--e}) implies that $[A^{-1}]_{kk} \geq 1/A_{kk}$, as stated in Eq.~(\ref{Eq-AA}). For a given $k$, the equality holds only when $ [\sqrt{A}]_{kj}= c_k [\sqrt{A^{-1}}]_{jk} $ for all $j$ and some constant $c_k$ (which may be different for each $k$). This is true when the $k$\textsuperscript{th} row and column of $A$ have zero entries everywhere except on the diagonal, $A_{kk}$, as then $ [\sqrt{A}]_{kj} =A_{kk}  [\sqrt{A^{-1}}]_{jk} $ for all $j$ (all the values are zero except the case of $j=k$). Moreover, it cannot hold in any other case - this can be inferred from the results in Appendix~\ref{App:Finequality-2} (see the discussion below Eq.~(\ref{vec-mat-eq})). This completes the proof.


\section{}\label{app:ancillas}
In this appendix we explain how the derivation of Section~\ref{sec:scalar-inv} to Section~\ref{sec:opt-for-p} can be adapted to show that a global estimation strategy is still detrimental to the estimation precision even when ancillary sensors are allowed. We are considering the enlarged Hilbert space $\mathcal{H}_{\mathbb{S}} \to  \mathcal{H}_{\mathbb{S}} \otimes \mathcal{H}_{\mathbb{A}}$, with $\mathcal{H}_{\mathbb{A}}$ the Hilbert space of some ancillary sensor(s), and a unitary evolution imprinting the parameters given by 
\begin{equation} U(\bs{\phi}) =\exp(-i \bs{\phi}^T \hat{\bs{H}}) \otimes \mathds{1}_{|\mathbb{A}|},
\end{equation}
 with $\bs{\hat{H}}=(\hat{H}_1,\dots,\hat{H}_d)$ still obeying Eq.~(\ref{H_k-com}). The resource operator is extended to the larger Hilbert space, by assumption, via 
  \begin{equation}
   \hat{R} \to \hat{R} \otimes \mathds{1} + \mathds{1} \otimes \hat{R}_{\mathbb{A}},
   \end{equation}
   where $\hat{R}_{\mathbb{A}}$ is a positive operator (meaning that it has non-negative eigenvalues, e.g., a number operator). 

For any state of this enlarged system, the mapping to a separable state of the probe sensors given by Eq.~(\ref{eq:psi-s}) still produces a separable state of the probes with a lower bound on the estimator uncertainty (with this bound still saturable), noting that this separable state is only of the original probe's Hilbert space and does not prescribe any state of the ancillas. As the resource operator on the ancillary probes is a positive operator, the original state including the ancillary sensors must contain an equal or greater amount of resources. Hence, the separable state without ancillas has the same amount of resources and a lower precision bound. The remainder of the derivation, i.e., showing that the optimal measurement is local, follows as before.

\section{}\label{App:Finequality-2}
Consider a $d\times d$ matrix $M$. Following the terminology of the main text, for such a $d\times d$ matrix, $M$, and a given `partitioning' of $d$ into $d=d_1+\dots+d_m$ then we denote by $M_{[jk]}$ the sub-matrix of $M$ obtained by removing the elements that are not both in rows $1+d_{<j}$ to $d_j+d_{<j}$ and columns $1+d_{<k}$ to $d_k+d_{<k}$, where $d_{<l}= \sum_{q<l} d_{q}$. Hence, 
\begin{equation}
M =  \begin{pmatrix}
  M_{[11]} & M_{[12]} & \cdots & M_{[1m]} \\
  M_{[21]} & M_{[22]} & \cdots & M_{[2n]} \\
  \vdots  & \vdots  & \ddots & \vdots  \\
  M_{[m1]} & M_{[m2]} & \cdots & M_{[mm]}
 \end{pmatrix}.
\end{equation}
Note that the parenthesis in the subscripts are used to denote that these are sub-matrices of $M$ and not just the matrix elements of $M$ ($M_{kk}=M_{[kk]}$ for all $k$ only if the partitioning is such that $d_j=1$ for all $j=1,\dots m$ with $m=d$). 

Using the analogous notation, we may write a $d$-dimensional vector, $\bs{v}$, in terms of the given partitioning as an $m$-dimensional vector of vectors with the natural notation that $v_{[j]}$ denotes the column vector consisting of the $1+d_{<j}$ to $d_j+d_{<j}$ elements of $\bs{v}$. Hence
 \begin{equation}
\bs{v} =\begin{pmatrix}
  v_{[1]}  \\
  v_{[2]} \\
  \vdots   \\
      v_{[m]}
 \end{pmatrix}.
\end{equation}

Consider any finite-dimensional $d \times d$ real, symmetric and positive definite matrix, $A$, along with a partitioning of $d=d_1+\dots + d_m$. In this appendix it is shown that for any such $A$ and any partitioning we have the inequality 
 \begin{equation} [A^{-1}]_{[kk]} \geq \left[A_{[kk]}\right]^{-1} ,
 \end{equation}
for all $k=1,,\dots,m$. To be clear, the matrix on the left hand side of this inequality is a sub-matrix of $A^{-1}$ and the matrix on the right hand side is the inverse of a sub-matrix of $A$, and hence this inequality is non-trivial. Furthermore, we will show that the equality is obtained if and only if $A_{[jk]}=A_{[kj]}=0$ for all $j \neq k$ (where $0$ denotes the matrix of all zeros). This is a generalization of the scalar relation $ [A^{-1}]_{kk} \geq A^{-1}_{kk}$ proven in Appendix~\ref{App:Finequality}.

\emph{Proof:} For a given $d\times d$ matrix $A>0$ and a partitioning $d=d_1+\dots + d_m$, consider the $d\times d$ matrix $P_k$ defined by the action on an arbitrary vector:
 \begin{equation}
P_k \begin{pmatrix}
  v_{[1]}  \\
  \vdots   \\
  v_{[m-2]} \\
    v_{[m-1]} \\
      v_{[m]}
 \end{pmatrix} =\begin{pmatrix}
  v_{[1]}  \\
  \vdots   \\
  v_{[m-1]} \\
    v_{[m]} \\
      v_{[k]}
 \end{pmatrix}.
\end{equation}
 $P_k$ is a permutation matrix and hence $P_kP^T_k=\mathds{1}$. Consider the matrix $\tilde{A}(k)=P_{k} A P_k^T$. This $\tilde{A}(k)$ matrix is symmetric as $A$ is symmetric. For any $s \times s$ matrix, $C$, and $t \times s$ matrix, $B$, then
  \begin{equation} C>0 \implies BCB^T \geq 0, \label{cdsd} \end{equation}
 and if $B$ is a (square) invertible matrix then $BCB^T > 0$ \footnote{As $C>0$ then by definition $\bs{v}^TC\bs{v} > 0$ for any vector $\bs{v}$ except when $\bs{v}=0$. We have that $\bs{v}^TBCB^T\bs{v}=\bs{w}^TC\bs{w} >0$ unless $\bs{w}=0$, where $\bs{w}=B^T\bs{v}$. Hence $BCB^T\geq 0$. For general $B$, $BCB^T$ is not guaranteed to be positive definite as we can have $\bs{w}=0$ for $\bs{v}\neq 0$. However, if $B$ is a (square) invertible matrix then  $\bs{w}=0$ only if $\bs{v}= 0$ and so $BCB^T> 0$.}. Hence $\tilde{A}(k)>0$ because $A>0$ and $P_k$ is invertible. It may be confirmed that
 \begin{equation} \tilde{A}(k)= \begin{pmatrix} A_{[\neq k]} & A_{k}^T \\ A_{k} & A_{[kk]} \end{pmatrix},
 \end{equation}
 where $A_{[\neq k]}$ is a positive definite matrix consisting of those $A_{[mn]}$ matrices with $m \neq k $ and $n \neq k$ (its exact form is irrelevant) and $A_k=(A_{[k1]},A_{[k2]},\dots,A_{[km]})$ where the second label in the subscripts here takes each value sequentially except that it misses out $k$.

 Consider any matrix $M$ that is symmetric, positive definite and has the form
\begin{equation} M= \begin{pmatrix} a & b^T \\ b & c \end{pmatrix},\label{matrix-22}
\end{equation}
 where $a$ and $c$ are square matrices of any sizes and $b$ is of the appropriate dimensions to make this a valid matrix. $M>0$ implies that $a>0$ and $c>0$. The inverse of $M$ exists and is given explicitly by
\begin{equation} M^{-1}= \begin{pmatrix} a^{-1}+a^{-1}b^Tg^{-1}ba^{-1} & -a^{-1}b^Tg^{-1} \\ -g^{-1}ba^{-1} & g^{-1} \end{pmatrix},
\label{apbeq2r}
\end{equation}
 where $g=c-ba^{-1}b^T$. It follows that $ba^{-1}b^T \geq 0$ because $a^{-1}>0$ (see Eq.~(\ref{cdsd})) and therefore $c \geq g$, which implies that $c^{-1} \leq g^{-1}$. 

When $b=0$ (i.e., $M$ is block diagonal) then $c=g$ which implies that $c^{-1}=g^{-1}$. Now, 
\begin{equation}
\left[ba^{-1}b^T\right]_{kk}=\bs{b}(k)^Ta^{-1}\bs{b}(k),
\label{vec-mat-eq}
\end{equation}
 where $b^T=(\bs{b}(1),\bs{b}(2),\dots)$, i.e., we have written $b^T$ as a row vector of column vectors. As $a^{-1}>0$, and via Eq.~(\ref{vec-mat-eq}) and the definition of a positive definite matrix, then if $\bs{b}(k)\neq 0$ it follows that $[ba^{-1}b^T]_{kk}>0$. This implies that $ba^{-1}b^T=0$ only if $b=b^T=0$. Hence, because obviously $c \neq g$ if and only if $ba^{-1}b^T \neq 0$ then $c \neq g$ if and only if $b\neq 0$. Therefore, we have shown that the inverse of the  bottom right diagonal matrix in $M$, $c^{-1}$, is less than or equal to the bottom right diagonal matrix in $M^{-1}$ with the equality obtained only when $M$ is block-diagonal.

Now, by noting that $\tilde{A}(k)$ has been written in the form of the matrix in Eq.~(\ref{matrix-22}), and satisfies the conditions demanded of it ($\tilde{A}(k)>0$), we may then infer that
\begin{equation} 
  [\tilde{A}(k)^{-1}]_{\text{br}} \geq \left[ A_{[kk]} \right]^{-1} , \label{Eq-dsfs}
 \end{equation}
 where $[\tilde{A}(k)^{-1}]_{\text{br}}$ is the $d_k \times d_k$ sub-matrix of $\tilde{A}(k)^{-1}$ in the bottom right corner of $\tilde{A}(k)^{-1}$. Furthermore, the equality only holds when $A_k=0$, implying that $A_{[kj]}=0$ for all $j \neq k$, and as $A$ is symmetric this implies that $A_{[jk]}=0$ for all $j \neq k$. Now $\tilde{A}(k)^{-1}=P_{k} A^{-1} P^T_k$, which implies that $  [\tilde{A}(k)^{-1}]_{\text{br}}=[A^{-1}]_{[kk]} $. Hence, by putting this into Eq.~(\ref{Eq-dsfs}) this leads us to the final conclusion that
   \begin{equation} [A^{-1}]_{[kk]} \geq \left[ A_{[kk]} \right]^{-1} ,
 \end{equation}
with the equality obtained if and only if $A_{[jk]}=A_{[kj]}=0$ for all $j \neq k$. This completes the proof.

\section{}\label{app:POVMs}

Consider a density operator, $\rho$, on some Hilbert space, $\mathcal{H}$, with dimension $q$. Now consider any purification of $\rho$ into $\mathcal{H} \otimes \mathcal{H} $, denoted $\ket{\Psi_1}$, and another purification of $\rho$ into a Hilbert space $\ket{\Psi_2} \in \mathcal{H} \otimes \mathcal{H}'$, where $\mathcal{H}'$ is of dimension $q'\geq q$. Consider the states obtained by enacting the local unitary $u$ on the `original' Hilbert space, i.e., the states 
\begin{align}
\ket{\Psi_1(u)} &= (u \otimes \mathds{1}_{q} ) \ket{\Psi_1},\\
\ket{\Psi_2(u)} &= (u \otimes \mathds{1}_{q'}) \ket{\Psi_2}.
\end{align}
Here we prove the fairly obvious result that, via only $u$-independent unitary transformations and partial traces, we may map $\ket{\Psi_1(u)} \otimes \ket{\text{fid}'} \to \ket{\Psi_2(u)} $, where $\ket{\text{fid}'}$ is some fiducial state in $\mathcal{H}'$. We will then explain why this implies the claim made in Section~\ref{sec:non-comm-proof}.

It is always possible to express $\ket{\Psi_1(u)}$ as
\begin{align} 
\ket{\Psi_1(u)} = \sum_{k=1}^q \alpha_k  \ket{\gamma_k^u} \otimes \ket{\varphi_k}, 
\end{align}
where the $\ket{\gamma_k^u}$ and $\ket{\varphi_k}$ states form orthonormal bases for $\mathcal{H}$, and only the $\ket{\gamma_k^u}$ depend on $u$. Because $\ket{\Psi_2(u)}$ is also a purification of $\rho$ it must be possible to express it in the similar form
\begin{align} 
 \ket{\Psi_2(u)} = \sum_{k=1}^q \alpha_k  \ket{\gamma_k^u} \otimes \ket{\vartheta_k}, 
\end{align}
where the $\ket{\vartheta_k}$ are $q$ states from an orthonormal basis of $\mathcal{H}'$ (that is, $\ket{\vartheta_k}$ for $k=1,\dots,q'$ is an orthonormal basis for $\mathcal{H}'$).

Now consider any unitaries, $U_k'$, such that $U_k' \ket{\text{fid}'} =  \ket{\vartheta_k}$ for $k=1,\dots,q$. Note that this doesn't fully define any of the unitaries. Using any such unitaries, we may construct the (controlled) unitary
\begin{align} 
 \Lambda_A  =\mathds{1} \otimes  \sum_{k=1}^q \ket{\varphi_k} \bra{\varphi_k} \otimes U_k' ,
\end{align}
which acts on $\mathcal{H}_{T} = \mathcal{H} \otimes \mathcal{H} \otimes \mathcal{H}'$. For any such $\Lambda_A$ it follows that
\begin{align} 
 \Lambda_A( \ket{\Psi_1(u)} \otimes\ket{\text{fid}'}) = \sum_{k=1}^q\alpha_k  \ket{\gamma_k^u} \otimes \ket{\varphi_k} \otimes \ket{\vartheta_k}.   
\end{align}
In essentially the same fashion we have that
\begin{align*} 
\Lambda_B\Lambda_A( \ket{\Psi_1(u)} \otimes\ket{\text{fid}'}) = \sum_{k=1}^q \alpha_k  \ket{\gamma_k^u} \otimes \ket{\text{fid}} \otimes \ket{\vartheta_k},
\end{align*}
where $ \Lambda_B$ is a unitary on $\mathcal{H}_{T}$ defined by
\begin{align} 
 \Lambda_B  =\mathds{1} \otimes   \sum_{k=1}^{q'} U_k^{\dagger} \otimes \ket{\vartheta_k} \bra{\vartheta_k} , 
\end{align}
where $U_k$ are any unitaries with the action $U_k \ket{\text{fid}} = \ket{\varphi_k}$ for $k=1,\dots,q$, where $ \ket{\text{fid}}$ is some fixed state in $\mathcal{H}$, and $U_k$ may have any arbitrary action for $k=q+1,\dots, q'$. Therefore, denoting
\begin{align} 
\ket{\xi(u)} = \Lambda_B\Lambda_A (\ket{\Psi_1(u)} \otimes \ket{\text{fid}'} ),
\end{align}
we have that 
\begin{align} 
 \ket{\Psi_2(u)} \bra{\Psi_2(u)} = \text{Tr}_{2} \left(\ket{\xi(u)} \bra{\xi(u)}\right),
 \end{align}
where the trace operation is over the second Hilbert space in $\mathcal{H}_{T}=\mathcal{H} \otimes \mathcal{H} \otimes \mathcal{H}'$. Hence, we can map $\ket{\Psi_1(u)}$ to $\ket{\Psi_2(u)}$ using only $u$-independent unitary transformations and a partial trace -- this is as we claimed above.

In Section~\ref{sec:non-comm-proof} we consider the two $\phi_{[l]}$-encoded states
\begin{align} 
\ket{\Psi_{\rho}^l} &=(\mathds{1}   \otimes \dots \otimes U_l(\phi_{[l]})    \otimes \dots \otimes \mathds{1} ) \ket{\Psi_{\rho}}, \\
\ket{\Psi_{s}^l} &=    \ket{\Psi_{1}} \otimes \dots \otimes  \ket{\Psi_{l}^l} \otimes \dots \otimes \ket{\Psi_{|\mathbb{S}|}},
\end{align}
where $\ket{\Psi_{l}^l}=(U_l(\phi_{[l]}) \otimes \mathds{1}) \ket{\Psi_{l}}$, and we claimed that using only $\phi_{[l]}$-independent unitary operations and partial traces (on an extended Hilbert space) we may map $\ket{\Psi_{s}^l} \to \ket{\Psi_{\rho}^l}$. Both $\ket{\Psi_l}$ and $\ket{\Psi_{\rho}}$ are purifications of the same density operator $\rho$. In particular, $\ket{\Psi_l^l}$ is a purification into the doubled Hilbert space and $\ket{\Psi_{\rho}}$ a purification into a larger Hilbert space (see main text). Furthermore, $\ket{\Psi_{l}^l} $ and $\ket{\Psi_{\rho}^l}$ are simply evolved by some unitary that is local to the `original' Hilbert space. As such, it is clear that our derivation above implies that there is a mapping $\ket{\Psi_{s}^l}\to \ket{\Psi_{l}^l} \to \ket{\Psi_{\rho}^l}$ which uses only $\phi_{[l]}$-independent unitary operations and partial traces (on an extended Hilbert space).

\section{}\label{App:refute-tim-conjecture}

In this appendix we consider a particular case of the problem of estimating $d$ linearly independent functions of a $d$-dimensional vector $\bs{\phi}$ (one of the problems considered in Section~\ref{m>1func}). Specifically, we consider the case of a 2-dimensional $\bs{\phi}=(\phi_1,\phi_2)$ with $\phi_1$ encoded into sensor 1 and $\phi_2$ encoded into sensor 2. That is, we wish to estimate some linear functions of $\bs{\phi}$, where $\bs{\phi}$  is imprinted on a probe state via 
\begin{equation}
U(\bs{\phi})= u_1(\phi_1) \otimes u_2(\phi_2),
\end{equation}
for some unitaries $u_1$ and $u_2$. When the aim is to estimate both functions with an equal importance weighting ($W\propto\mathds{1}$), we will show that for such problems the optimal estimation strategy is -- in at least some cases -- a global estimation strategy. This is because the ideal probe state is entangled between sensors 1 and 2. In the main text we have show that, for any $d$, when $\bs{\theta}= M\bs{\phi}$ with $M$ orthogonal and $W\propto\mathds{1}$, the optimal strategy is always a local estimation strategy. Therefore, the results of this appendix imply that this conclusion does \emph{not} extend to general non-orthogonal $M$. 

Considering the $2$-dimensional problem we have introduce above, in some such problems there exist probe states that have a QFIM of
\begin{equation} 
\mathcal{F}(\bs{\phi}) = \nu \begin{pmatrix} 1 & x \\  x & 1\end{pmatrix} ,
\label{F-ex-A}
\end{equation}
where $-1 \leq x \leq 1$ and $\nu>0$ is a constant. For example, consider a two-qubit problem with 
\begin{equation}
U(\bs{\phi})= \exp(-i \phi_1 \sigma_z/2) \otimes  \exp(-i \phi_2 \sigma_z/2),
\label{qubits-U2}
\end{equation}
where $\sigma_z$ is the Pauli operator $\sigma_z= (\begin{smallmatrix}1 & 0 \\ 0 & -1 \end{smallmatrix})$ (as in the main text, we denote the $+1$ and $-1$ normalized eigenvectors of $\sigma_z$ by $\ket{\uparrow }$ and $\ket{\downarrow }$, respectively). The QFIM of Eq.~(\ref{F-ex-A}) with $\nu=1$ is obtained by the probe state
\begin{equation} 
\ket{\psi} = \mathcal{N} \big( \ket{\downarrow \downarrow}  + \gamma(\ket{\downarrow \uparrow}+\ket{\uparrow \downarrow}) +\ket{\uparrow \uparrow}\big) ,
\end{equation}
where $\gamma=\sqrt{1-x}/\sqrt{1+x}$ and $\mathcal{N}=1/\sqrt{(2+2\gamma^2)}$, as may be confirmed using Eq.~(\ref{QFIM-CovH}). A QFIM of the form given in Eq.~(\ref{F-ex-A}) may also be obtained with two multi-atom magnetic field sensors or in an optics problem in a similar fashion. For simplicity, we will explicitly consider the 2-qubit problem below (and hence $\nu=1$).

The most general pair of normalized linear functions of $\bs{\phi}$ are given by
\begin{align}
 \theta_1 &= \phi_1 \cos\alpha+  \phi_2\sin\alpha, \\
 \theta_2 &=  \phi_1 \sin\beta+ \phi_2\cos\beta  ,
\end{align}
and they are linearly independent under the condition that $\cos (\alpha + \beta)\neq 0$, which we assume from now on. The Jacobian matrix, $M$, for this reparameterization is simply given by
\begin{equation} 
M =  \begin{pmatrix} \cos\alpha & \sin\alpha\\\sin\beta & \cos\beta \end{pmatrix} .
\end{equation}
The inverse of $\mathcal{F}(\bs{\phi}) $, as given by Eq.~(\ref{F-ex-A}) with $\nu=1$, exists for $|x| \neq 1$. It is given by
\begin{equation} 
\mathcal{F}(\bs{\phi})^{-1} =\frac{1}{1-x^2} \begin{pmatrix} 1 & -x \\  -x & 1\end{pmatrix} .
\end{equation}
Therefore, by using $\mathcal{F}(\bs{\theta})^{-1}=M \mathcal{F}(\bs{\phi})^{-1}M^T$, it may be shown that
\begin{equation} 
E_{\bs{\Theta}}(x) \geq \text{Tr}(\mathcal{F}(\bs{\theta})^{-1} )  = \frac{2-g(\alpha,\beta)x}{1-x^2} ,
\end{equation}
where $g(\alpha,\beta) \equiv \sin (2\alpha) + \sin(2\beta)$. 

For $g(\alpha,\beta) \neq 0$ and using the restriction that $ -1<x <1$, it may be confirmed that $E_{\bs{\Theta}}(x)$ is minimized by
\begin{equation}
x_{\min} = \frac{2-\sqrt{4 - g(\alpha,\beta)^2}}{g(\alpha,\beta)}.
\end{equation}  
This implies that $E_{\bs{\Theta}}(x)$ is minimized by some non-zero $x_{\min} $ for all valid $\alpha$ and $\beta$ for which $g(\alpha,\beta) \neq 0$ (as $0<g(\alpha,\beta)^2 < 2$). For those $\alpha$ and $\beta$ with $g(\alpha,\beta) =0$, the minimum is $x_{\min}=0$.


The state with $x=0$ is the optimal local estimation strategy for all values of $\alpha$ and $\beta$, and $x\neq 0$ implies an entangled probe state (for pure states). Therefore, for almost all $\alpha$ and $\beta$, the probe state which is optimal for estimating $\bs{\theta}$ (i.e., the state which minimizes $E_{\bs{\Theta}}$) is entangled. Hence, in most cases a global estimation strategy is optimal. There are certain special cases where the optimal probe state is separable: the optimal state for estimating $\theta_1= \phi_1$ and $\theta_2= \phi_2$, and for estimating $\theta_1 \propto \phi_1 + \phi_2$ and $\theta_2 \propto \phi_1 - \phi_2$, is the separable state given by $x=0$ (as in both cases  $g(\alpha,\beta) = 0$). These two cases are associated with an orthogonal $M$, and hence this is consistent with the results of the main text. 

\section{}
\label{QFIM-pss-inverse}

In this appendix we derive the QCRB bounds for sensor-symmetric states in a network in which a single parameter is encoded into each sensor with the generating operator $\hat{g}$ (Section~\ref{general-QCRB-ss}). In the particular case where each sensors is an optical mode, and $\hat{g}$ is the number operator $\hat{n}$, we obtain the formulas stated in Section~\ref{sec:optics}. In Section~\ref{QCRB-gnss} we then use the equations of Section~\ref{general-QCRB-ss} to derive QCRBs for estimation problems with the ``atomic and optical GNSs'' of Section~\ref{genGNSs}.

\subsection{QCRBs for symmetric states}
\label{general-QCRB-ss}
Consider a $D$-sensor sensing network in which a single parameter is encoded into each of the first $d$ sensors with the Hermitian generating operator $\hat{g}$, and where the remaining sensors are ancillas. Denote the unknown vector of parameters by $\bs{\theta}=(\theta_1,\dots,\theta_d)$. Consider a pure state $\Psi$ of the sensing network which is symmetric with respect to the exchange of any of the first $d$ sensors, but otherwise arbitrary. In this appendix we derive a simple equation for the QCRB for $\text{Var}(\Theta_i)$ and
\begin{equation} 
E_{\bs{\theta}} = \frac{1}{d}\sum_{i=1}^{d}\text{Var}(\Theta_i),
\label{scalrun}
\end{equation}
 for any such probe state $\Psi$, where $\Theta_i$ is the estimate of $\theta_i$. 

The symmetry condition on $\Psi$ implies that the QFIM for any such probe state is given by
\begin{equation} 
\mathcal{F} = 4\left( (v- c)  \mathds{1}+ c  \mathcal{I} \right),
 \end{equation}
where $\mathds{1}$ is the $d \times d$ identity matrix, $\mathcal{I}$ is the $d \times d$ matrix of all ones, $v$ is the variance of $\hat{g}$ in any of the first $d$ modes, and $c$ is the covariance of $\hat{g}$ between any of the first $d$ modes. Specifically
 \begin{align}
 v &= \langle g_i^2 \rangle - \langle g_i\rangle^2,\\
 c &= \langle g_ig_j \rangle - \langle g_i\rangle\langle g_j\rangle,
 \end{align}
for arbitrary $i,j \in [0,\dots, d]$ with $i\neq j$, where $\hat{g}_i$ denotes the generator acting on the $i$\textsuperscript{th} sensor. 

The inverse of any matrix of the form $M =  \lambda (\mathds{1} + \omega \mathcal{I}) $ is given by
 \begin{equation}
 M^{-1} = \frac{1}{\lambda} \left( \mathds{1} - \frac{\omega}{1+\omega d} \mathcal{I} \right),
 \end{equation}
for $\omega \neq -1/d$ and $\lambda \neq 0$. This may be confirmed directly by noting that $\mathcal{I}^2 = d \mathcal{I}$. Therefore, as long as $\mathcal{F}$ is invertible, its inverse is given by
\begin{equation}
 \mathcal{F}^{-1} = \frac{1}{4(v- c)} \left( \mathds{1} -  \frac{c}{v +(d-1)c } \mathcal{I} \right).
\label{p-qwer}
  \end{equation}
From the QCRB, the saturable bound on the scalar estimation uncertainty $E_{\bs{\Theta}}$ of Eq.~(\ref{scalrun}) is $E_{\bs{\Theta}} \geq \text{Tr}[\mathcal{F}^{-1} ] /\mu d$, and $\text{Var}(\Theta_i) \geq [\mathcal{F}^{-1} ]_{ii} /\mu d$. Using Eq.~(\ref{p-qwer}) and simple algebra, it may then be shown that
\begin{align}
E_{\bs{\theta}} \geq   \frac{ g(\Psi) }{4\mu v },\hspace{1cm} \text{Var}(\Theta_i) \geq   \frac{ g(\Psi) }{4\mu v },
\label{EVar--}
\end{align}
where $ g(\Psi)$ is a function of $d$ and $\mathcal{J} \equiv c/v$ (and hence, the probe state $\Psi$) defined by
\begin{equation} 
g(\Psi) = \frac{1+(d-2)\mathcal{J}}{(1-\mathcal{J})(1+(d-1)\mathcal{J})}. 
\end{equation}
This is the formula which is used in the main text. Specifically, to obtain Eq.~(\ref{Egpsi}) we let $d \to d'$. 


\subsection{QCRBs for GNSs}
\label{QCRB-gnss}
In Section~\ref{genGNSs}, and in particular Eq.~(\ref{eq:atom-opticsGES}), we introduced the $d$ probe sensor and 1 ancillary sensor entangled state
\begin{multline}
\ket{\Psi_{\textsc{gns}}(N)} = \frac{1}{\sqrt{d}}\big( \ket{\kappa_N,0, \dots,0} +  \ket{0,\kappa_N,\dots,0} +  \dots \\ \dots +  \ket{0,0,\dots,\kappa_N} \big),
\end{multline}
where $\ket{0}$ denotes the vacuum state, and $\ket{\kappa_N}= \ket{\lambda_{\max}(N)}$ if  $|\lambda_{\max}| \geq |\lambda_{\min}|$ and  $\ket{\kappa_N}=\ket{\lambda_{\min}(N)}$ otherwise. The notation used here is introduced in Section~\ref{sec:optics-atoms}, and this state is termed a generalized NOON state (GNS). As in the main text, assume that the $d$-dimensional $\bs{\phi}$ is imprinted onto a GNS via a unitary $U(-i\phi_i\hat{g}_i)$ on each probe sensor, where $\hat{g}_i$ is the generic generator with linearly-spaced eigenvalues defined in Section~\ref{sec:optics-atoms} (see Eq.~(\ref{gen-g})). We now derive the QCRBs for $E_{\bs{\Phi}}=\text{avg}_i[\text{Var}(\Phi_i)]$ and $\text{Var}(\Phi_i)$ with the $N_{\max}$ particles GNS. 

This state is symmetric with respect to the $d$ probe sensors, so we may use the formulas in Eq.~(\ref{EVar--}) for the QCRB for both of these quantities. To calculate these quantities we need the variance $v$, with respect to any probe sensor, and the covariance $c$, with respect to any pair of probe sensors, of the generating operator $\hat{g}$. It is easily confirmed that for the $N_{\max}$ particle GNS
\begin{align}
v &= \frac{ dN^2 \max\{\lambda_{\max}^2,\lambda_{\min}^2\}}{4(d+1)^2},
\\ c &= -\frac{N^2\max\{\lambda_{\max}^2,\lambda_{\min}^2\}}{4(d+1)^2}.
\end{align}
This implies that $\mathcal{J} = -1/d$, and hence  $g(\Psi_{\textsc{gns}}(N_{\max})) = 2d/(d+1)$. Using Eq.~(\ref{EVar--}) we then arrive at the precision bounds
\begin{align}
E_{\bs{\Phi}} &\geq    \frac{ d+1 }{\mu N^2\max\{\lambda_{\max}^2,\lambda_{\min}^2\}},\\
\text{Var}(\Phi_i) &\geq    \frac{ d+1 }{\mu N^2\max\{\lambda_{\max}^2,\lambda_{\min}^2\}}.
\end{align}
These are the bounds stated, or implicitly used, in Section~\ref{genGNSs}.

\section{}
\label{app-m>1}
In Section~\ref{m>1func} we analyzed a significant subset of those estimation problems in which the aim is to estimate some linear functions of $\bs{\phi}$. In particular, we have discussed the case of (1) estimating a single arbitrary linear function of $\bs{\phi}$, (2) estimating $d$ linear functions of $\bs{\phi}$. However, this leaves a range of cases which we have not addressed. Therefore, in this appendix we discuss the fully general case. Consider $\bs{\theta} = M\bs{\phi}$ where $M$ is an arbitrary invertible and real $d \times d$ matrix -- this is the most general set of linear (and linearly independent) functions of the $\phi_k$. Furthermore, consider a weighting matrix, $W$, with arbitrary weightings, and hence the figure of merit for the estimation problem is $E_{\bs{\Theta}} =\text{Tr}(W\text{Cov}(\Theta))$ for general $W$. Let us consider only some finite-dimensional sub-space of the entire Hilbert space for the sensing problem of interest, which we denote $\mathcal{B}$. For example, we may again consider our general atomic and optical sensing formalism and restrict ourselves to the ``$N_{\max}$ particles or fewer'' sub-space $\mathcal{S}(N_{\max})$. In this setting we can derive a generalization of the bounds in Eqs.~(\ref{ineq-f-1} -- \ref{ineq-f-4}).

In this general estimation problem we wish to estimate $d'\in [1,d]$ elements of $\bs{\theta}$, and each state in $\mathcal{B}$ is associated with a reduced vector $\tilde{\bs{\theta}}$, which is of dimension $s$ for some $s \in [d',d]$ (which depends on the state). Now for \emph{any} probe state $\psi$ in  $\mathcal{B}$, with an invertible QFIM for its reduced vector $\bs{\theta}$, $E_{\bs{\Theta}} $ is bounded by
\begin{align}
E_{\bs{\Theta}} 
&\geq \frac{1}{\mu}\sum_{k=1}^{s(\psi)} \tilde{W}_{kk} [\mathcal{F}(\tilde{\bs{\theta}})^{-1}]_{kk} ,  \label{ineq-f-1--}
\\&  \geq \frac{1}{\mu} \sum_{k=1}^{s(\psi)}  \frac{\tilde{W}_{kk}}{ \mathcal{F}(\tilde{\bs{\theta}})_{kk} }, \label{ineq-f-2--}
\\&= \frac{1}{\mu} \sum_{k=1}^{s(\psi)}  \frac{\tilde{W}_{kk}}{4 \text{Var} (\psi,\tilde{H}_{k}')},\label{ineq-f-3--}
\\& \geq \frac{1}{\mu} \min_{\Psi \in \tilde{\mathcal{B}}}\left[ \sum_{k=1}^{s(\Psi)}  \frac{\tilde{W}_{kk}}{4 \text{Var} (\Psi,\tilde{H}_{k}')} \right],  \label{ineq-f-4--}
\end{align}
where $\bs{\tilde{H}'}$ is the permutation of $\bs{\hat{H}'}= (M^{-1})^T \bs{\hat{H}}$ so that the ordering is consistent with the ordering of $\tilde{\bs{\theta}}$ obtained from the `reduction' process, $\tilde{\mathcal{B}}$ is the sub-space of $\mathcal{B}$ containing states for which the reduced QFIM is invertible, the dependence of $s$ on $\psi$ is to make it clear that the dimensionality of the reduced vector will depend on the probe state, and $\mu$ is the number of repeats of the experiment. 

Beyond those differences already noted, the derivation of these bounds is completely analogous to derivation of Eqs.~(\ref{ineq-f-1} -- \ref{ineq-f-4}). As before, if a state saturates all of these bounds it is guaranteed to be the probe state with the minimal $E_{\bs{\Theta}}$ for the estimation problem (whether it is also the optimal probe state per resource depends on what the resource is, and we have assumed nothing about resources here).

The bounds of Eqs.~(\ref{ineq-f-1--} -- \ref{ineq-f-4--}) are a fairly simple generalization of the bounds we used earlier (Eqs.~(\ref{ineq-f-1} -- \ref{ineq-f-4})), to find optimal strategies for estimating single linear functions. However, it is not clear how to draw any conclusions from them for the completely general problem we are considering here. Despite this, it is likely that they will be useful for analyzing particular cases of interest. 

\end{document}